\documentclass[a4paper,11pt]{article}
\pdfoutput=1 % to allow pdflatex compilation in JCAP
\usepackage{jcappub}
\usepackage{amsmath}
\usepackage{mathrsfs} 
\usepackage{amsfonts}
\usepackage{amssymb}
\usepackage{amsthm}
\usepackage{mathtools}
\usepackage{float}
\usepackage{graphicx}
\usepackage{latexsym}
\usepackage{xspace}
\usepackage{hyperref} 
\usepackage{bm}
\usepackage{relsize}
\usepackage{tabularx}
\usepackage{multirow}
\usepackage{mathdots}
\usepackage{amssymb}
\usepackage{wasysym}
\usepackage[normalem]{ulem}
\usepackage{lscape}
\usepackage[most]{tcolorbox}
\usepackage{braket}
\usepackage[utf8]{inputenc}
\usepackage[english]{babel}
\usepackage{slashed}
\usepackage{empheq}
\usepackage[makeroom]{cancel}

\usepackage[margin=1in,includefoot]{geometry}
\usepackage{xfrac} %Allows for slanted fractions
\usepackage{tikz-feynman}

\allowdisplaybreaks

\makeatletter
\gdef\@fpheader{}
\g@addto@macro\bfseries{\boldmath}
\makeatother

\makeatletter
\DeclareRobustCommand*\uell{{\mathpalette\@uell\relax}}
\newcommand*\@uell[2]{
	\setbox0=\hbox{#1\ell#1\ell}
	\setbox1=\hbox{\rotatebox{10}{#1\ell#1\ell}}
	\dimen0=\wd0 \advance\dimen0 by -\wd1 \divide\dimen0 by 2
	\mathord{\lower 0.1ex \hbox{\kern\dimen0\unhbox1\kern\dimen0}}
}

\newcommand{\OSlin}{{\widetilde{O}^{\mathcal{S}}_{\mathrm{lin}}}}
\newcommand{\OSnon}{{\widetilde{O}^{\mathcal{S}}_{\mathrm{NL}}}}
\newcommand{\KSlin}{{  \bar{\mathcal{K}}_{\mathcal{S},\hspace{0.3mm}\mathrm{lin}}^{(2)} }}
\newcommand{\KSnon}{{  \bar{\mathcal{K}}_{\mathcal{S},\hspace{0.3mm}\mathrm{NL}}^{(2)} }}
\newcommand{\Slin}{{ S^{\mathrm{lin}}_2 }}
\newcommand{\Snon}{{ S^{\mathrm{NL}}_2 }}
\newcommand{\SEE}{S_{\mathrm{EE}}}

\newcommand{\ie}{\textsl{i.e.}~}

\newcommand{\eg}{\textsl{e.g.}\xspace}

\newcommand{\etc}{\textsl{etc.}\xspace}

\newcommand{\order}[1]{\mathcal{O}\!\left(#1\right)}

\newcommand{\dd}{\mathrm{d}}
\newcommand{\ee}{e}

\newcommand{\sss}[1]{{\scriptscriptstyle{#1}}}

\newcommand{\uPl}{\mathrm{Pl}}
\newcommand{\uin}{\mathrm{in}}

\newcommand{\usssPl}{\sss{\uPl}}

\newcommand{\Rea}{\Re \mathrm{e}\,}

\newcommand{\cs}{c_{_\mathrm{S}}}

\newcommand{\Mp}{M_\usssPl}

\newcommand{\efolds}{$e$-folds~}

\newcommand{\beq}{\begin{equation}}
	\newcommand{\eeq}{\end{equation}}
\newcommand{\bea}{\begin{equation}\begin{aligned}}
		\newcommand{\eea}{\end{aligned}\end{equation}}
\newlength{\wsingfig}
\newlength{\wdblefig}
\newlength{\wquadfig}
\newlength{\wtriplefig}
\newcommand{\Eq}[1]{Eq.~(\ref{#1})}
\newcommand{\Eqs}[1]{Eqs.~(\ref{#1})}
\newcommand{\Fig}[1]{Fig.~{\ref{#1}}}
\newcommand{\Figs}[1]{Figs.~{\ref{#1}}}
\newcommand{\Refa}[1]{Ref.~{\cite{#1}}}
\newcommand{\Refs}[1]{Refs.~{\cite{#1}}}
\newcommand{\Sec}[1]{Sec.~\ref{#1}}

\newcommand{\App}[1]{Appendix~\ref{#1}}

\newcommand{\bs}[1]{\boldsymbol{#1}}

\usepackage{dsfont}

\newcommand{\exd}{{\hbox{d}}}

\def\bmx{{\boldsymbol{x}}}

\def\bmk{{\boldsymbol{k}}}
\def\bmp{{\boldsymbol{p}}}
\def\bmq{{\boldsymbol{q}}}

\newcommand*{\Scale}[2][4]{\scalebox{#1}{$#2$}}

\subheader{}

\title{In-in formalism for the entropy of quantum fields in curved spacetimes}

\author[a,b]{Thomas Colas,}
\author[b]{Julien Grain,}
\author[c,d]{Greg Kaplanek,}
\author[e]{Vincent Vennin}

\affiliation[a]{Department of Applied Mathematics and Theoretical Physics, University of Cambridge, Wilberforce Road, Cambridge, CB3 0WA, UK}

\affiliation[b]{Universit\'e Paris-Saclay, CNRS, Institut d'Astrophysique Spatiale, 91405, Orsay, France}

\affiliation[c]{Theoretical Physics, Blackett Laboratory, Imperial College, London, SW7 2AZ, UK}

\affiliation[d]{Perimeter Institute for Theoretical Physics, Waterloo, Ontario, N2L 2Y5, Canada}

\affiliation[e]{Laboratoire de Physique de l'\'Ecole Normale Sup\'erieure, ENS, Universit\'e PSL, CNRS, Sorbonne Universit\'e, Universit\'e Paris Cit\'e, F-75005 Paris, France}

\begin{document}
	\sloppy

	\abstract{We show how to compute the purity and entanglement entropy for quantum fields in a systematic perturbative expansion. To that end, we generalize the in-in formalism to non-unitary dynamics (\ie accounting for the presence of an environment) and to the calculation of quantum information measures, which are not observables in the usual sense. This allows us to reduce the problem to one involving standard correlation functions, and to organize their computation in a diagrammatic expansion for which we construct the corresponding Feynman rules. As an illustration, we apply the formalism to a cosmological setting inspired by the effective field theory of inflation. We find that at late times, non-linear loop corrections share the same time behavior as the linear contribution, and only yield a slight redressing of the purity. In particular, when the environment is heavy compared to the Hubble scale, the phenomenon of recoherence previously encountered is robust to the class of non-linear extensions considered. Bridging the gap between perturbative quantum field theory and open quantum systems paves the way to a better understanding of renormalization and resummation in open effective field theories. It also enables a more systematic exploration of quantum information properties in field theoretic settings.
}
 		
	\maketitle
	
	\section{Introduction}
	
The interplay between quantum information theory and quantum field theory has raised an ever-increasing interest in recent years~\cite{Koks:1996ga, Rosenhaus:2014woa, Boyanovsky:2018fxl, Cheung:2023hkq, Subba:2024mnl}. Being able to describe and quantify quantum correlations for relativistic fields is of particular importance in cosmology, where all structures are expected to arise from the gravitational amplification of vacuum quantum fluctuations during an era of accelerated expansion called inflation. How those fluctuations acquire classical properties remains an open issue~\cite{Polarski:1995jg, Kiefer:1998qe, Kiefer:2006je, Kiefer:2008ku, Sudarsky:2009za, Burgess:2014eoa, Martin:2015qta, Martin:2021znx, Chandran:2023ogt}, together with the possibility of genuine signatures of the quantum origin of cosmic structures~\cite{Campo:2005sv, Maldacena:2015bha, Martin:2016tbd, Choudhury:2016cso, Choudhury:2016pfr, Martin:2017zxs, Ando:2020kdz, Espinosa-Portales:2022yok, Tejerina-Perez:2024opu, Sou:2024tjv}. A thorough investigation of these aspects requires a better characterization of the entanglement acquired by quantum fields in accelerating space-times~\cite{Fukuma:2013uxa, Choudhury:2017qyl, Boyanovsky:2018soy, Akhtar:2019qdn, Kaplanek:2020iay, Martin:2021xml, Martin:2021qkg, Kaplanek:2021fnl, Brahma:2021mng, Brahma:2023hki, Belfiglio:2023moe}, as well as an understanding of the role played by quantum decoherence in cosmology~\cite{Brandenberger:1990bx, Barvinsky:1998cq, Lombardo:2004fr, Lombardo:2005iz, Martineau:2006ki, Prokopec:2006fc, Burgess:2006jn, Sharman:2007gi, Campo:2008ju, Anastopoulos:2013zya, Nelson:2016kjm, Martin:2018zbe, Martin:2018lin, Danielson:2022tdw, Danielson:2022sga, Oppenheim:2022xjr, Colas:2022kfu, DaddiHammou:2022itk, Burgess:2022nwu, Sharifian:2023jem, Ning:2023ybc, Biggs:2024dgp, Danielson:2024yru}.

Decoherence~\cite{Zurek:1981xq, Zurek:1982ii, Joos:1984uk} is the process by which, when a system couples to an environment that remains observationally inaccessible, the reduced state of the system transits from a pure state to a statistical mixture where quantum coherence between the pointer states is lost. When a quantum field is coupled to an environment, it is understood as an open system undergoing non-unitary evolution. Describing this evolution necessitates the development of open quantum field theory~\cite{breuerTheoryOpenQuantum2002, Burgess:2007pt, Burgess:2022rdo, Colas:2023wxa, Colas:2024lse}. These tools (influence functionals, master equations and their stochastic unravelling, \etc) have been applied to cosmological or black hole backgrounds in various works, see \eg \Refs{Boyanovsky:2015xoa, Boyanovsky:2015jen, Boyanovsky:2015tba, Hollowood:2017bil, Shandera:2017qkg, Choudhury:2018ppd, Burrage:2018pyg, Cheung:2018cwt, Brahma:2020zpk, Rai:2020edx, Burgess:2021luo, Banerjee:2021lqu, Brahma:2022yxu, Kaplanek:2022xrr, Kaplanek:2022opa, Burgess:2022rdo, Cao:2022kjn, Prudhoe:2022pte, Kading:2022jjl, Kading:2022hhc, Colas:2023wxa, Alicki:2023tfz, Alicki:2023rfv, Kading:2023mdk, Creminelli:2023aly, Pelliconi:2023ojb, Colas:2024xjy, Keefe:2024cia, Bowen:2024emo, Bhattacharyya:2024duw, Salcedo:2024smn, Belfiglio:2024qsa, Colas:2024lse}. They have been used to efficiently resum secular late-time effects in a number of cases \cite{Burgess:2015ajz, Kaplanek:2019vzj, Kaplanek:2019dqu, Colas:2022hlq, Chaykov:2022zro, Burgess:2024eng}, which makes them particularly well-suited to the description of inflationary fluctuations.

In practice however, these methods often rely on approximation schemes (\eg Markovianity) that substantially differ from the traditional perturbative framework employed in asymptotically non-flat space-times, known as the in-in formalism. In this setup, a systematic expansion in the system-environment coupling is performed, which allows one to rely on the whole suite of Quantum Field Theory (QFT) perturbative tools, such as Feynman diagrams, renormalization methods, \etc. So far the in-in formalism has been developed for unitary theories mostly~\cite{Donath:2024utn} (see however \Refa{Salcedo:2024smn}), hence it cannot be directly employed in open QFTs. In particular, quantum purity, which measures how much a quantum field decoheres, cannot be computed using existing in-in results.

The goal of the present work is to bridge this gap, and show how quantum information measures of an open quantum field can be computed in the in-in approach. This has several advantages. First, it allows one to rely on the perturbative QFT tools mentioned above, which greatly help in the treatment of non-linearities and the associated divergences. Second, being expressed in the language of cosmological correlators, it makes it possible to relate quantum information properties to summary statistics of the system. Third, in order to better understand resummation in open QFTs (and what exactly is being resummed), perturbative evolution should be understood as a first step.

There are two main obstacles when dealing with quantum purity or other entropy measures in the in-in framework. First, as already mentioned, in-in methods have been developed in the context of unitary theories, whereas the loss of purity is inherently a non-unitary phenomenon. Second, the in-in framework is constructed to deliver correlation functions, or more generally the expectation value of observables. However, because purity is not an observable\footnote{Purity cannot be expressed as the expectation value of a Hermitian function of the phase-space operators and as such it would require a full state tomography to be assessed \cite{breuerTheoryOpenQuantum2002}.}, extending the formalism becomes necessary.\\

In practice, the reduced state of a system $\mathcal{S}$ is obtained by partially tracing the full density matrix $\widehat{\rho}$ over the Hilbert space of the environment $\mathcal{E}$,
	\begin{equation} \label{red_rho_def}
	\widehat{\rho}_{\mathrm{red}} \equiv \mathrm{Tr}_{\mathcal{E}} \left(\widehat{\rho}\right) . 
	\end{equation}
In the cosmological context, the system is usually comprised of a set of observable degrees of freedom, \eg the curvature perturbation $\zeta$ on large scales, while the environment can encompass the set of unobservable modes of $\zeta$ concealed behind the de-Sitter horizon~\cite{Brahma:2020zpk, Burgess:2022nwu, Brahma:2023hki}, separate regions in physical space \cite{Martin:2021qkg}, or other matter fields (as we later consider in this work as an application). The computation of correlation functions of system operators relies on the reduced density matrix only, and since $\widehat{\rho}_{\mathrm{red}}$ is Hermitian and trace normalized it can always be decomposed as
	\begin{equation}
	\widehat{\rho}_{\mathrm{red}} = \sum_{j} p_j |\psi_j \rangle \langle \psi_j |
	\end{equation}
where $\{ | \psi_j \rangle  \}$ are pointer states and $p_j$ are non-negative coefficients that sum up to $1$. If the system is in a pure state, all $p_j$ coefficients vanish except one, and $\widehat{\rho}_{\mathrm{red}} = \widehat{\rho}_{\mathrm{red}}^2$ is a projector. Otherwise, $\widehat{\rho}_{\mathrm{red}}$ describes a statistical mixture. This arises when the system becomes entangled with the environment, and one possible measure of that entanglement is the quantum purity~\cite{Serafini:2003ke}
	\begin{equation} \label{purity_gamma_def}
	\gamma \equiv \mathrm{Tr}_{\mathcal{S}} \left(\widehat{\rho}_{\mathrm{red}}^2\right)  .
	\end{equation}
	If the system's Hilbert space has finite dimension $D$, then $1/D\leq \gamma\leq 1$. Pure states have $\gamma=1$, and maximal decoherence corresponds to $\gamma\to 1/D$, which simply means $\gamma \to 0$ in the case of a quantum field. Let us stress that purity is not only a relevant measure of the system's decoherence, it also captures the importance of non-unitary effects and thus signals the need to go beyond unitary treatments. Indeed, if $\widehat{\rho}_{\mathrm{red}}$ evolves unitarily, 
	\begin{align} \label{purity_conserved}
	\frac{\dd \widehat{\rho}_{\mathrm{red}}}{\dd t} = - i \left[\widehat{H}, \widehat{\rho}_{\mathrm{red}}\right]
	\end{align}
where $\widehat{H}$ is a self-adjoint Hamiltonian, then the purity is conserved by cyclicity of the trace (and likewise for any other entropy tracer). Deviations of the purity from 1 can thus be seen as a measure of departure from \Eq{purity_conserved}. 
	
	\paragraph{Summary of the main results:} In this paragraph we provide a summary of the main results. While not all relevant quantities and concepts are fully defined here -- detailed definitions will be provided later -- the goal is to give a first flavour of the results derived in this work. A more formal summary is presented in the concluding section.
	
Consider an interaction Hamiltonian of the form
	\begin{align}
		g \widehat{H}_{\mathrm{int}}(\eta) = g \int  \frac{\dd^3\bmk}{(2\pi)^3} \;   \widehat{\mathcal{O}}^{\mathcal{S}}_\bmk(\eta) \otimes \widehat{\mathcal{O}}^{\mathcal{E}}_{-\bmk}(\eta)
	\end{align} 
	made of (possibly non-linear) system and environment operators $\widehat{\mathcal{O}}_{\mathcal{S}}$ and $\widehat{\mathcal{O}}_{\mathcal{E}}$ and controlled by a coupling constant $g$. The Hamiltonian is expanded in Fourier space where the $\bm{k}$ and $-\bm{k}$ arguments merely result from the isotropy of the background, and $\eta$ labels time. We perform a systematic expansion of the purity in powers of the interaction Hamiltonian,
	\begin{align}
		\gamma = \sum_{n=0}^{\infty} g^n \gamma^{(n)}.
	\end{align}
Assuming that the system is initially in a pure state, one has $\gamma^{(0)} = 1$, and the first non-unitary correction reads
	\begin{align}
	\label{eq:gamma2:intro}
		\gamma^{(2)}(\eta) &= - 2 \int_{-\infty}^\eta \dd \eta_1  \int_{-\infty}^{\eta} \dd \eta_2 \int  \frac{\dd^3\bmk_1}{(2\pi)^3} \int  \frac{\dd^3\bmk_2}{(2\pi)^3}
		\; \bar{\mathcal{K}}_{\mathcal{S}}^{(2)}(\bmk_1,\bmk_2, \eta_1,\eta_2) \bar{\mathcal{K}}_{\mathcal{E}}^{(2)}(\bmk_1,\bmk_2, \eta_1,\eta_2) \, .
	\end{align}
It involves the unequal-time, centred, two-point functions
	\begin{align}
		\bar{\mathcal{K}}_{\mathcal{S}/\mathcal{E}}^{(2)}(\bmk_1,\bmk_2, \eta_1,\eta_2) &\equiv \left<\widetilde{\mathcal{O}}^{\mathcal{S}/\mathcal{E}}_{\pm\bmk_1} (\eta_1)  \widetilde{\mathcal{O}}^{\mathcal{S}/\mathcal{E}}_{\pm\bmk_2} (\eta_2) \right>- \left< \widetilde{\mathcal{O}}^{\mathcal{S}/\mathcal{E}}_{\pm\bmk_1} (\eta_1)  \right>  \left<   \widetilde{\mathcal{O}}^{\mathcal{S}/\mathcal{E}}_{\pm\bmk_2} (\eta_2) \right> 
	\end{align}
that encode the system and environment statistics in the free theory. 

If $\widehat{\mathcal{O}}_{\mathcal{S}}$ is linear in the phase-space fields of the system, the above reduces to
	\begin{align}
	\label{eq:gamma2:Gauss:intro}
		\gamma^{(2)}(\eta) = - 4  {\int_{\bmk\in\mathbb{R}^{3+}}\dd^3\bmk} \; \mathrm{det}^{(2)} \bs{\mathrm{Cov}}(k,\eta)\, ,
	\end{align}
where $\mathrm{det}^{(2)} \bs{\mathrm{Cov}}(k,\eta)$ is the determinant of the covariance matrix at second order in $\widehat{H}_{\mathrm{int}}$. We will show that this formula remains correct even when $\widehat{\mathcal{O}}_{\mathcal{S}}$ is quadratic in the phase-space variables (up to an additional factor $1/2$ in the right-hand side), in spite of the system being in a non-Gaussian state in that case. This is because, as we will find, the perturbation to the determinant of the covariance matrix can be related to that of the expectation value of the particle number operator. The expression~\eqref{eq:gamma2:Gauss:intro} is also independent of the form of $\widehat{\mathcal{O}}_{\mathcal{E}}$, hence it holds true even in the presence of highly non-linear environment operators.

\begin{figure}[tbp]
		\centering
		\includegraphics[width=0.8\textwidth]{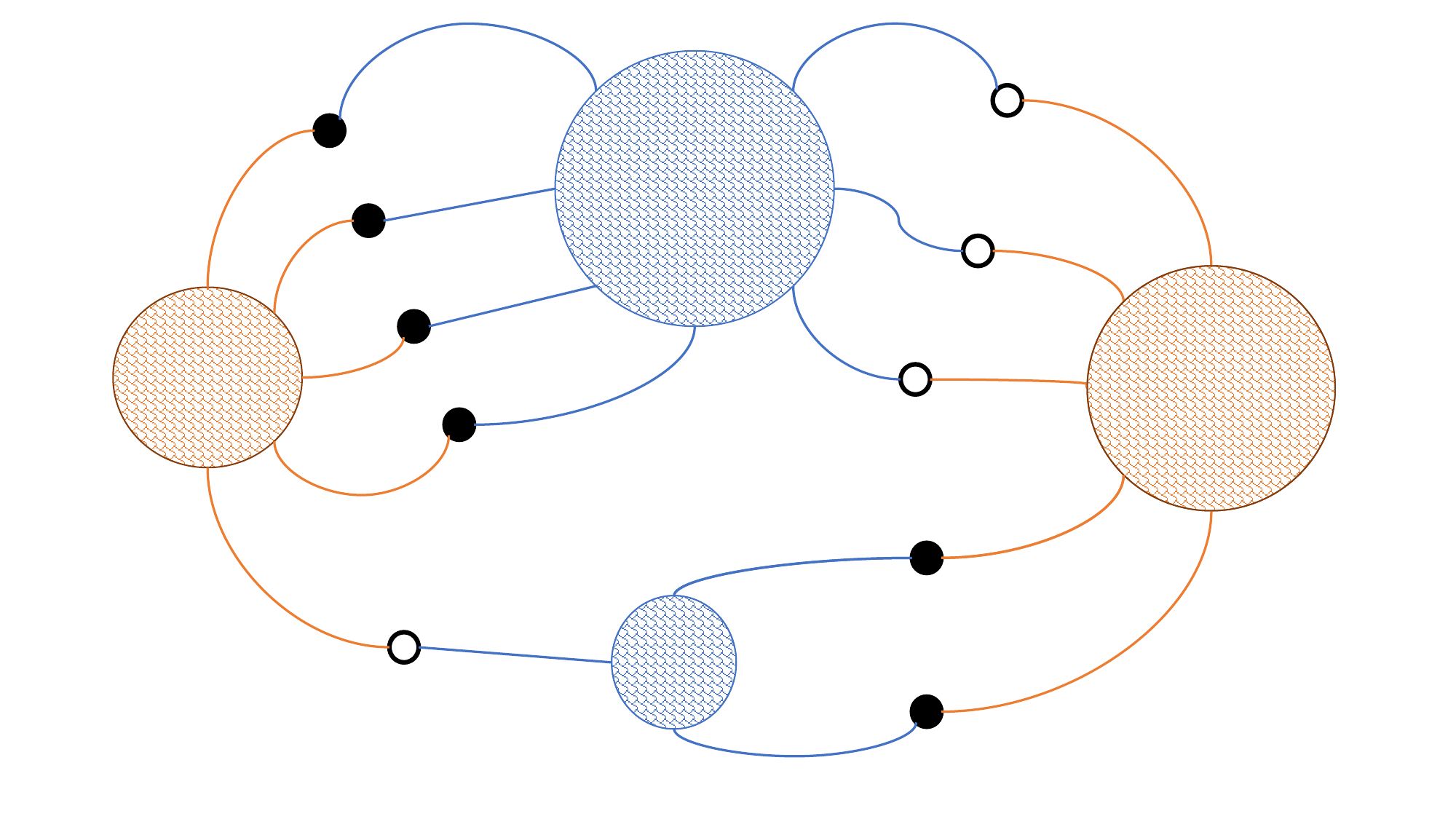}
		\caption{Example of a diagram appearing in the computation of $\gamma^{(n)}$. Blue and orange blobs represent unequal-time correlators of the system and the environment respectively. To each conversion vertex $\Circle, \CIRCLE$ is associated a power of the coupling constant $g$, an integral over time and over three-momenta, and the insertion of $\widehat{\mathcal{O}}^{\mathcal{S}}_{\bmk_i}(\eta_i) \otimes \widehat{\mathcal{O}}^{\mathcal{E}}_{-\bmk_i}(\eta_i)$.}
		\label{fig:gammaintro}
	\end{figure}

The formula~\eqref{eq:gamma2:intro} can be generalized to arbitrary order $n$, where we will find $n$-nested time integrals involving products of unequal-time correlators of both the system and the environment. Their calculation can be assisted with a set of Feynman rules to compute diagrams representing the various terms appearing in $\gamma^{(n)}$. An example of such diagrams is given in \Fig{fig:gammaintro}, the detailed meaning of which will become clearer below.
	
	We will then apply these techniques to a model inspired by the EFT of Inflation (EFToI) construction~\cite{Cheung:2007st}, in which the pseudo-Goldstone mode $\pi_c$ interacts with a hidden scalar sector $\sigma$ through~\cite{Tolley:2009fg,Chen:2009zp,Assassi:2013gxa,Jazayeri:2022kjy,Jazayeri:2023xcj}     
	\begin{align}
	S_{\mathrm{int}} & = - \int \dd \eta\; \dd^3\bm{x} \; g \left[ a^3 \pi'_c \sigma  - \frac{1}{\Lambda_1} a^2 (\partial_i \pi_c)^2 \sigma - \frac{1}{\Lambda_2} a^2 \pi_c^{\prime 2} \sigma \right] ,
	\end{align} 
	where $a$ is the scale factor and $g$,  $\Lambda_1$ and $\Lambda_2$ are Wilson coefficients. At super-Hubble scales (\ie in the late-time limit) we will find that the purity of a given mode $k$ is given by
	\begin{align}
		\gamma^{(2)}_{\mathrm{NL}}(k,\eta) = \gamma^{(2)}_{\mathrm{lin}}(k,\eta) \left( 1 - \frac{1}{(2\pi)^9} \frac{3}{8} \frac{\Delta_{\zeta}^2}{\cs^4} \right)
	\end{align}
	with $\cs $ the speed of sound and $\Delta_{\zeta}^2 \sim 10^{-9}$ the amplitude of the primordial power spectrum. The second term in the square brackets corresponds to non-linear contributions (\ie loop corrections), which are found to scale in time in the exact same way as the linear contribution $\gamma^{(2)}_{\mathrm{lin}}$, yielding only a slight dressing of the late-time purity. In particular, if the environmental field $\sigma$ is heavy compared to the Hubble mass, we will find that the system recoheres at late time, and that the non-linear interactions considered in this work do not jeopardize the recoherence process. In the opposite limit where $\sigma$ is light, we will find that $\gamma^{(2)}_{\mathrm{lin}}$ scales as $a^2$, and decoherence is very efficient.
		
	\paragraph{Outline:} The article is organized as follow. In \Sec{sec:methods}, we develop the in-in formalism to compute quantum information properties of cosmological fields. In \Sec{subsec:inin}, we present the perturbative expansion underlying the in-in treatment. In \Sec{subsec:pp}, we apply it to the computation of the system's purity. We introduce the concept of spectral purity in \Sec{sec:Spectral:Purity}, to deal with the case where non-linear interactions make different Fourier modes couple, and the reduced density matrix does not factorize in Fourier space. We consider the particular case of Gaussian systems in \Sec{sec:Gaussian:Purity}, where we show that the relation between the purity and the covariance matrix also holds for quadratic interactions at leading order. After commenting on the role of resummation in \Sec{sec:resummation}, we derive Feynman diagrammatic rules in \Sec{subsubsec:NLO}. Lastly, in \Sec{sec:vN} we take a brief detour to discuss other measures of mixedness, specifically the entanglement entropy. We then apply these techniques to a model of phenomenological interest in \Sec{sec:applic}, which provides a non-linear extension of the setup considered in \Refa{Colas:2022kfu}. In \Sec{subsec:model}, we motivate the model from an EFT perspective and discuss its regime of validity. In \Sec{subsec:ppapp}, we compute the second-order perturbative purity and study different cases of interest. In particular, we find that recoherence occurs if the environment is heavy compared to the Hubble scale, and that this conclusion is not altered by the presence of non-linearities. We discuss various prospects this work opens in \Sec{sec:outlook} and present our main conclusions in \Sec{sec:conclu}, which is followed by a few appendices to which some technical details are deferred. 
	
	\section{Methods}\label{sec:methods}
	
Quantum purity is a measure of decoherence, or ``mixedness'', of a given [possibly reduced, see \Eq{red_rho_def}] quantum state $\widehat{\rho}_{\mathrm{red}}$. It is not an observable in the sense of the definition given in \Eq{purity_gamma_def}, since it is clear that it cannot be written as $\mathrm{Tr}_{\mathcal{S}} [\widehat{\rho}_{\mathrm{red}} \widehat{O}]$ for some Hermitian operator $\widehat{O}$. This is why the standard tools of the in-in formalism cannot be used directly to compute the purity. However, they can be extended, which is the goal of this section. 
		
	\subsection{In-in formalism}\label{subsec:inin}
	
	The \textit{in-in formalism} \cite{Weinberg:2005vy, Weinberg:2006ac, Chen:2017ryl} is a well-established technique aiming at computing perturbatively the expectation value of a given equal-time quantum operator $\langle \widehat{Q} (\eta) \rangle$ where
	\begin{align}
	\label{eq:Q:O}
		\widehat{Q} (\eta) \equiv \widehat{\mathcal{O}}_1(\eta,\bmx_1) \widehat{\mathcal{O}}_2(\eta,\bmx_2) \cdots \widehat{\mathcal{O}}_n(\eta,\bmx_n) \, .
	\end{align}
	The operators $\widehat{\mathcal{O}}_i(\eta,\bmx_i)$ are constructed locally out of the field operators appearing in the Lagrangian, and the expectation value is taken with respect to the initial Bunch-Davies vacuum state $\ket{\mathrm{BD}}$. Given an interaction Hamiltonian $g \widehat{H}_{\mathrm{int}}$, the expectation value $\langle \widehat{Q} (\eta) \rangle$ can be written in the interaction picture 
	\begin{align}\label{eq:ininstart}
		\left< \widehat{Q} (\eta) \right> &= \bra{\mathrm{BD}} \left[\overline{\mathcal{T}} \ee^{i g \int_{-\infty(1-i\epsilon)}^\eta \dd{\eta}' \;  \widetilde{H}_{\mathrm{int}}({\eta}')}\right]\widetilde{Q} (\eta) \left[ \mathcal{T}\ee^{-i g \int_{-\infty(1+i\epsilon)}^\eta \dd{\eta}'\; \widetilde{H}_{\mathrm{int}}({\eta}') }\right]\ket{\mathrm{\mathrm{BD}}}
	\end{align}
	where $\mathcal{T}$ and $\overline{\mathcal{T}}$ represent time and anti-time ordering respectively and tildes denote quantities in the interaction picture, which we introduce properly below in \Eqs{eq:rho:Hint} and \eqref{eq:Hintintpic}. The $i \epsilon$ deformation is here to ensure the projection of the adiabatic vacuum of the interacting theory onto the vacuum of the free theory in the asymptotic past, see \Refs{Adshead:2009cb,Kaya:2018jdo, Albayrak:2023hie} for in-depth discussions. In practice, the correlation functions of the theory are computed perturbatively, at a given order in $\widehat{H}_{\mathrm{int}}$. Defining  $\langle \widehat{Q} (\eta) \rangle^{(n)}$ the $n^{\mathrm{th}}$ order of the expansion, one has
	\begin{align}\label{eq:pertinin}
		\left< \widehat{Q} (\eta) \right>^{(n)} &=(ig)^n  \int_{-\infty}^{\eta}\dd \eta_n\int_{-\infty}^{\eta_n}\dd \eta_{n-1} \cdots \int_{-\infty}^{\eta_{2}}\dd \eta_1 \\
		& \bra{\mathrm{BD}}\left[\widetilde{H}_{\mathrm{int}}(\eta_1),\left[\widetilde{H}_{\mathrm{int}}(\eta_{2}), \cdots \left[\widetilde{H}_{\mathrm{int}}(\eta_n),\widetilde{Q} (\eta) \right]\cdots \right] \right]\ket{\mathrm{BD}}  \nonumber
	\end{align}
	where the appropriate $i \epsilon$ prescription must be used depending on which branch of \Eq{eq:ininstart} provides the time integration. This approach has been used in a variety of problems in primordial cosmology to compute cosmological correlators of scalar and tensor perturbations.
	
	\paragraph{Derivation:} Let us review the derivation of the above expressions. In a unitary setting, the wavefunction $\ket{\Psi}$ and the density matrix $\widehat{\rho}=\ket{\Psi}\bra{\Psi}$ respectively obey the Schr\"odinger equation and the Liouville-von-Neumann equation,
\begin{align}\label{eq:eom:Schrodinger}
		\frac{\dd \ket{\Psi(\eta)} }{\dd \eta} = - i \widehat{H}({\eta})\ket{\Psi(\eta)} \quad  \text{and} \quad  \frac{\dd \widehat{\rho}(\eta) }{\dd \eta} = - i \left[\widehat{H}({\eta}), \widehat{\rho}(\eta)\right]\, ,
\end{align}
where $\widehat{H}({\eta})$ is the Hamiltonian. These equations can be solved formally as
	\begin{align}\label{eq:evolgen}
		\ket{\Psi(\eta)} = \widehat{\mathcal{U}}(\eta,- \infty)  \ket{\mathrm{BD}} \quad \quad \text{and} \quad \quad
		\widehat{\rho}(\eta) =  \widehat{\mathcal{U}}(\eta,- \infty)  \ket{\mathrm{BD}} \bra{\mathrm{BD}} \widehat{\mathcal{U}}^\dag(\eta,- \infty)\, ,
	\end{align}
where $\ket{\mathrm{BD}}$ is the initial Bunch-Davies vacuum state and we have introduced the evolution operator 
	\begin{align}
	\label{eq:U:H}
		\widehat{\mathcal{U}}(\eta, - \infty)= \mathcal{T} \exp\left[{-i \int_{- \infty}^\eta \dd{\eta}'\;  \widehat{H}({\eta}')}\right] .
	\end{align}
This is the so-called Schr\"odinger picture, where the state (\ie $\ket{\Psi}$ and $ \widehat{\rho}$) evolve with $\widehat{\mathcal{U}}$ and observables $\widehat{O}$ evolve only through their explicit dependence on time, if any.
	
We the  divide the Hamiltonian into a free part and an interaction part,
	\begin{align}
		\widehat{H}(\eta)= \widehat{H}_0(\eta) +  g \widehat{H}_{\mathrm{int}} (\eta)
	\end{align}
	and introduce the free evolution operator $\widehat{\mathcal{U}}_0$, defined as in \Eq{eq:U:H} where $\widehat{H}$ is replaced by $\widehat{H}_0$. It is often convenient to work in the so-called interaction picture to perform perturbative expansions. In this picture, quantum states evolve with the interaction Hamiltonian $g\widehat{H}_{\mathrm{int}}$ and operators evolve with the free Hamiltonian $\widehat{H}_0$. The link between the Schr\"odinger and the interaction picture is given by 
	\begin{align}
		\label{eq:rho:Hint}
		\widetilde{\rho}(\eta) = \widehat{\mathcal{U}}^{\dag}_0(\eta,-\infty) \widehat{\rho}(\eta)\widehat{\mathcal{U}}_0(\eta,-\infty)  \, , 
	\end{align}
where tildes denote quantities evaluated in the interaction picture. From \Eq{eq:eom:Schrodinger} it is easy to show that the state evolves according to
	\begin{align}\label{eq:eom:IntPict}
		\frac{\dd \widetilde{\rho}}{\dd \eta}=
		-ig \left[ \widetilde{H}_{\mathrm{int}}(\eta),\widetilde{\rho}(\eta) \right] \, ,
	\end{align}
	where
	\begin{align}\label{eq:Hintintpic}
		\widetilde{H}_{ \mathrm{int}} (\eta) = \widehat{\mathcal{U}}^{\dag}_0(\eta,-\infty) \widehat{H}_{\mathrm{int}}(\eta) \widehat{\mathcal{U}}_0(\eta,-\infty)
	\end{align}
	and we have used that the evolution operator is Hermitian, \ie $\widehat{\mathcal{U}}_0 \widehat{\mathcal{U}}_0^\dag = \widehat{\mathcal{U}}_0^\dag \widehat{\mathcal{U}}_0 = \mathrm{Id}$, since $\widehat{H}_0=\widehat{H}_0^\dagger$. One can then introduce a perturbative scheme, where 
	\begin{align}\label{eq:gexp:rho}
		\widetilde{\rho} = \sum_{n=0}^\infty g^n \widetilde{\rho}^{(n)}
	\end{align}	
	with $\widetilde{\rho}^{(0)}=\ket{\mathrm{BD}}\bra{\mathrm{BD}}$. By inserting \Eq{eq:gexp:rho} into \Eq{eq:eom:IntPict} and identifying terms of the same order in $g$, one finds
	\begin{align}
		\frac{\dd \widetilde{\rho}^{(n)}}{\dd \eta}=
		-ig \left[ \widetilde{H}_{\mathrm{int}}(\eta),\widetilde{\rho}^{(n-1)}(\eta) \right] \,.
	\end{align}
	This can be solved iteratively as
	\begin{align}
		\label{eq:rho:iterative:SPT}
		\widetilde{\rho}^{(n)}(\eta) &= (-i)^n \int_{-\infty}^{\eta}\dd \eta_1\int_{-\infty}^{\eta_1}\dd \eta_2 \cdots \int_{-\infty}^{\eta_{n-1}}\dd \eta_n  \\
		&             \left[\widetilde{H}_{\mathrm{int}}(\eta_1),\left[ \widetilde{H}_{\mathrm{int}}(\eta_2),\cdots \left[\widetilde{H}_{\mathrm{int}}(\eta_n),\ket{\mathrm{BD}}\bra{\mathrm{BD}} \right] \cdots \right] \right].\nonumber 
	\end{align}
	One can use this expression to compute $\langle \widehat{Q} (\eta) \rangle^{(n)}$ from Eq.~(\ref{eq:pertinin}) since expectation values are computed in the interaction picture via
	\begin{align}
		\left<\widehat{Q}(\eta)\right>=\mathrm{Tr}\left[\widetilde{Q}(\eta)\widetilde{\rho}(\eta)\right].
	\end{align}
        Inserting \Eq{eq:rho:iterative:SPT} into this expression and making iterative use of the relation $\mathrm{Tr}(\widetilde{A}[\widetilde{B},\widetilde{C}])= \mathrm{Tr}([\widetilde{A},\widetilde{B}]\widetilde{C})$  (a mere consequence of the trace cyclicity) one recovers \Eq{eq:pertinin}.     
	
	\subsection{Perturbative purity}\label{subsec:pp}
	
	Let us now partition the physical setup into two subsystems, $\mathcal{S}$ (system) and $\mathcal{E}$ (environment), and consider the reduced density matrix $\widehat{\rho}_{\mathrm{red}} = \mathrm{Tr}_{\mathcal{E}}\left(\widehat{\rho}\right)$, see \Eq{red_rho_def}, and its purity parameter $\gamma = \mathrm{Tr}_{\mathcal{S}}\left(\widehat{\rho}_{\mathrm{red}}^2\right)$, see \Eq{purity_gamma_def}. Our goal is to compute this quantity in the perturbative scheme introduced above. We assume that the free Hamiltonian does not entangle the system and the environment, \ie that it is of the form\footnote{Note that this amounts to treating linear mixing between system and environment as an interaction, which is indeed commonly done in the context of primordial cosmology, see \eg \Refa{Chen:2017ryl}.}
	\begin{align}
		\widehat{H}_0 = \widehat{H}_{\mathcal{S}} \otimes \mathrm{Id}_{\mathcal{E}} + \mathrm{Id}_{\mathcal{S}} \otimes \widehat{H}_{\mathcal{E}}\, .
	\end{align}
	In this case the free evolution operator factorizes as $\widehat{\mathcal{U}}_0 = \widehat{\mathcal{U}}_{0, \mathcal{S}}\otimes\widehat{\mathcal{U}}_{0, \mathcal{E}}$, and using partial trace cyclicity this implies that 
	\begin{align}
		\widetilde{\rho}_{\mathrm{red}}(\eta)=  \widehat{\mathcal{U}}^{\dag}_{0,\mathcal{S}}(\eta,-\infty) \widehat{\rho}_{\mathrm{red}}(\eta)\widehat{\mathcal{U}}_{0,\mathcal{S}}(\eta,-\infty) .
	\end{align}
As a consequence, \Eq{purity_gamma_def} leads to
	\bea
	\gamma=\mathrm{Tr}_{\mathcal{S}}\left(\widetilde{\rho}_{\mathrm{red}}^2\right)
	\eea
and the purity can be computed directly in the interaction picture. In general, one can expand
	\begin{align}\label{eq:gammanorder}
		\gamma = \sum_{n=0}^{\infty} g^n \gamma^{(n)} \qquad \mathrm{with} \qquad \gamma^{(n)}=\sum_{m=0}^n \mathrm{Tr}_{\mathcal{S}}\left[\widetilde{\rho}_{\mathrm{red}}^{(m)}\, \widetilde{\rho}_{\mathrm{red}}^{(n-m)} \right]
	\end{align}
	and 
	\begin{equation} \label{rhored_n}
	\widetilde{\rho}_{\mathrm{red}}^{(n)} \equiv \mathrm{Tr}_{\mathcal{E}}\left[\widetilde{\rho}^{(n)}\right] .
	\end{equation}

	\subsubsection*{Leading orders}\label{subsubsec:LO}
	
	Let us compute the first few terms in this expansion. At order $\mathcal{O}(g^0)$, one finds $\widetilde{\rho}_{\mathrm{red}}^{(0)}=\mathrm{Tr}_{\mathcal{E}}\left[\ket{\mathrm{BD}}\bra{\mathrm{BD}}\right]$. Since both the system and the environment start out in their vacuum states, there is no initial cross-correlation and the state is initially separable, 
	\begin{align} \label{IC_start}
		\ket{\mathrm{BD}}\bra{\mathrm{BD}} = \ket{\mathrm{BD}}\bra{\mathrm{BD}}_{\mathcal{S}} \otimes \ket{\mathrm{BD}}\bra{\mathrm{BD}}_{\mathcal{E}}
	\end{align}
	hence $\gamma^{(0)}=1$.       
	At order $\mathcal{O}(g)$, \Eq{eq:rho:iterative:SPT} reduces to
	\begin{align}
		\widetilde{\rho}^{(1)}(\eta) = -i  \int_{-\infty}^\eta \dd \eta_1 \left[\widetilde{H}_{\mathrm{int}}(\eta_1),\ket{\mathrm{BD}}\bra{\mathrm{BD}}\right].
	\end{align}
	For explicitness, assuming local interactions, we decompose the interaction Hamiltonian in the tensorial basis of local operators 
	\begin{align}\label{eq:HintFourier}
		\widehat{H}_{\mathrm{int}}(\eta) = \int \dd^3 \bmx \; \widehat{\mathcal{O}}^{\mathcal{S}}(\eta, \bmx) \otimes \widehat{\mathcal{O}}^{\mathcal{E}}(\eta, \bmx)  = \int  \frac{\dd^3\bmk}{(2\pi)^3} \;   \widehat{\mathcal{O}}^{\mathcal{S}}_\bmk(\eta) \otimes \widehat{\mathcal{O}}^{\mathcal{E}}_{-\bmk}(\eta)  \,,  
	\end{align}
	where $\widehat{\mathcal{O}}^{\mathcal{S}}(\eta, \bmx)$ and
        $\widehat{\mathcal{O}}^{\mathcal{E}}(\eta, \bmx)$ act on the   Hilbert space of the system and the environment respectively. If the operators $\widehat{\mathcal{O}}^{\mathcal{S}}$ and $\widehat{\mathcal{O}}^{\mathcal{E}}$ are linear in the phase-space variables, the interaction only couples modes of opposite wavevectors. However, if they involve non-linear combinations of the phase-space variables, $\widehat{\mathcal{O}}^{\mathcal{S}}_\bmk$ and $\widehat{\mathcal{O}}^{\mathcal{E}}_{-\bmk}$ are given by convolution products in Fourier space and all Fourier modes couple.
	
        Recalling that $\widetilde{\rho}^{(0)}=\ket{\mathrm{BD}}\bra{\mathrm{BD}}_{\mathcal{S}} \otimes \ket{\mathrm{BD}}\bra{\mathrm{BD}}_{\mathcal{E}}$, this gives rise to
	\begin{align}
		\widetilde{\rho}^{(1)}(\eta) = -i \int_{-\infty}^\eta \dd \eta_1 \int  \frac{\dd^3\bmk}{(2\pi)^3} & \left[  \widetilde{\mathcal{O}}^{\mathcal{S}}_\bmk(\eta_1) \ket{\mathrm{BD}}\bra{\mathrm{BD}}_{\mathcal{S}}  \otimes \widetilde{\mathcal{O}}^{\mathcal{E}}_{-\bmk}(\eta_1)\ket{\mathrm{BD}}\bra{\mathrm{BD}}_{\mathcal{E}}  \right. \nonumber \\ 
		& - \left. \ket{\mathrm{BD}}\bra{\mathrm{BD}}_{\mathcal{S}} \widetilde{\mathcal{O}}^{\mathcal{S}}_\bmk(\eta_1)  \otimes \ket{\mathrm{BD}}\bra{\mathrm{BD}}_{\mathcal{E}} \widetilde{\mathcal{O}}^{\mathcal{E}}_{-\bmk}(\eta_1) \right] .
		\label{eq:rho1:expanded}
	\end{align}
	As a consequence, tracing over the environment Hilbert space, we obtain
	\begin{align}
		\widetilde{\rho}^{(1)}_{\mathrm{red}}(\eta) = -i \int_{-\infty}^\eta \dd \eta_1 \int  \frac{\dd^3\bmk}{(2\pi)^3} \left[\widetilde{\mathcal{O}}^{\mathcal{S}}_\bmk(\eta_1) ,\ket{\mathrm{BD}}\bra{\mathrm{BD}}_{\mathcal{S}}\right] \mathcal{K}_{\mathcal{E}}^{(1)}(\bmk, \eta_1)
	\end{align}
	where 
	\begin{align}\label{eq:K1:def}
		\mathcal{K}_{\mathcal{S}/\mathcal{E}}^{(1)}(\bmk, \eta_1) \equiv \mathrm{Tr}_{\mathcal{S}/\mathcal{E}}\left[ \widetilde{\mathcal{O}}^{\mathcal{S}/\mathcal{E}}_{\pm\bmk}(\eta_1)  \ket{\mathrm{BD}}\bra{\mathrm{BD}}_{\mathcal{S}/\mathcal{E}} \right] = \bra{\mathrm{BD}}  \widetilde{\mathcal{O}}^{\mathcal{S}/\mathcal{E}}_{\pm\bmk} (\eta_1) \ket{\mathrm{BD}}_{\mathcal{S}/\mathcal{E}} \, .
	\end{align}
In this expression, $\mathcal{K}_{\mathcal{S}}^{(1)}(\bm{k}, \eta_1)$, which is introduced for later convenience, is the vacuum expectation value of $ \widetilde{\mathcal{O}}^{\mathcal{S}}_{\bm{k}}(\eta_1)$, while $\mathcal{K}_{\mathcal{E}}^{(1)}(\bm{k}, \eta_1)$ is the vacuum expectation value of $ \widetilde{\mathcal{O}}^{\mathcal{E}}_{-\bm{k}}(\eta_1)$. Since $\widehat{O}^{\mathcal{S}}(\bm{x})$ is a Hermitian operator, one has $\widehat{O}^{\mathcal{S}}_{-\bm{k}}=(\widehat{O}^{\mathcal{S}}_{\bm{k}})^\dagger$. If the free, linear evolution is isotropic, the free evolution operator $\widehat{\mathcal{U}}_0(\eta_1, - \infty)$ only depends on $k=\vert\bm{k}\vert $, hence one also has $\widetilde{O}^{\mathcal{S}}_{-\bm{k}}=(\widetilde{O}^{\mathcal{S}}_{\bm{k}})^\dagger$ and 
\begin{align}
\label{eq:K1:cc}
    \mathcal{K}_{\mathcal{S}}^{(1)}(-\bm{k}, \eta_1)= \left[\mathcal{K}_{\mathcal{S}}^{(1)}(\bm{k}, \eta_1)\right]^*,
\end{align}
and similarly for $\widetilde{O}^{\mathcal{E}}_{-\bm{k}}(\eta_1)$. Following \Eq{eq:gammanorder}, this gives rise to
	\begin{align}
		\gamma^{(1)} =& -2i \int_{-\infty}^\eta \dd \eta_1 \int  \frac{\dd^3\bmk}{(2\pi)^3} \mathrm{Tr}_{\mathcal{S}} \left\{ \ket{\mathrm{BD}}\bra{\mathrm{BD}}_{\mathcal{S}}  \left[\widetilde{\mathcal{O}}^{\mathcal{S}}_\bmk(\eta_1) ,\ket{\mathrm{BD}}\bra{\mathrm{BD}}_{\mathcal{S}}\right] \right\}\mathcal{K}_{\mathcal{E}}^{(1)}(\bmk, \eta_1) = 0 \label{eq:pert10}
	\end{align}
	since the trace operator is cyclic. We conclude that decoherence does not proceed at order $g$, a well-known fact indeed, see \eg \Refa{breuerTheoryOpenQuantum2002}. This can be understood as follows. At order $g$, the modification to the quantum state of the system, $\widetilde{\rho}^{(1)}_{\mathrm{red}}$, only involves the one-point function of the operators $\widetilde{O}^{\mathcal{S/E}}$ appearing in the interaction, see \Eq{eq:K1:def}. Upon performing the replacement 
     \begin{align}
         \widetilde{O}^{\mathcal{S/E}} = \mathcal{K}_{\mathcal{S/E}}^{(1)}\mathrm{Id}_{\mathcal{S/E}} + \delta\widetilde{O}^{\mathcal{S/E}}
     \end{align}
     in the Hamiltonian, and expressing its interacting part $\widehat{H}_{\mathrm{int}}$ in terms of the centred $\delta\widetilde{O}^{\mathcal{S/E}}$ operators, one obtains a setup in which the free-evolution terms  $\widehat{H}_{\mathcal{S}}$ and $\widehat{H}_{\mathcal{E}}$ are modified but the one-point functions $\mathcal{K}_{\mathcal{S}/\mathcal{E}}^{(1)}$ vanish, hence  $\widetilde{\rho}^{(1)}_{\mathrm{red}}=0$. Such a replacement leaves the definition of the system unchanged, hence its quantum state remains indeed pure at order $g$.
	
	Let us carry on the expansion and consider the order $\mathcal{O}(g^2)$. Expanding the commutators and tracing over the environment Hilbert space, \Eq{eq:rho:iterative:SPT} gives rise to 
	\begin{align}\label{eq:rho:red:2}
		\widetilde{\rho}^{(2)}_{\mathrm{red}}(\eta) &= - \int_{-\infty}^\eta \dd \eta_1  \int_{-\infty}^{\eta_1} \dd \eta_2 \int  \frac{\dd^3\bmk_1}{(2\pi)^3} \int  \frac{\dd^3\bmk_2}{(2\pi)^3}   \\
		\bigg\{& \left[\widetilde{\mathcal{O}}^{\mathcal{S}}_{\bmk_1}(\eta_1) \widetilde{\mathcal{O}}^{\mathcal{S}}_{\bmk_2}(\eta_2) \ket{\mathrm{BD}}\bra{\mathrm{BD}}_{\mathcal{S}} -  \widetilde{\mathcal{O}}^{\mathcal{S}}_{\bmk_2}(\eta_2) \ket{\mathrm{BD}}\bra{\mathrm{BD}}_{\mathcal{S}} \widetilde{\mathcal{O}}^{\mathcal{S}}_{\bmk_1}(\eta_1) \right]  \mathcal{K}_{\mathcal{E}}^{(2)}(\bmk_1,\bmk_2, \eta_1,\eta_2) \nonumber \\
		-&\left[ \widetilde{\mathcal{O}}^{\mathcal{S}}_{\bmk_1}(\eta_1)\ket{\mathrm{BD}}\bra{\mathrm{BD}}_{\mathcal{S}} \widetilde{\mathcal{O}}^{\mathcal{S}}_{\bmk_2}(\eta_2) - \ket{\mathrm{BD}}\bra{\mathrm{BD}}_{\mathcal{S}} \widetilde{\mathcal{O}}^{\mathcal{S}}_{\bmk_2}(\eta_2)\widetilde{\mathcal{O}}^{\mathcal{S}}_{\bmk_1}(\eta_1)  \right] \mathcal{K}_{\mathcal{E}}^{(2)}(\bmk_2,\bmk_1, \eta_2,\eta_1) \bigg\} \nonumber
	\end{align}
	where, similarly to \Eq{eq:K1:def}, we introduce
	\begin{align}
		\mathcal{K}_{\mathcal{S}/\mathcal{E}}^{(2)}(\bmk_1,\bmk_2, \eta_1,\eta_2) \equiv& \ \mathrm{Tr}_{\mathcal{S}/\mathcal{E}}\left[ \widetilde{\mathcal{O}}^{\mathcal{S}/\mathcal{E}}_{\pm\bmk_1}(\eta_1) \widetilde{\mathcal{O}}^{\mathcal{S}/\mathcal{E}}_{\pm\bmk_2}(\eta_2)  \ket{\mathrm{BD}}\bra{\mathrm{BD}}_{\mathcal{S}/\mathcal{E}} \right] \\
		=& \bra{\mathrm{BD}}  \widetilde{\mathcal{O}}^{\mathcal{S}/\mathcal{E}}_{\pm\bmk_1} (\eta_1)  \widetilde{\mathcal{O}}^{\mathcal{S}/\mathcal{E}}_{\pm\bmk_2} (\eta_2) \ket{\mathrm{BD}}_{\mathcal{S}/\mathcal{E}} \, .
		\label{eq:K2:def}
	\end{align}
        Note that, since $\widehat{O}^{\mathcal{S}/\mathcal{E}}_{-\bm{k}}=(\widehat{O}^{\mathcal{S}/\mathcal{E}}_{\bm{k}})^\dagger$ as mentioned above, one has
        \begin{align}
\label{eq:K2:cc}
            \mathcal{K}_{\mathcal{S}/\mathcal{E}}^{(2)}(-\bmk_2,-\bmk_1, \eta_2,\eta_1) = \left[\mathcal{K}_{\mathcal{S}/\mathcal{E}}^{(2)}(\bmk_1,\bmk_2, \eta_1,\eta_2)\right]^*.
        \end{align}
        Therefore, upon changing integration variables $\bmk_1 \rightarrow -\bmk_1$ and $\bmk_2 \rightarrow -\bmk_2$, the third line of \Eq{eq:rho:red:2} is nothing but the Hermitian conjugate of the second line, which is consistent with the fact that the reduced density matrix is Hermitian.
        In order to derive the second-order perturbative purity $\gamma^{(2)}$, from \Eq{eq:gammanorder} we have to compute two terms. The first one is
	\begin{align}
		\mathrm{Tr}_{\mathcal{S}} \left[   \left(\widetilde{\rho}^{(1)}_{\mathrm{red}}\right)^2 \right] 
		&=  \int_{-\infty}^\eta \dd \eta_1  \int_{-\infty}^\eta \dd \eta_2 \int  \frac{\dd^3\bmk_1}{(2\pi)^3}  \int  \frac{\dd^3\bmk_2}{(2\pi)^3} \bigg\{\mathcal{K}_{\mathcal{E}}^{(1)}(\bmk_1, \eta_1) \mathcal{K}_{\mathcal{E}}^{(1)}(\bmk_2, \eta_2) \nonumber \\
		&\qquad \qquad \left[ \bar{\mathcal{K}}_{\mathcal{S}}^{(2)}(\bmk_1,\bmk_2, \eta_1,\eta_2) + \bar{\mathcal{K}}_{\mathcal{S}}^{(2)}(\bmk_2,\bmk_1, \eta_2,\eta_1) \right] \bigg\} \label{eq:term1}
	\end{align}
        where we defined the centred two-point functions
	\begin{align} \label{centred}
		\bar{\mathcal{K}}_{\mathcal{S}/\mathcal{E}}^{(2)}(\bmk_1,\bmk_2, \eta_1,\eta_2)\equiv \mathcal{K}_{\mathcal{S}/\mathcal{E}}^{(2)}(\bmk_1,\bmk_2, \eta_1,\eta_2) - \mathcal{K}_{\mathcal{S}/\mathcal{E}}^{(1)}(\bmk_1, \eta_1) \mathcal{K}_{\mathcal{S}/\mathcal{E}}^{(1)}(\bmk_2, \eta_2).
	\end{align}
	When deriving the above expressions, we have used that $\left(\ket{\mathrm{BD}}\bra{\mathrm{BD}}_{\mathcal{S}}\right)^2=\ket{\mathrm{BD}}\bra{\mathrm{BD}}_{\mathcal{S}}$, \ie the fact that the system is initially placed in a pure state, as well as
	\begin{align}
		\mathrm{Tr}_{\mathcal{S}}\left[ \widetilde{\mathcal{O}}^{\mathcal{S}}_{\bmk_1}(\eta_1)  \ket{\mathrm{BD}}\bra{\mathrm{BD}}_{\mathcal{S}}  \widetilde{\mathcal{O}}^{\mathcal{S}}_{\bmk_2}(\eta_2)  \ket{\mathrm{BD}}\bra{\mathrm{BD}}_{\mathcal{S}} \right] = \mathcal{K}_{\mathcal{S}}^{(1)}(\bmk_1, \eta_1) \mathcal{K}_{\mathcal{S}}^{(1)}\left(\bmk_2, \eta_2\right) .
	\end{align}
	The second term we have to compute is given by
	\begin{align}\label{eq:pert20}
		\mathrm{Tr}_{\mathcal{S}} \left[   \widetilde{\rho}^{(0)}_{\mathrm{red}}  \widetilde{\rho}^{(2)}_{\mathrm{red}}\right] &=  - \int_{-\infty}^\eta \dd \eta_1  \int_{-\infty}^{\eta_1} \dd \eta_2 \int  \frac{\dd^3\bmk_1}{(2\pi)^3} \int  \frac{\dd^3\bmk_2}{(2\pi)^3}   \\
		\bigg[ &\bar{\mathcal{K}}_{\mathcal{S}}^{(2)}(\bmk_1,\bmk_2, \eta_1,\eta_2) \mathcal{K}_{\mathcal{E}}^{(2)}(\bmk_1,\bmk_2, \eta_1,\eta_2) 
		+ \bar{\mathcal{K}}_{\mathcal{S}}^{(2)}(\bmk_2,\bmk_1, \eta_2,\eta_1) \mathcal{K}_{\mathcal{E}}^{(2)}(\bmk_2,\bmk_1, \eta_2,\eta_1) \bigg] .\nonumber
	\end{align}      
The integrand is invariant under the $1 \leftrightarrow 2$ label exchange, hence it can be written as an un-nested time integral by replacing $\int_{-\infty}^\eta \dd \eta_1  \int_{-\infty}^{\eta_1} \dd \eta_2 \to \frac{1}{2}\int_{-\infty}^\eta \dd \eta_1  \int_{-\infty}^{\eta} \dd \eta_2 $. Making further use of \Eqs{eq:K1:cc} and~\eqref{eq:K2:cc}, the above results lead to
	\begin{tcolorbox}[%
		enhanced, 
		breakable,
		skin first=enhanced,
		skin middle=enhanced,
		skin last=enhanced
		]
		\paragraph{Second-order perturbative purity:}
		\begin{align}\label{eq:gamma:gen:final}
			\gamma^{(2)} &= - 2 \int_{-\infty}^\eta \dd \eta_1  \int_{-\infty}^{\eta} \dd \eta_2 \int  \frac{\dd^3\bmk_1}{(2\pi)^3} \int  \frac{\dd^3\bmk_2}{(2\pi)^3}
			\; \bar{\mathcal{K}}_{\mathcal{S}}^{(2)}(\bmk_1,\bmk_2, \eta_1,\eta_2) \bar{\mathcal{K}}_{\mathcal{E}}^{(2)}(\bmk_1,\bmk_2, \eta_1,\eta_2) \, .
		\end{align}
	\end{tcolorbox}
	\noindent
	 	
At this order, the purity involves two unequal-time correlators, one that describes the state the system in the free theory and one that describes the environment. If these correlators feature non-singular behavior in the coincident time limit, \Eq{eq:gamma:gen:final} implies that $\gamma$ always decreases initially. Indeed, denoting the initial time by $\eta_\uin$ (although in practice we let $\eta_\uin=-\infty$) one has
    \begin{align}\label{eq:deriv1}
         \frac{\dd\gamma^{(2)}}{\dd \eta}\bigg|_{\eta=\eta_\uin} = 0
    \end{align}
    and 
    \begin{align}\label{eq:deriv2}
        \frac{\dd^2\gamma^{(2)}}{\dd \eta^2}\bigg|_{\eta=\eta_\uin} & = - 4g^2  \int  \frac{\dd^3\bmk_1}{(2\pi)^3} \int  \frac{\dd^3\bmk_2}{(2\pi)^3}\; \Rea\Big[\bar{\mathcal{K}}_{\mathcal{S}}^{(2)}(\bmk_1,\bmk_2, \eta_\uin,\eta_\uin) \bar{\mathcal{K}}_{\mathcal{E}}^{(2)}(\bmk_1,\bmk_2, \eta_\uin,\eta_\uin) \Big] \, .
    \end{align}
    In this expression, $\bar{\mathcal{K}}_{\mathcal{S}}^{(2)}(\bmk_1,\bmk_2, \eta_\uin,\eta_\uin) $ and $\bar{\mathcal{K}}_{\mathcal{E}}^{(2)}(\bmk_1,\bmk_2, \eta_\uin,\eta_\uin) $ are nothing but the initial power spectra of the centred (and possibly composite) $\widehat{\mathcal{O}}^{\mathcal{S}}$ and $\widehat{\mathcal{O}}^{\mathcal{E}}$ operators. Once integrated over all Fourier modes, the right-hand-side of \Eq{eq:deriv2} can be shown to be non-negative. This implies that, close enough to the initial time, $\gamma\simeq 1- \frac{1}{2}\gamma''(\eta_\uin)  (\eta-\eta_\uin)^2$ necessarily decreases, and is below one. The fact that purity remains below one indicates that at order $g^2$, perturbation theory provides a \textit{complete-positive and trace-preserving} (CPTP) dynamical map \cite{breuerTheoryOpenQuantum2002}. If $\gamma<0$ is found at some later time, this simply indicates the breakdown of perturbation theory, but, under the above assumptions, there is always a finite time interval  around $\eta_\uin$ where it is valid. This expression also implies that \Eq{eq:deriv2} can be seen as the squared decay rate of purity initially, hence it provides a relevant time scale for the evolution of purity, which is simply made of the product of the power spectra of the system and environment operators involved in the interaction. For linear interactions, the correlators $\bar{\mathcal{K}}_{\mathcal{S}}^{(2)}$ and $\bar{\mathcal{K}}_{\mathcal{E}}^{(2)}$ are non singular in the time-coincident limit and the above considerations hold. However, as stressed in \Refa{Burgess:2024eng}, they do not apply in general when non-linear interactions are at play.

 Let us also note that under the form \eqref{eq:gamma:gen:final}, the purity is symmetric under exchanging the system and the environment, hence it is the same for both subsystems. This holds at all orders,\footnote{This can be shown using the Schmidt decomposition of the pure state describing the whole system-environment setup, $\ket{\Psi} = \sum_i \lambda_i \ket{i}_{\mathcal{S}}  \ket{i}_{\mathcal{E}} $, where $\lbrace \ket{i}_{\mathcal{S}} \rbrace$ and $\lbrace \ket{i}_{\mathcal{E}} \rbrace$ are orthonormal sets of states for the system and the environment respectively. This is a mere consequence of the singular value decomposition (the sum may be discrete or continuous). From this expression one can easily compute $\widehat{\rho}_{\mathrm{red}} = \mathrm{Tr}_{\mathcal{E}}(\hat{\rho}) = \sum_i \lambda_i^2  \ket{i}_{\mathcal{S}}  \bra{i}_{\mathcal{S}} $, hence the purity of the system reads $\gamma=\mathrm{Tr}_{\mathcal{S}}(\hat{\rho}_{\mathrm{red}}^2)=\sum_i \lambda_i^4$, and a similar calculation gives the same result for the purity of the environment.} and comes from the fact that purity is directly related to entropy measures such as linear entropy, the entanglement entropy \cite{Serafini:2003ke} (see \Sec{sec:vN}), or the R\'enyi-$2$ entropy \cite{renyi1961measures} as we shall now see. Since it quantifies the amount of information shared between the system and the environment, it must indeed be a symmetric quantity~\cite{Serafini:2003ke}.  

	\subsection{R\'enyi-2 entropy}
	\label{sec:Spectral:Purity}
	
In the case of linear interactions, the system remains in a Gaussian state whose Fourier modes are uncoupled, and the reduced density matrix can be factorized into
	\begin{align}
		\label{eq:Fourier:factorised}
		\widetilde{\rho}_{\mathrm{red}}=\bigotimes _{\bmk\in\mathbb{R}^{3+}} \widetilde{\rho}_{\mathrm{red}} (\bmk)\, .
	\end{align}
In this expression, the tensorial product is taken over $\mathbb{R}^{3+}$ to avoid double counting of the degrees of freedom. For each Fourier mode $\bm{k}$, one can compute $\mathrm{Tr}[\widetilde{\rho}_{\mathrm{red}}^2 (\bmk)]$ and assign a purity per mode $\bm{k}$. This is relevant in cosmology, where a particular emphasis is often given to how much decoherence proceeds at a given scale. However, if non-linearities are  present, this factorization cannot be performed anymore, and the purity cannot be computed mode by mode. This is why we introduce and propose the concept of ``spectral purity'' in this section, to quantify decoherence \textit{at a given scale} even in the presence of mode coupling. 

A first remark is that purity is a multiplicative (rather than additive) quantity. Indeed, if $\mathcal{A}$ and $\mathcal{B}$ denote two Hilbert spaces in which $\widetilde{\rho}_A$ and $\widetilde{\rho}_B$ lie, the purity of the whole state $\widetilde{\rho}=\widetilde{\rho}_A\otimes\widetilde{\rho}_B$ is given by $\mathrm{Tr}(\widetilde{\rho}^2)=\mathrm{Tr}(\widetilde{\rho}_A^2\otimes \widetilde{\rho}_B^2)=\mathrm{Tr}_{\mathcal{A}}(\widetilde{\rho}_A^2)\mathrm{Tr}_{\mathcal{B}}(\widetilde{\rho}_B^2)$, hence one obtains the product of the individual purities. In order to deal with additive quantities instead, it is convenient to introduce the logarithm of the purity, which is nothing but the \textit{R\'enyi-2 entropy} \cite{renyi1961measures,Serafini:2003ke,Kudler-Flam:2022zgm,Kudler-Flam:2023kph}\footnote{This quantity is called ``entropy'' in the sense that it provides a measure of the correlations between two subsystems $A$ and $B$, through the \textit{R\'enyi-2 mutual information} \cite{renyi1961measures,Martin:2021znx, Martin:2022kph, Kudler-Flam:2022zgm,Kudler-Flam:2023kph}
		\begin{align} 
			\mathcal{I}_2\left(A,B\right) = S_2\left(\widehat{\rho}_{A}\right) + S_2\left(\widehat{\rho}_{B}\right) - S_2\left( \widehat{\rho}_{AB}\right) ,
		\end{align}
		which is the relative R\'enyi-2 entropy between the full state $\widehat{\rho}_{AB}$ and its lowest-order factorized form $\widehat{\rho}_{A} \otimes \widehat{\rho}_{B}$. }
	\begin{align}\label{eq:Renyi2}
		S_2\left(\widehat{\rho}_{\mathrm{red}}\right)  \equiv -  \ln(\gamma).
	\end{align}
This quantity is naturally additive. In the absence of mode coupling, when the factorization~\eqref{eq:Fourier:factorised} is valid, one can define its Fourier version as
	\begin{align}
		\label{eq:Gamma:def}
  \delta(\bm{0})
		 \ln \mathrm{Tr}\left[\widetilde{\rho}_{\mathrm{red}} (\bmk) \widetilde{\rho}_{\mathrm{red}}(\bmk')\right] = - S_2(k) \delta(\bmk-\bmk')\, .
	\end{align}
The presence of the volume factor $\delta(\bm{0})$ arises when taking the continuum limit in Fourier space~\cite{Burgess:2022nwu, Colas:2024xjy}.
In this expression, the Dirac distribution arises from the fact that if $\bm{k}\neq\bm{k}'$ then $\widetilde{\rho}_{\mathrm{red}} (\bmk)$ and $\widetilde{\rho}_{\mathrm{red}}(\bmk')$ live in different Hilbert spaces, hence $ \mathrm{Tr}\left[\widetilde{\rho}_{\mathrm{red}} (\bmk) \widetilde{\rho}_{\mathrm{red}}(\bmk')\right]= \mathrm{Tr}\left[\widetilde{\rho}_{\mathrm{red}} (\bmk) ]\mathrm{Tr}[\widetilde{\rho}_{\mathrm{red}}(\bmk')\right]=1$. 
Combining the above expressions, one obtains, in the factorized state~\eqref{eq:Fourier:factorised}
\bea
\label{eq:S2:S2k}
S_2\left(\widetilde{\rho}_{\mathrm{red}}\right) =  - \ln \prod_{\bmk\in\mathbb{R}^{3+}}\mathrm{Tr}\left[\widetilde{\rho}_{\mathrm{red}}^2 (\bmk) \right] 
= &-  \delta(\bm{0})\int_{\bmk\in\mathbb{R}^{3+}}\dd^3\bmk\;\ln\mathrm{Tr}\left[\widetilde{\rho}_{\mathrm{red}}^2 (\bmk) \right]\\
= &\; \delta(\bm{0})\int_{\bmk\in\mathbb{R}^{3+}}\dd^3\bmk \; S_2(k)\, .
\eea

Even though \Eq{eq:S2:S2k} was derived in the factorized state~\eqref{eq:Fourier:factorised}, it is always the case that $S_2$ can be cast in the form 
\begin{align}
\label{eq:S2:S2k:gen}
S_2\left(\widetilde{\rho}_{\mathrm{red}}\right)=\delta(\bm{0})\int_{\bmk\in\mathbb{R}^{3+}}\dd^3\bmk \; S_2(k)\, ,
\end{align}
even in the presence of mode coupling, as we will see below. Our strategy is therefore to express the purity as an integral over Fourier space, and to identify the integrand with the one in the above expression to read off the spectral purity $S_2(k)$. In this way, the spectral purity can be defined for any type of interaction. In practice, it is convenient to introduce the reduced spectral purity $\mathcal{S}_2(k)=2\pi k^3 S_2(k)$, such that $S_2(\widetilde{\rho}_{\mathrm{red}})=\delta(\bm{0})\int\dd\ln k \, \mathcal{S}_2(k)$. Scale invariance of the reduced spectral purity thus corresponds to volume-law entropy production.

At order $g^2$ in the perturbative expansion, the purity is given by \Eq{eq:gamma:gen:final}. In this expression, since the background is homogeneous and isotropic, the system and environment unequal-time two-point functions $\bar{\mathcal{K}}_{\mathcal{S}}^{(2)}(\bmk_1,\bmk_2, \eta_1,\eta_2)$ and $\bar{\mathcal{K}}_{\mathcal{E}}^{(2)}(\bmk_1,\bmk_2, \eta_1,\eta_2)$ must be of the form
	\begin{align}
	\label{eq:def:Kreduced}
		\bar{\mathcal{K}}_{\mathcal{S}/\mathcal{E}}^{(2)}(\bmk_1,\bmk_2, \eta_1,\eta_2)  \equiv  \bar{\mathcal{K}}_{\mathcal{S}/\mathcal{E}}^{(2)}(k_1, \eta_1,\eta_2)  \delta\left(\bmk_1 + \bmk_2\right) ,
	\end{align}
	which defines $\bar{\mathcal{K}}_{\mathcal{S}/\mathcal{E}}^{(2)}(k_1, \eta_1,\eta_2)$. One can thus rewrite \Eq{eq:gamma:gen:final} as
	\begin{align}
		\gamma^{(2)} &= -2 \left(2\pi\right)^{-3}  \int_{-\infty}^\eta \dd \eta_1  \int_{-\infty}^{\eta} \dd \eta_2 \int  \frac{\dd^3\bmk}{(2\pi)^3} 
		\; \bar{\mathcal{K}}_{\mathcal{S}}^{(2)}(k, \eta_1,\eta_2) \bar{\mathcal{K}}_{\mathcal{E}}^{(2)}(k, \eta_1,\eta_2) \delta(\bm{0})\, .
	\end{align}
At order $g^2$, one has $S_2(\widetilde{\rho}_{\mathrm{red}})=-g^2 \gamma^{(2)}$, and by identification with \Eq{eq:S2:S2k:gen} one obtains for the spectral purity
	\begin{align}
	\label{eq:S2:KS:KE}
		S_2^{(2)}(k) = 4 (2\pi)^{-6} \int_{-\infty}^\eta \dd \eta_1  \int_{-\infty}^{\eta} \dd \eta_2  \; 
		\bar{\mathcal{K}}_{\mathcal{S}}^{(2)}(k, \eta_1,\eta_2) \bar{\mathcal{K}}_{\mathcal{E}}^{(2)}(k, \eta_1,\eta_2) .
	\end{align}
	
\subsection{Gaussian case}
\label{sec:Gaussian:Purity}

As mentioned above, when interactions are linear the system remains in a Gaussian state that factorizes according to \Eq{eq:Fourier:factorised}, where each $\widetilde{\rho}_{\mathrm{red}} (\bmk)$ describes a Gaussian state. Gaussian states are fully described by their covariance matrix~\cite{Serafini:2003ke, Colas:2021llj, Martin:2022kph}
	\begin{align}
	\label{eq:cov:def:matrix}
	\delta(\bm{0})	\bs{\mathrm{Cov}}(k,\eta) \equiv \frac{1}{2}\mathrm{Tr}_{\mathcal{S}}\left[ \left\{\widetilde{\bm{z}}(\bmk,\eta),\widetilde{\bm{z}}^\dag(\bmk,\eta) \right\} \widetilde{\rho}_{\mathrm{red}}(\eta) \right]
	\end{align}
where $\widetilde{\bm{z}}(\bmk,\eta)$ is a vector gathering all phase-space coordinates of the Fourier mode $\bmk$ for the system. In particular, $\mathrm{Tr}[\widetilde{\rho}_{\mathrm{red}}^2 (\bmk)]$ is given by the determinant of the covariance matrix,
	\begin{align} \label{logdetpurity}
	\ln \mathrm{Tr}[\widetilde{\rho}_{\mathrm{red}}^2 (\bmk)]=\ln\left[\frac{1}{4 \det \bs{\mathrm{Cov}}(k,\eta) } \right]  ,
	\end{align}
	where for explicitness we have assumed that the system is made of a single scalar degree of freedom, hence $\widetilde{\bm{z}}(\bmk,\eta)$ has two entries. By comparison with \Eq{eq:Gamma:def}, this leads to
\begin{align}
\label{eq:S2(k):Gauss}
 S_2(k) = \ln\left[4 \det \bs{\mathrm{Cov}}(k,\eta) \right] .
\end{align}
When non-linear interactions are at play, the system is not in a Gaussian state and the above expression does not apply. In that case, the purity is not only sensitive to the two-point functions but to all higher-order correlation functions. However, in \App{app:proof}, we show that \Eq{eq:S2(k):Gauss} remains valid at order $g^2$ for interactions that are quadratic in the system's phase-space variables, with an additional factor $1/2$. In other words, we show that, for linear \emph{and} quadratic interactions,
\begin{align}
\label{eq:S2(k):2}
 S_2^{(2)}(k) =\frac{4}{n} \mathrm{det}^{(2)}  \bs{\mathrm{Cov}}(k,\eta) \, ,
\end{align}
where $n=1$ for linear interactions (the result then coincides with \Eq{eq:S2(k):Gauss} when expanded at order $g^2$) and $n=2$ for quadratic interactions. The fact that the perturbed purity is still entirely determined by the determinant of the covariance matrix even for quadratic interactions is rather remarkable, since as mentioned above non-linear interactions probe all correlation functions in principle. In the context of cosmological perturbation theory, the dominant interaction terms are cubic~\cite{Maldacena:2002vr}, hence they are either linear or quadratic in the system's variables, and the above result implies that the power spectra of the system are enough to determine its perturbed purity.

\subsection{Partial resummation}
\label{sec:resummation}

For Gaussian states, \Eq{eq:S2(k):Gauss} is an exact statement and it is therefore valid to all orders in perturbation theory. This is of substantial practical use, since perturbation theory generically breaks down at late times. Indeed, the in-in formalism expands unitary time evolution operators (schematically) as $e^{- i ( H_0 +  g H_{\mathrm{int}} ) t } \simeq e^{- i H_0 t }(1 - i g H_{\mathrm{int}} t +\ldots )$, which is a good approximation for the true dynamics only at relatively early times where $g t$ is small compared to timescales set by the interaction. Secular growth does not occur in computations of scattering amplitudes since interactions are taken to be turned off in the asymptotic past and future (where momentum eigenstates are taken to be free) -- however, in virtually any other setting where the interactions stay turned on,
secular growth is a persistent issue. This is the case, for example, in the calculation of correlators in flat-space \cite{Burgess:2018sou,Akhmedov:2020haq,Chaykov:2022zro}, cosmological space-times~\cite{Tsamis:1993ub,Starobinsky:1994bd,Onemli:2002hr,Tsamis:2005hd,Riotto:2008mv,Burgess:2009bs,Marolf:2010zp,Burgess:2010dd, Burgess:2015ajz, Gorbenko:2019rza, Kaplanek:2019dqu, Kaplanek:2019vzj, Green:2020txs, Colas:2022hlq, Cespedes:2023aal}, or even black-hole backgrounds~\cite{Akhmedov:2015xwa}. 
   
In general, at second-order in perturbation theory, one finds that $\det \bs{\mathrm{Cov}} \sim 1/4+ g^2 t^m$ (for some power $m$). The perturbative treatment is therefore a priori valid only for early times $g^2 t^m \ll 1$. The exact relation~\eqref{logdetpurity} provides a simple yet powerful way to resum late-time breakdowns. It leads to $\gamma \sim g^{-2} t^{-m}$, which remains reliable even at late times where $g^2 t^m \gg 1$. It is the non-perturbative nature of \Eq{eq:S2(k):Gauss} that allows one to trust this approximation out to much later times, since the dynamics have been resummed to all orders in $g^2 t^m$. This type of a resummation was used in \Refs{Colas:2022hlq, Colas:2022kfu, Colas:2024xjy, Burgess:2024eng} for the calculation of $\gamma$. In \Sec{sec:applic}, we illustrate the use of such a resummation (see \Fig{fig:gamma_full}), and otherwise leave more detailed investigations (for instance, involving master equations) of late-time resummations for future work.
		
	\subsection{Higher orders}\label{subsubsec:NLO}
	
Beyond the lowest order correction to the purity of the system, it might be of interest to derive higher-order corrections, for instance in cosmological applications to relate the evolution of the purity to higher-order statistics (bispectrum, trispectrum and so on). We now explain how the above formalism can be extended to higher orders. This will lead us to diagrammatic rules for the calculation of the purity, allowing for systematic expansions similar to those commonly employed in (unitary) perturbative QFT.
	
	\subsubsection{Kraus representation}
	\label{sec:Kraus}	

At order $n$, the integrand of \Eq{eq:rho:iterative:SPT} involves the interaction Hamiltonian evaluated at $n$ different times $\eta_1,\, \cdots,\, \eta_n$. Let us introduce the $\widetilde{\bs{\mathcal{K}}}^{(i)}$ vectors, which contain all products of $i$ interaction Hamiltonians, where the indices of the time arguments are ordered (which corresponds to anti-ordering the time arguments, since in \Eq{eq:rho:iterative:SPT}, $\eta_1>\eta_2>\cdots>\eta_n$). More explicitly, 
		\begin{align}
\label{eq:Kij:def:in}
		\widetilde{\bs{\mathcal{K}}}^{(0)} &\equiv \left[ \mathbb{I} \right] \\
		\widetilde{\bs{\mathcal{K}}}^{(1)} &\equiv \left[ \widetilde{H}_{\mathrm{int}}(\eta_1), \widetilde{H}_{\mathrm{int}}(\eta_2), \cdots , \widetilde{H}_{\mathrm{int}}(\eta_n)\right] \\
		\widetilde{\bs{\mathcal{K}}}^{(2)}  &\equiv \left[\widetilde{H}_{\mathrm{int}}(\eta_1) \widetilde{H}_{\mathrm{int}}(\eta_2), \cdots, \widetilde{H}_{\mathrm{int}}(\eta_i) \widetilde{H}_{\mathrm{int}}(\eta_j) ,\cdots\right] \qquad \mathrm{with}\qquad i < j \\
		\widetilde{\bs{\mathcal{K}}}^{(3)}  &\equiv \left[\widetilde{H}_{\mathrm{int}}(\eta_1) \widetilde{H}_{\mathrm{int}}(\eta_2)\widetilde{H}_{\mathrm{int}}(\eta_3), \cdots, \widetilde{H}_{\mathrm{int}}(\eta_i) \widetilde{H}_{\mathrm{int}}(\eta_j) \widetilde{H}_{\mathrm{int}}(\eta_k) ,\cdots\right] \quad \mathrm{with}\quad i < j < k \\ 
		&\qquad \qquad \cdots \nonumber \\
		\widetilde{\bs{\mathcal{K}}}^{(n)}  &\equiv \left[ \widetilde{H}_{\mathrm{int}}(\eta_1) \widetilde{H}_{\mathrm{int}}(\eta_2) \cdots \widetilde{H}_{\mathrm{int}}(\eta_n)\right]\, .
\label{eq:Kij:def:end}
	\end{align}
	
Each $\widetilde{\bs{\mathcal{K}}}^{(i)}$ vector contains $\binom{n}{i}=n!/[i! (n-i)!]$ elements, which corresponds to choosing $i$ instants $\eta_j$ amongst the $n$ ones. When expanding the commutator in \Eq{eq:rho:iterative:SPT}, one finds
\bea
\label{eq:Krauss:n}
\left[\widetilde{H}_{\mathrm{int}}(\eta_1),\left[ \widetilde{H}_{\mathrm{int}}(\eta_2),\cdots \left[\widetilde{H}_{\mathrm{int}}(\eta_n),\ket{\mathrm{BD}}\bra{\mathrm{BD}} \right] \cdots \right] \right] = 
\sum_{i=0}^n (-1)^{n-i} \sum_{j=1}^{\binom{n}{i}} \widetilde{\bs{\mathcal{K}}}^{(i)}_j \ket{\mathrm{BD}}\bra{\mathrm{BD}}\left( \widetilde{\bs{\mathcal{K}}}^{(n-i)}_{\binom{n}{i}-j+1}\right)^\dagger  .
\eea
This can be shown by induction and the detailed proof,  which fixes the ordering of the elements in each vector $\widetilde{\bs{\mathcal{K}}}^{(i)}$, is given in \App{sec:unnest}. As a consequence, \Eq{eq:rho:iterative:SPT} can be written as
	\begin{tcolorbox}[%
		enhanced, 
		breakable,
		skin first=enhanced,
		skin middle=enhanced,
		skin last=enhanced
		]
	\begin{align}
		\label{eq:Kraus}
		\widetilde{\rho}^{(n)}(\eta) &= (-i)^n \int_{-\infty}^{\eta}\dd \eta_1\int_{-\infty}^{\eta_1}\dd \eta_2 \cdots \int_{-\infty}^{\eta_{n-1}}\dd \eta_n  \sum_{i=0}^n (-1)^{n-i} \sum_{j=1}^{\binom{n}{i}} \widetilde{\bs{\mathcal{K}}}^{(i)}_j \ket{\mathrm{BD}}\bra{\mathrm{BD}}\left( \widetilde{\bs{\mathcal{K}}}^{(n-i)}_{\binom{n}{i}-j+1}\right)^\dagger .
	\end{align}	
\end{tcolorbox}
\noindent Note that taking $i\to n-i$ and $j\to \binom{n}{i}-j+1$ gives the Hermitian conjugate of the above expression, which is consistent with the fact that the density matrix is Hermitian at every order. This also means that, in practice, it is enough to compute the terms for $i$ up to $\lfloor \frac{n}{2}\rfloor$ (the floor of $n/2$) and take the Hermitian part of the result. 
	
We are now in a position where we can trace over the environmental degrees of freedom. For local interactions of the type~\eqref{eq:HintFourier}, one obtains
\begin{align}
\widetilde{\rho}^{(n)}_{\mathrm{red}}(\eta) =& (-i)^n \int_{-\infty}^{\eta}\dd \eta_1 \int_{-\infty}^{\eta_1}\dd \eta_2 \cdots \int_{-\infty}^{\eta_{n-1}}\dd \eta_n \int \frac{\dd^3\bmk_1}{(2\pi)^3} \int \frac{\dd^3\bmk_2}{(2\pi)^3}  \cdots \int \frac{\dd^3\bmk_n}{(2\pi)^3} 
\nonumber \\ &
\sum_{i=0}^n (-1)^{n-i} \sum_{j=1}^{\binom{n}{i}}
\bs{\mathcal{K}}_{\mathcal{E},i,j} 
\widetilde{\bs{\mathcal{K}}}^{(i)}_{\mathcal{S},j} 
\ket{\mathrm{BD}}
\bra{\mathrm{BD}}_{\mathcal{S}}\left( \widetilde{\bs{\mathcal{K}}}^{(n-i)}_{\mathcal{S},\binom{n}{i}-j+1}\right)^\dagger ,
\label{eq:rhored:Kraus}
\end{align}	
where
\begin{align}
\bs{\mathcal{K}}_{\mathcal{E},i,j}  = 
\bra{\mathrm{BD}}\left( \widetilde{\bs{\mathcal{K}}}^{(n-i)}_{\mathcal{E},\binom{n}{i}-j+1}\right)^\dagger
\widetilde{\bs{\mathcal{K}}}^{(i)}_{\mathcal{E},j} 
 \ket{\mathrm{BD}}_{\mathcal{E}}\, .
 \label{eq:correl:env:gen:n}
\end{align}	
Here, $\widetilde{\bs{\mathcal{K}}}^{(i)}_{\mathcal{S},j} $ are defined as in \Eq{eq:Kij:def:in}-\eqref{eq:Kij:def:end}, where each $\widetilde{H}_{\mathrm{int}}(\eta_\ell)$ is replaced with $\widetilde{O}^{\mathcal{S}}_{\bmk_\ell}(\eta_\ell)$, with $\ell=1\cdots i$. Likewise, $\widetilde{\bs{\mathcal{K}}}^{(i)}_{\mathcal{E},j} $ are defined as in \Eqs{eq:Kij:def:in}-\eqref{eq:Kij:def:end} where each $\widetilde{H}_{\mathrm{int}}(\eta_\ell)$ is replaced with $\widetilde{O}^{\mathcal{E}}_{-\bmk_\ell}(\eta_\ell)$. With $n=2$, one recovers \Eq{eq:rho:red:2}. With $n=3$, one finds
	\begin{align}\label{eq:rhored3}
		\widetilde{\rho}^{(3)}_{\mathrm{red}}(\eta) &= i \int_{-\infty}^{\eta}\dd \eta_1\int_{-\infty}^{\eta_1}\dd \eta_2  \int_{-\infty}^{\eta_2}\dd \eta_3   \int \frac{\dd^3\bmk_1}{(2\pi)^3} \int \frac{\dd^3\bmk_2}{(2\pi)^3}  \int \frac{\dd^3\bmk_3}{(2\pi)^3} \nonumber \\
		\bigg\{&\mathcal{K}^{(3)}_{\mathcal{E}} \left( \bmk_1, \bmk_2, \bmk_3;\eta_1,\eta_2,\eta_3\right)  \left[ \widetilde{\mathcal{O}}^{\mathcal{S}}_{\bmk_1} (\eta_1) ,\widetilde{\mathcal{O}}^{\mathcal{S}}_{\bmk_2} (\eta_2) \widetilde{\mathcal{O}}^{\mathcal{S}}_{\bmk_3} (\eta_3)  \ket{\mathrm{BD}}\bra{\mathrm{BD}}_{\mathcal{S}} \right] \\
		-&\mathcal{K}^{(3)}_{\mathcal{E}} \left( \bmk_2, \bmk_1, \bmk_3;\eta_2,\eta_1,\eta_3\right)  \left[   \widetilde{\mathcal{O}}^{\mathcal{S}}_{\bmk_1} (\eta_1)   ,\widetilde{\mathcal{O}}^{\mathcal{S}}_{\bmk_3} (\eta_3) \ket{\mathrm{BD}}\bra{\mathrm{BD}}_{\mathcal{S}}   \widetilde{\mathcal{O}}^{\mathcal{S}}_{\bmk_2} (\eta_2) \right]        \bigg\}  + \mathrm{h.c.} \nonumber
	\end{align}
where we have introduced 
\begin{align}
		\mathcal{K}^{(3)}_{\mathcal{E}} \left( \bmk_1, \bmk_2, \bmk_3;\eta_1,\eta_2,\eta_3\right)  \equiv  \bra{\mathrm{BD}}  \widetilde{\mathcal{O}}^{\mathcal{E}}_{-\bmk_1} (\eta_1)  \widetilde{\mathcal{O}}^{\mathcal{E}}_{-\bmk_2} (\eta_2)  \widetilde{\mathcal{O}}^{\mathcal{E}}_{-\bmk_3} (\eta_3) \ket{\mathrm{BD}}_{\mathcal{E}} \, ,
\end{align}
and higher orders can be computed similarly. When $n$ gets large, the list of terms to write down becomes long: there are $2^n$ correlators of the type~\eqref{eq:correl:env:gen:n}, although using trace cyclicity and Hermiticity under index relabelling as mentioned below \Eq{eq:Kraus} leaves $2^n/4$ correlators to compute (hence $1$ correlator when $n=2$ and 2 correlators when $n=3$, this is consistent with \Eqs{eq:rho:red:2} and~\eqref{eq:rhored3} respectively). The number of terms still grows rapidly, which is why we now provide diagrammatic rules to assist the calculation. These will allow us to organize the computation of $\widetilde{\rho}_{\mathrm{red}}^{(n)}$, and of the terms $\mathrm{Tr}_{\mathcal{S}}[\widetilde{\rho}_{\mathrm{red}}^{(m)}\, \widetilde{\rho}_{\mathrm{red}}^{(n-m)} ]$ that appear in the calculation of $\gamma^{(n)}$ in \Eq{eq:gammanorder}.
	
	\subsubsection{Diagrammatic rules}\label{subsubsec:diagram}
	
\begin{figure}[tbp]
		\centering
		\includegraphics[width=0.7\textwidth]{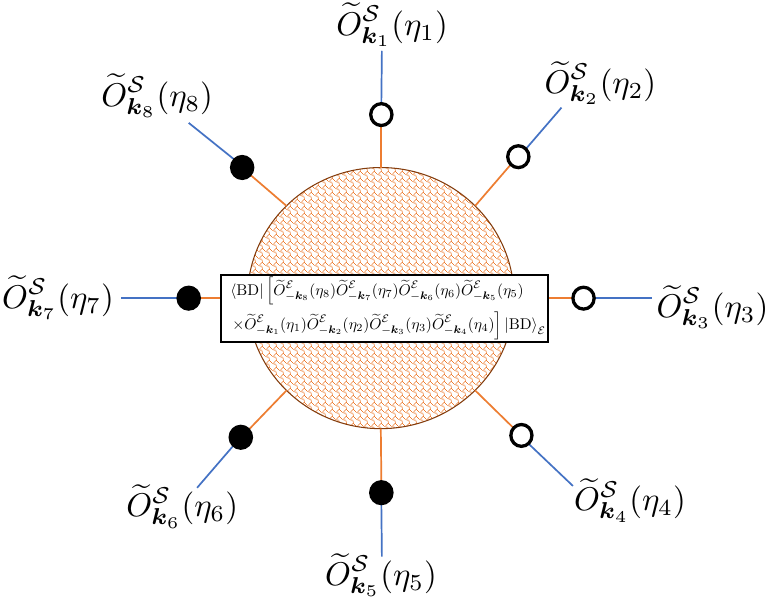}
		\caption{One of the terms appearing in the computation of $\widetilde{\rho}^{(8)}_{\mathrm{red}}$. The system is represented in blue and the environment in orange. This diagram has $n=8$ vertices, labeled by $\ell$, with $i=4$ of them being $\Circle$ vertices.}
		\label{fig:rhored}
	\end{figure}

From \Eq{eq:rhored:Kraus}, the Feynman rules for the $n^{\mathrm{th}}$-order perturbative purity are given as follows.
	\begin{tcolorbox}[%
		enhanced, 
		breakable,
		skin first=enhanced,
		skin middle=enhanced,
		skin last=enhanced
		]
		\paragraph{Feynman rules for $\widetilde{\rho}^{(n)}_{\mathrm{red}}$} (an example is given in \Fig{fig:rhored}):
		\vspace{0.1in}
		\begin{enumerate}
			\item Draw a diagram with one environment \textit{blob} connected to $n$ environment propagators leading to $n$ conversion vertices, themselves connected to $n$ system propagators.  This diagram corresponds to a fixed $i$ and $j$ in \Eq{eq:rhored:Kraus}, the vertices are labelled by $\ell$.
			\item To each vertex is associated a 
			\begin{align}
				(-i) \int_{-\infty}^\eta \dd \eta_{\ell} \int \frac{\dd^3 \bmk_{\ell}}{(2\pi)^3}.
			\end{align}
			The diagram is time-ordered such that there is an overall
			\begin{align}
				\Theta(\eta_{\ell_1} - \eta_{\ell_2}) \qquad \forall ~\ell_1 < \ell_2\, .
			\end{align}
Amongst the $n$ vertices, $i$ vertices are ``coloured'' with $\Circle \big.$, and $n-i$ with $\CIRCLE$ which contributes a factor $(-1)$. To the $\Circle$ vertices are attached $\widetilde{O}_{\bmk_\ell}^{\mathcal{S}}(\eta_\ell)$ (system propagator) and $\widetilde{O}_{-\bmk_\ell}^{\mathcal{E}}(\eta_\ell)$ (environment propagator), and to the $\CIRCLE$ vertices are attached $[\widetilde{O}_{\bmk_\ell}^{\mathcal{S}}(\eta_\ell)]^\dagger$ and $[\widetilde{O}_{-\bmk_\ell}^{\mathcal{E}}(\eta_\ell)]^\dagger$. 
			\item The amplitude part of the diagram is controlled by 
			\begin{align}
				\bra{\mathrm{BD}}  \widetilde{\mathcal{T}} \left[ \CIRCLE_{\mathcal{E}}\right] \mathcal{T} \left[ \Circle_{\mathcal{E}}\right]\ket{\mathrm{BD}}_{\mathcal{E}} 
			\end{align}
			where $\mathcal{T}[ \CIRCLE_{\mathcal{E}}]$ corresponds to the time-ordering of the environment operators connected to the $\CIRCLE$ vertices, and $\widetilde{\mathcal{T}}[ \Circle_{\mathcal{E}}]$ corresponds to the anti-time-ordering of the environment operators connected to the $\Circle$ vertices.
			\item The operator part of the diagram is controlled by
			\begin{align}
				\mathcal{T} \left[ \Circle_{\mathcal{S}}\right]\ket{\mathrm{BD}} \bra{\mathrm{BD}}_{\mathcal{S}} \widetilde{\mathcal{T}} \left[ \CIRCLE_{\mathcal{S}}\right]
			\end{align}
	with similar notations.
			\item Sum over all possibilities for the number $i$ of $\Circle$ vertices and their arrangement in the diagram (labeled by $j$). 
		\end{enumerate}             
	\end{tcolorbox}
	
		The diagrams for $\widetilde{\rho}^{(n)}_{\mathrm{red}}$ constitute the building blocks for the ones of $\gamma^{(n)}$. The latter follow from \Eq{eq:gammanorder} and are given as follows.
	
\begin{figure}[tbp]
		\centering
		\includegraphics[width=0.8\textwidth]{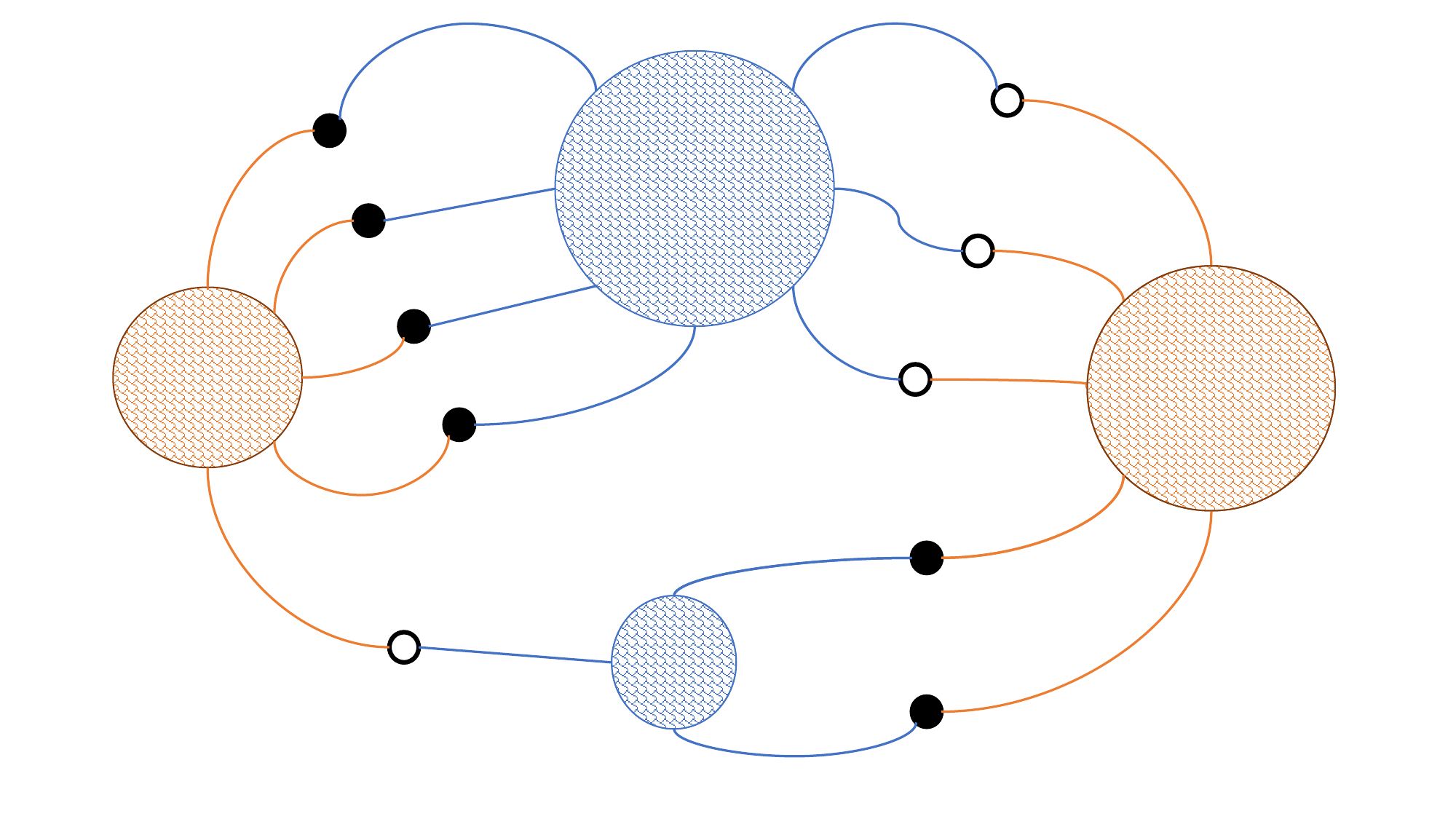}
		\caption{One of the terms appearing in the computation of $\gamma^{(10)}$ for $m=5$.}
		\label{fig:gamma}
	\end{figure}

	\begin{tcolorbox}[%
		enhanced, 
		breakable,
		skin first=enhanced,
		skin middle=enhanced,
		skin last=enhanced
		]
		\paragraph{Feynman rules for $\gamma^{(n)}$} (an example is given in \Fig{fig:gamma}):
		\vspace{0.1in}
		\begin{enumerate}
			\item Draw one of the diagrams appearing in $\widetilde{\rho}^{(m)}_{\mathrm{red}}$ (called \textit{left cluster}) and one of the diagrams appearing in $\widetilde{\rho}^{(n-m)}_{\mathrm{red}}$ called \textit{right cluster}) for $m$ between $0$ and $ \lfloor \frac{n}{2}\rfloor $.
			\item Add two system blobs between the clusters: the first one is connected to all $\CIRCLE$ vertices of the left cluster and all $\Circle$ vertices of the right cluster, while the second one is connected to all $\Circle$ vertices of the left cluster and all $\CIRCLE$ vertices of the right cluster.
			\item Each system blob contributes to 
			\begin{align}
				\bra{\mathrm{BD}}  \widetilde{\mathcal{T}} \left[ \CIRCLE_{\mathcal{S}}\right] \mathcal{T} \left[ \Circle_{\mathcal{S}}\right]\ket{\mathrm{BD}}_{\mathcal{S}} \, ,
			\end{align}
                which is further multiplied by the amplitude parts of the environment blobs.
			\item Sum over $m$ from $0$ to $ \lfloor \frac{n}{2}\rfloor$. There is a prefactor $2$ if $m\neq n/2$. 
		\end{enumerate}             
	\end{tcolorbox}
	
\begin{figure}[tbp]
		\centering
		\includegraphics[width=0.8\textwidth]{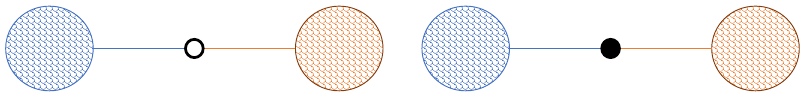}
		\caption{The two terms appearing in the computation of $\gamma^{(1)}$.}
		\label{fig:pert10}
	\end{figure}

In order to get more familiar with the above diagrammatic representation, let us employ it to compute perturbative purity up to order $g^4$.
\paragraph{$\gamma^{(1)}$:}
 At first order, there are only two diagrams, displayed in \Fig{fig:pert10}. Using the Feynman rules derived above, we recover the result given in \Eq{eq:pert10}, that is 
	\begin{align}
		\gamma^{(1)} = -2i \int_{-\infty}^\eta \dd \eta_1 \int \frac{\dd^3 \bmk}{(2\pi)^3} \left[\mathcal{K}_{\mathcal{S}}^{(1)}(\bmk, \eta_1)\mathcal{K}_{\mathcal{E}}^{(1)}(\bmk, \eta_1) - \mathcal{K}_{\mathcal{S}}^{(1)}(\bmk, \eta_1)\mathcal{K}_{\mathcal{E}}^{(1)}(\bmk, \eta_1) \right]= 0.
	\end{align}
The first non-trivial contribution thus arises at second order, as already mentioned above.

\begin{figure}[tbp]
		\centering
		\includegraphics[width=0.8\textwidth]{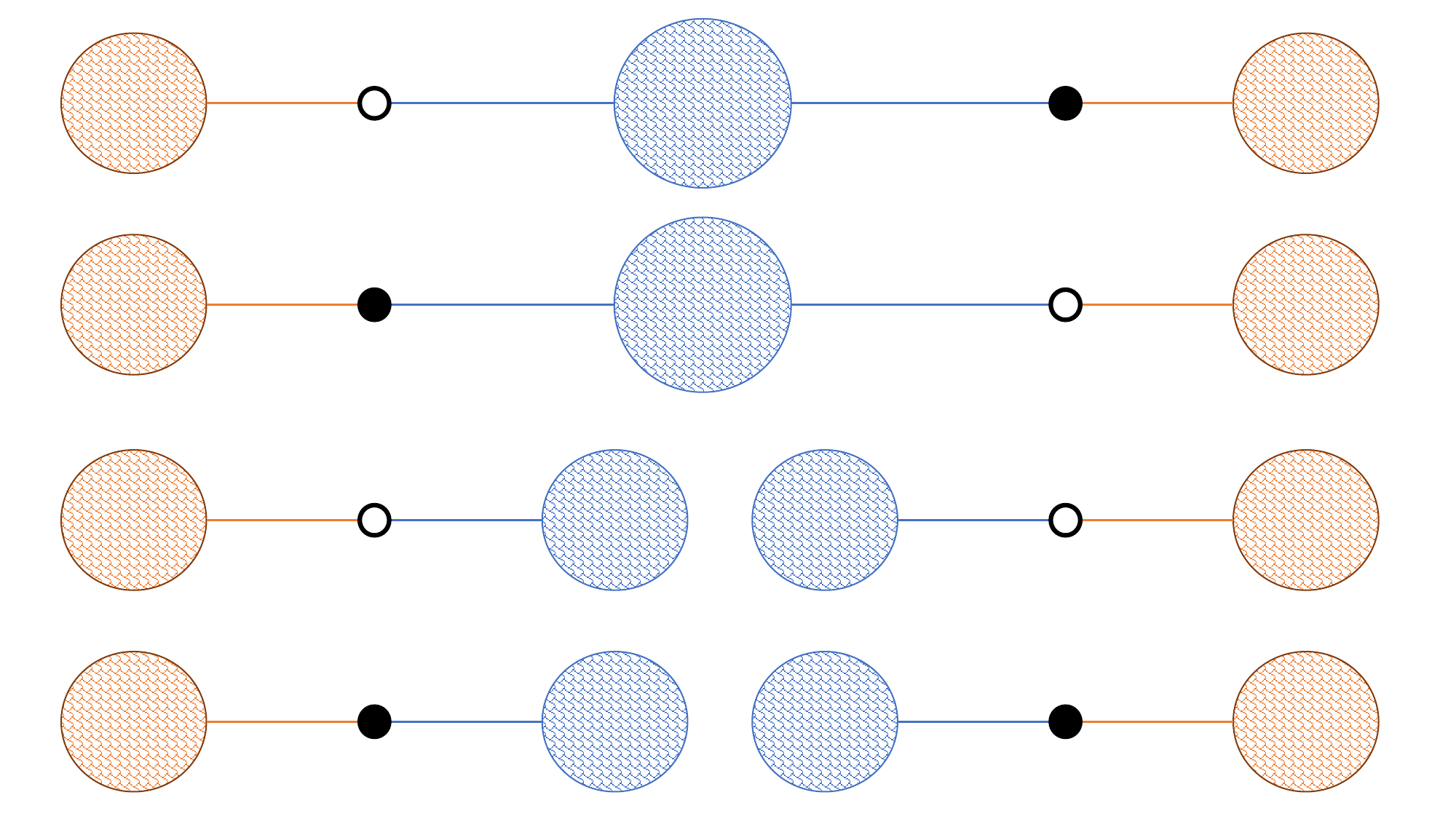}
		\caption{Diagrams appearing in the computation of $\gamma^{(1,1)}$.}
		\label{fig:pert11}
	\end{figure}

\begin{figure}[tbp]
		\centering
		\includegraphics[width=0.8\textwidth]{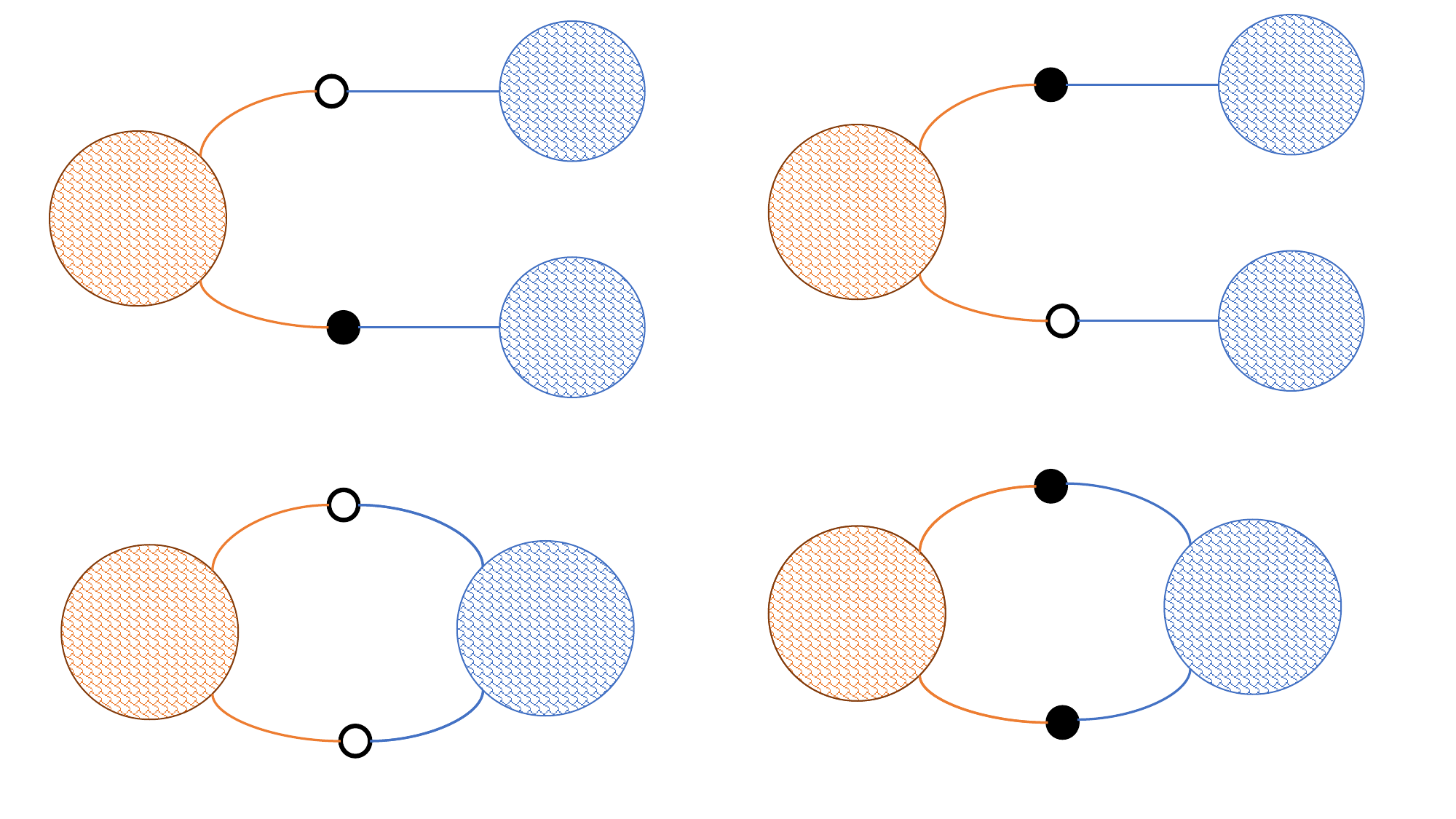}
		\caption{Diagrams appearing in the computation of $\gamma^{(2,0)}$.}
		\label{fig:pert20}
	\end{figure}

\paragraph{$\gamma^{(2)}$:} Second-order purity is made of two kinds of diagrams that come from $m=0$ and $m=1$. To simplify the discussions, let us introduce 
	\begin{align}
		\gamma^{(n-m,m)} \equiv 2^{1-\delta_{n,m-n}}\mathrm{Tr}_{\mathcal{S}}\left[\widetilde{\rho}_{\mathrm{red}}^{(m)}\, \widetilde{\rho}_{\mathrm{red}}^{(n-m)} \right].
	\end{align}
The diagrams for $\gamma^{(1,1)}$ are given in \Fig{fig:pert11} and the ones for $\gamma^{(2,0)}$ in \Fig{fig:pert20}. Let us first consider $\gamma^{(1,1)}$. Following the Feynman rules given above, we recover \Eq{eq:term1}, namely
	\begin{align}\label{eq:gamma11}
		\gamma^{(1,1)}
		&= (-i)^2 \int_{-\infty}^\eta \dd \eta_1  \int_{-\infty}^{\eta} \dd \eta_2 \int  \frac{\dd^3\bmk_1}{(2\pi)^3}  \int  \frac{\dd^3\bmk_2}{(2\pi)^3} \\
		\bigg[-& \mathcal{K}_{\mathcal{S}}^{(2)}(\bmk_1,\bmk_2, \eta_1,\eta_2)\mathcal{K}_{\mathcal{E}}^{(1)}(\bmk_1, \eta_1) \mathcal{K}_{\mathcal{E}}^{(1)}(\bmk_2, \eta_2)
		- \mathcal{K}_{\mathcal{S}}^{(2)}(\bmk_2,\bmk_1, \eta_2,\eta_1)\mathcal{K}_{\mathcal{E}}^{(1)}(\bmk_1, \eta_1) \mathcal{K}_{\mathcal{E}}^{(1)}(\bmk_2, \eta_2) \nonumber \\
		+&\mathcal{K}_{\mathcal{S}}^{(1)}(\bmk_1, \eta_1) \mathcal{K}_{\mathcal{S}}^{(1)}(\bmk_2, \eta_2)\mathcal{K}_{\mathcal{E}}^{(1)}(\bmk_1, \eta_1) \mathcal{K}_{\mathcal{E}}^{(1)}(\bmk_2, \eta_2)
		+\mathcal{K}_{\mathcal{S}}^{(1)}(\bmk_1, \eta_1) \mathcal{K}_{\mathcal{S}}^{(1)}(\bmk_2, \eta_2)\mathcal{K}_{\mathcal{E}}^{(1)}(\bmk_1, \eta_1) \mathcal{K}_{\mathcal{E}}^{(1)}(\bmk_2, \eta_2)\bigg] \, .  \nonumber
	\end{align} 
	Similarly for $\gamma^{(2,0)}$, one can recover from \Fig{fig:pert20} the result of \Eq{eq:pert20}, that is
	\begin{align}\label{eq:gamma20}
		\gamma^{(2,0)} &=  2(-i)^2\int_{-\infty}^\eta \dd \eta_1  \int_{-\infty}^{\eta_1} \dd \eta_2 \int  \frac{\dd^3\bmk_1}{(2\pi)^3} \int  \frac{\dd^3\bmk_2}{(2\pi)^3}  \nonumber \\
		\bigg[ -&\mathcal{K}_{\mathcal{S}}^{(1)}(\bmk_1, \eta_1) \mathcal{K}_{\mathcal{S}}^{(1)}(\bmk_2, \eta_2) \mathcal{K}_{\mathcal{E}}^{(2)}(\bmk_1,\bmk_2, \eta_1,\eta_2) \nonumber-\mathcal{K}_{\mathcal{S}}^{(1)}(\bmk_1, \eta_1) \mathcal{K}_{\mathcal{S}}^{(1)}(\bmk_2, \eta_2) \mathcal{K}_{\mathcal{E}}^{(2)}(\bmk_2,\bmk_1, \eta_2,\eta_1) \Big.\nonumber \\
		+& \mathcal{K}_{\mathcal{S}}^{(2)}(\bmk_1,\bmk_2, \eta_1,\eta_2) \mathcal{K}_{\mathcal{E}}^{(2)}(\bmk_1,\bmk_2, \eta_1,\eta_2)  +\mathcal{K}_{\mathcal{S}}^{(2)}(\bmk_2,\bmk_1, \eta_2,\eta_1) \mathcal{K}_{\mathcal{E}}^{(2)}(\bmk_2,\bmk_1, \eta_2,\eta_1)\bigg].
	\end{align}
	
	
\begin{figure}[tbp]
		\centering
		\includegraphics[width=0.8\textwidth]{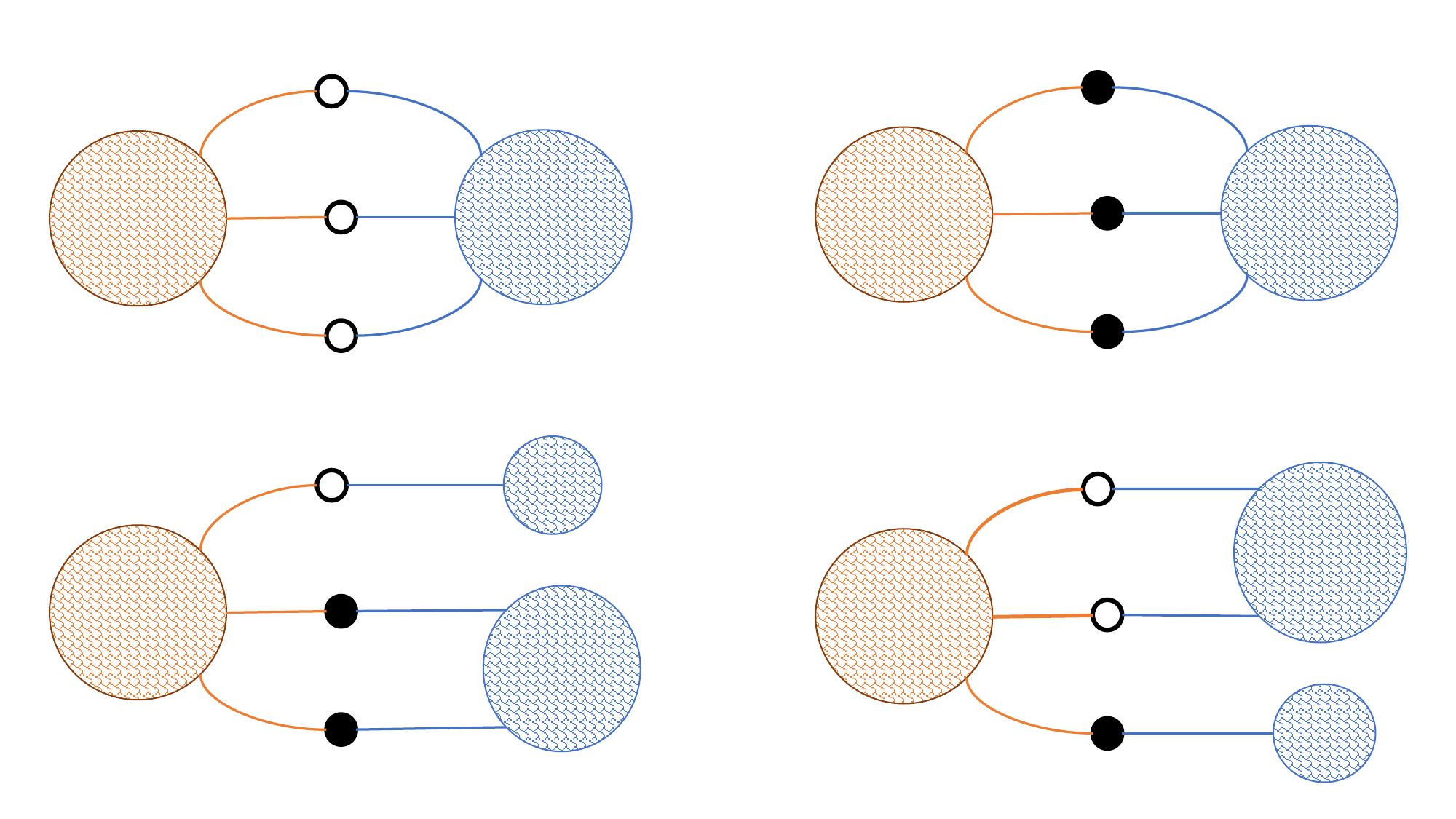}
		\caption{Diagrams appearing in the computation of $\gamma^{(3,0)}$.}
		\label{fig:pert30}
	\end{figure}

\begin{figure}[tbp]
		\centering
		\includegraphics[width=1\textwidth]{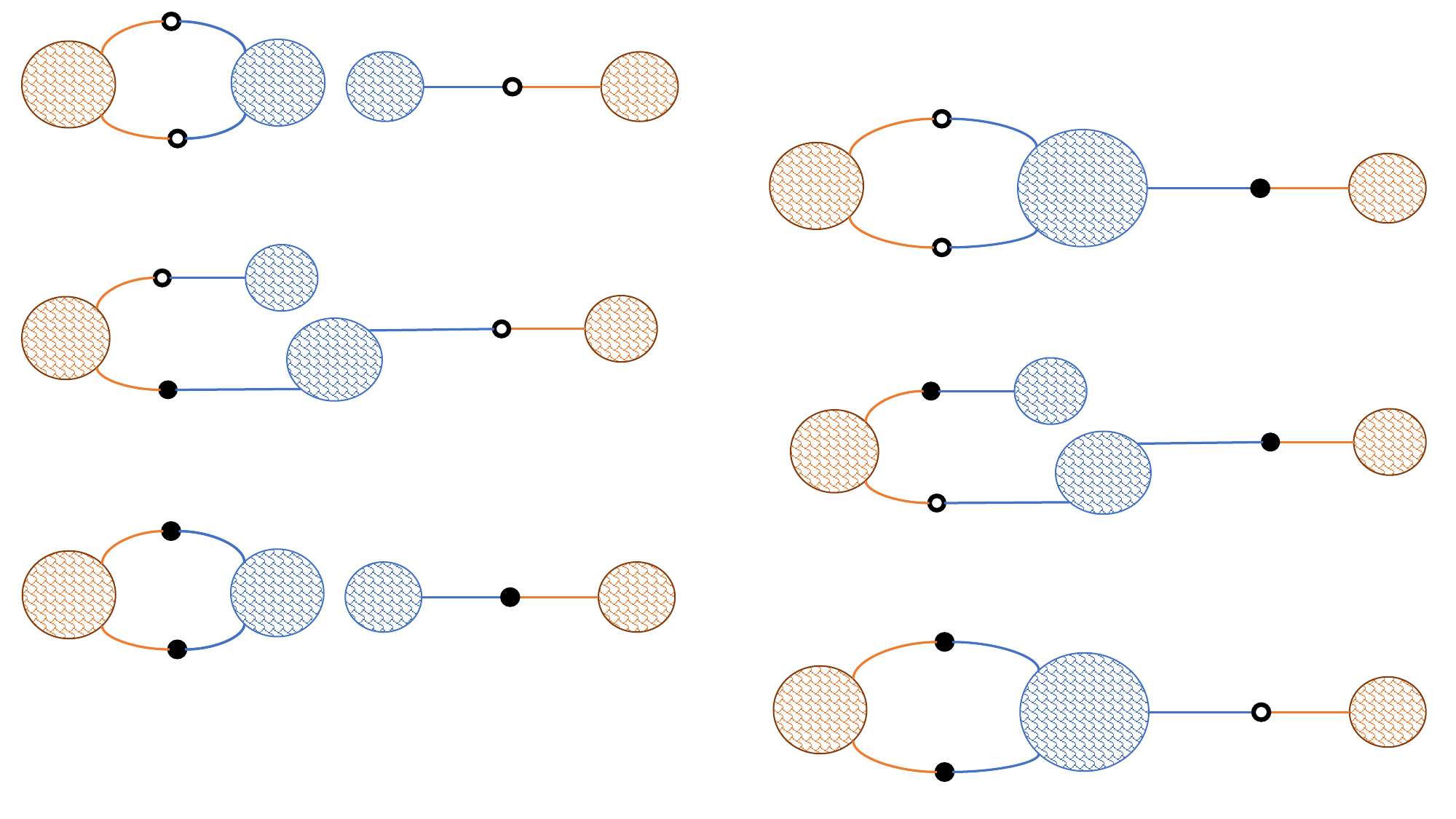}
		\caption{Diagrams appearing in the computation of $\gamma^{(2,1)}$.}
		\label{fig:pert21}
	\end{figure}

	\paragraph{$\gamma^{(3)}$:} 
	We have to compute diagrams for $\gamma^{(3,0)}$ and $\gamma^{(2,1)}$, which are represented in \Figs{fig:pert30} and \ref{fig:pert21} respectively. From the diagrammatic representation, one has
	\begin{align}
		\gamma^{(3,0)} =& 2(-i)^3\int_{-\infty}^\eta \dd \eta_1  \int_{-\infty}^{\eta_1} \dd \eta_2 \int_{-\infty}^{\eta_2} \dd \eta_3 \int  \frac{\dd^3\bmk_1}{(2\pi)^3} \int  \frac{\dd^3\bmk_2}{(2\pi)^3} \int  \frac{\dd^3\bmk_3}{(2\pi)^3} \nonumber \\
		\bigg[ &\mathcal{K}_{\mathcal{S}}^{(3)}(\bmk_1,\bmk_2,\bmk_3, \eta_1,\eta_2,\eta_3) \mathcal{K}_{\mathcal{E}}^{(3)}(\bmk_1,\bmk_2,\bmk_3, \eta_1,\eta_2,\eta_3) \nonumber\\
		&- \mathcal{K}_{\mathcal{S}}^{(3)}(\bmk_3,\bmk_2,\bmk_1, \eta_3,\eta_2,\eta_1) \mathcal{K}_{\mathcal{E}}^{(3)}(\bmk_3,\bmk_2,\bmk_1, \eta_3,\eta_2,\eta_1) \Big. \nonumber \\
		&+\mathcal{K}_{\mathcal{S}}^{(1)}(\bmk_1,\eta_1) \mathcal{K}_{\mathcal{S}}^{(2)}(\bmk_3,\bmk_2,\eta_3,\eta_2) \mathcal{K}_{\mathcal{E}}^{(3)}(\bmk_3,\bmk_2,\bmk_1, \eta_3,\eta_2,\eta_1)  \Big. \nonumber\\
		&+\mathcal{K}_{\mathcal{S}}^{(1)}(\bmk_2,\eta_2) \mathcal{K}_{\mathcal{S}}^{(2)}(\bmk_3,\bmk_1,\eta_3,\eta_1) \mathcal{K}_{\mathcal{E}}^{(3)}(\bmk_3,\bmk_1,\bmk_2, \eta_3,\eta_1,\eta_2)  \Big. \nonumber\\
		&+\mathcal{K}_{\mathcal{S}}^{(1)}(\bmk_3,\eta_3) \mathcal{K}_{\mathcal{S}}^{(2)}(\bmk_2,\bmk_1,\eta_2,\eta_1) \mathcal{K}_{\mathcal{E}}^{(3)}(\bmk_2,\bmk_1,\bmk_3, \eta_2,\eta_1,\eta_3)  \Big. \nonumber\\
		&-\mathcal{K}_{\mathcal{S}}^{(1)}(\bmk_1,\eta_1) \mathcal{K}_{\mathcal{S}}^{(2)}(\bmk_2,\bmk_3,\eta_2,\eta_3) \mathcal{K}_{\mathcal{E}}^{(3)}(\bmk_1,\bmk_2,\bmk_3, \eta_1,\eta_2,\eta_3)  \Big. \nonumber\\
		&-\mathcal{K}_{\mathcal{S}}^{(1)}(\bmk_2,\eta_2) \mathcal{K}_{\mathcal{S}}^{(2)}(\bmk_1,\bmk_3,\eta_1,\eta_3) \mathcal{K}_{\mathcal{E}}^{(3)}(\bmk_2,\bmk_1,\bmk_3, \eta_2,\eta_1,\eta_3)  \Big. \nonumber\\
		&-\mathcal{K}_{\mathcal{S}}^{(1)}(\bmk_3,\eta_3) \mathcal{K}_{\mathcal{S}}^{(2)}(\bmk_1,\bmk_2,\eta_1,\eta_2) \mathcal{K}_{\mathcal{E}}^{(3)}(\bmk_3,\bmk_1,\bmk_2, \eta_3,\eta_1,\eta_2)  
		\bigg] \, .
	\end{align}
This coincides with what can be directly obtained from \Eq{eq:rhored3}. Similarly, we find 
	\begin{align}
		\gamma^{(2,1)} =& 2(-i)^3\int_{-\infty}^\eta \dd \eta_1  \int_{-\infty}^{\eta_1} \dd \eta_2 \int_{-\infty}^{\eta} \dd \eta_3 \int  \frac{\dd^3\bmk_1}{(2\pi)^3} \int  \frac{\dd^3\bmk_2}{(2\pi)^3} \int  \frac{\dd^3\bmk_3}{(2\pi)^3} \nonumber \\
		\bigg[ &\mathcal{K}_{\mathcal{S}}^{(1)}(\bmk_3,\eta_3) \mathcal{K}_{\mathcal{S}}^{(2)}(\bmk_1,\bmk_2,\eta_1,\eta_2) \mathcal{K}_{\mathcal{E}}^{(1)}(\bmk_3,\eta_3) \mathcal{K}_{\mathcal{E}}^{(2)}(\bmk_1,\bmk_2,\eta_1,\eta_2)\nonumber\\
		&-\mathcal{K}_{\mathcal{S}}^{(3)}(\bmk_3,\bmk_1,\bmk_2,\eta_3,\eta_1,\eta_2) \mathcal{K}_{\mathcal{E}}^{(1)}(\bmk_3,\eta_3) \mathcal{K}_{\mathcal{E}}^{(2)}(\bmk_1,\bmk_2,\eta_1,\eta_2)\Big.\nonumber\\
		&-\mathcal{K}_{\mathcal{S}}^{(1)}(\bmk_1,\eta_1) \mathcal{K}_{\mathcal{S}}^{(2)}(\bmk_2,\bmk_3,\eta_2,\eta_3) \mathcal{K}_{\mathcal{E}}^{(1)}(\bmk_3,\eta_3) \mathcal{K}_{\mathcal{E}}^{(2)}(\bmk_2,\bmk_1,\eta_2,\eta_1)\Big.\nonumber\\
		&-\mathcal{K}_{\mathcal{S}}^{(1)}(\bmk_2,\eta_2) \mathcal{K}_{\mathcal{S}}^{(2)}(\bmk_1,\bmk_3,\eta_1,\eta_3) \mathcal{K}_{\mathcal{E}}^{(1)}(\bmk_3,\eta_3) \mathcal{K}_{\mathcal{E}}^{(2)}(\bmk_1,\bmk_2,\eta_1,\eta_2)\Big.\nonumber\\
		&+\mathcal{K}_{\mathcal{S}}^{(1)}(\bmk_1,\eta_1) \mathcal{K}_{\mathcal{S}}^{(2)}(\bmk_3,\bmk_2,\eta_3,\eta_2) \mathcal{K}_{\mathcal{E}}^{(1)}(\bmk_3,\eta_3) \mathcal{K}_{\mathcal{E}}^{(2)}(\bmk_1,\bmk_2,\eta_1,\eta_2)\Big.\nonumber\\
		&+\mathcal{K}_{\mathcal{S}}^{(1)}(\bmk_2,\eta_2) \mathcal{K}_{\mathcal{S}}^{(2)}(\bmk_3,\bmk_1,\eta_3,\eta_1) \mathcal{K}_{\mathcal{E}}^{(1)}(\bmk_3,\eta_3) \mathcal{K}_{\mathcal{E}}^{(2)}(\bmk_2,\bmk_1,\eta_2,\eta_1)\Big.\nonumber\\ 
		&-\mathcal{K}_{\mathcal{S}}^{(1)}(\bmk_3,\eta_3) \mathcal{K}_{\mathcal{S}}^{(2)}(\bmk_2,\bmk_1,\eta_2,\eta_1) \mathcal{K}_{\mathcal{E}}^{(1)}(\bmk_3,\eta_3) \mathcal{K}_{\mathcal{E}}^{(2)}(\bmk_2,\bmk_1,\eta_2,\eta_1)\Big.\nonumber\\
		&+\mathcal{K}_{\mathcal{S}}^{(3)}(\bmk_2,\bmk_1,\bmk_3,\eta_2,\eta_1,\eta_3) \mathcal{K}_{\mathcal{E}}^{(1)}(\bmk_3,\eta_3) \mathcal{K}_{\mathcal{E}}^{(2)}(\bmk_2,\bmk_1,\eta_2,\eta_1)\bigg]\, .
	\end{align}
	
	
\begin{figure}[tbp]
		\centering
		\includegraphics[width=0.7\textwidth]{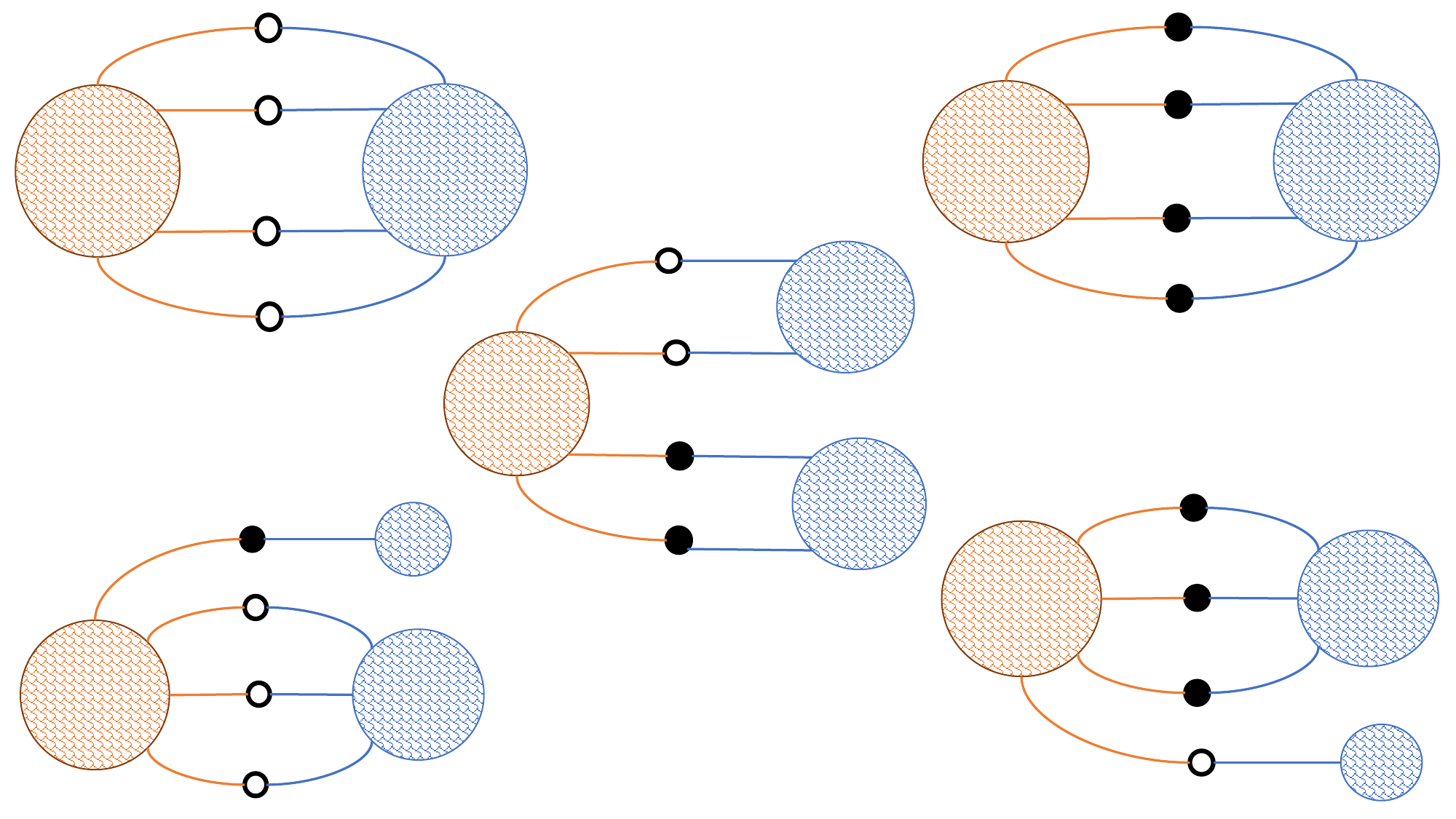}
		\caption{Diagrams appearing in the computation of $\gamma^{(4,0)}$.}
		\label{fig:pert40}
	\end{figure}

\begin{figure}[tbp]
		\centering
		\includegraphics[width=1\textwidth]{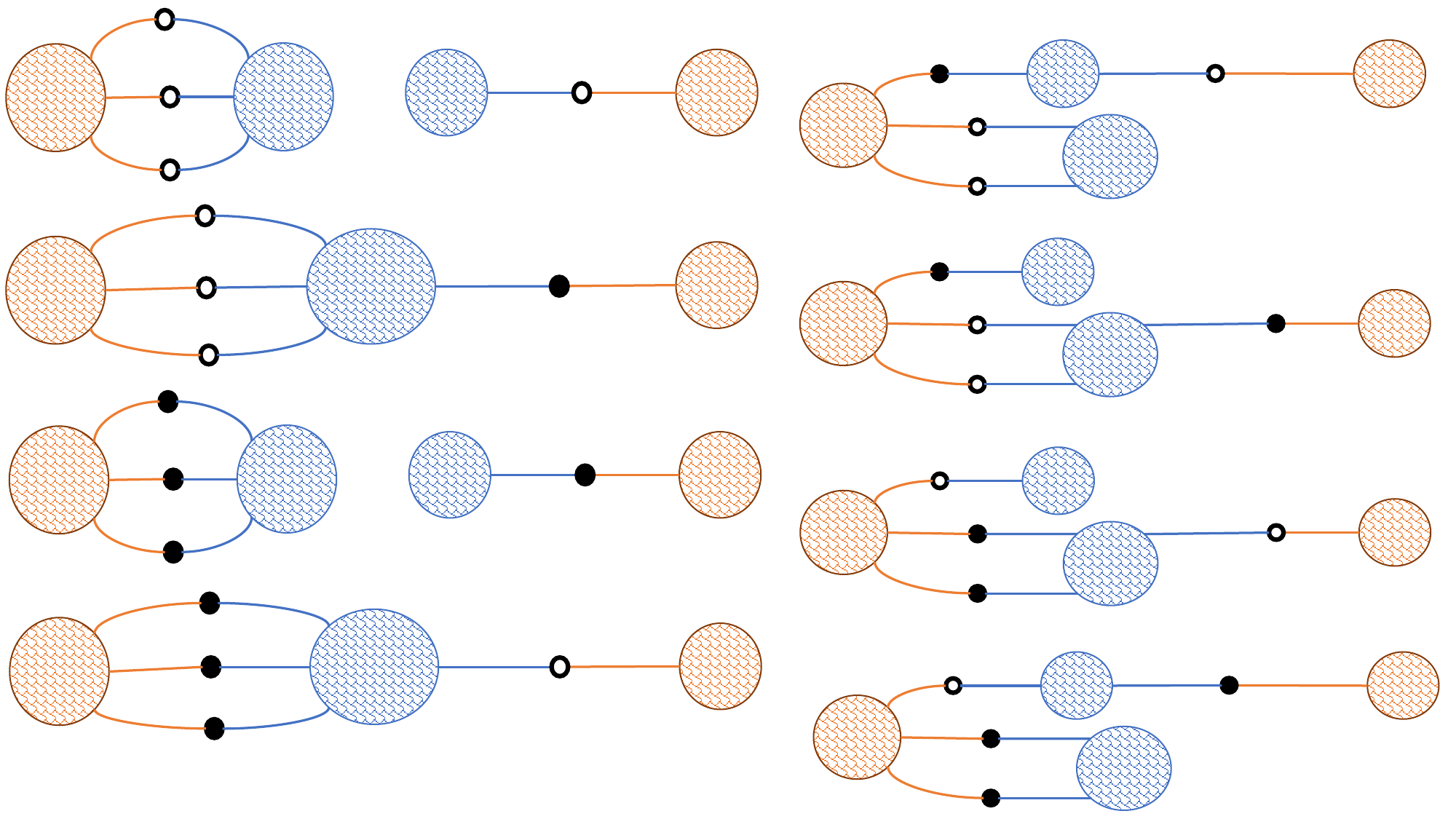}
		\caption{Diagrams appearing in the computation of $\gamma^{(3,1)}$.}
		\label{fig:pert31}
	\end{figure}

\begin{figure}[tbp]
		\centering
		\includegraphics[width=1\textwidth]{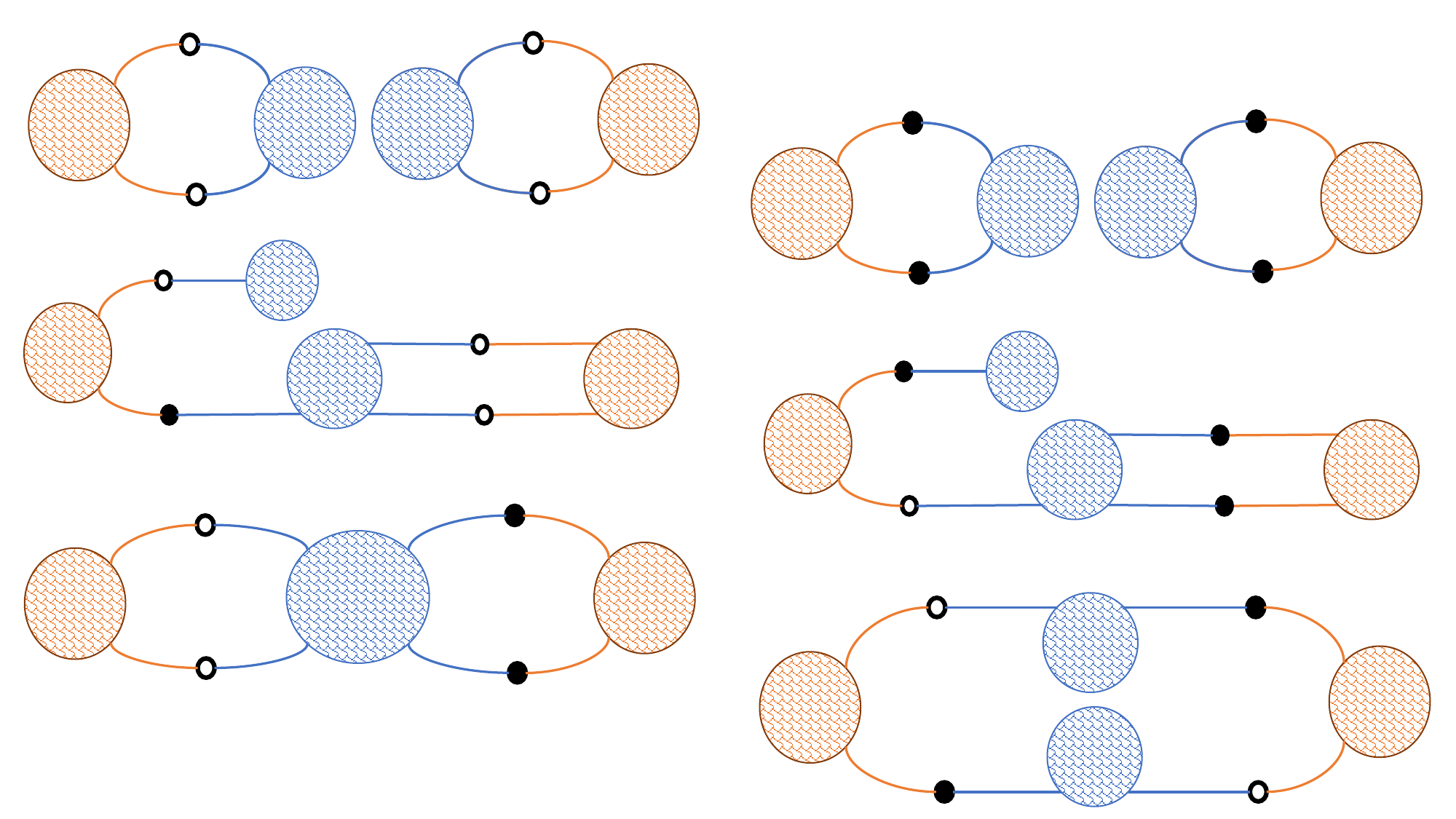}
		\caption{Diagrams appearing in the computation of $\gamma^{(2,2)}$.}
		\label{fig:pert22}
	\end{figure}

	\paragraph{$\gamma^{(4)}$:}  In \Figs{fig:pert40}, \ref{fig:pert31} and \ref{fig:pert22}, we draw the diagrams needed to compute $\gamma^{(4)}$. They correspond to $\gamma^{(4,0)}$, $\gamma^{(3,1)}$ and $\gamma^{(2,2)}$ respectively. We do not write the corresponding expressions here, for the sake of brevity, but they can be obtained straightforwardly following similar lines.
	
Ultimately, the computation boils down to the evaluation of the blobs representing the various unequal-time correlators of the system and the environment in a model of interest. If the theory does not contain any system and environment non-linearities (that is if $\widehat{H}_{\mathcal{S}}$ and $\widehat{H}_{\mathcal{E}}$ are quadratic), the expectation values $\langle  \widetilde{\mathcal{T}} \left[ \CIRCLE_{\mathcal{S}}\right] \mathcal{T} \left[ \Circle_{\mathcal{S}}\right] \rangle_{\mathcal{S}}$ and $\langle  \widetilde{\mathcal{T}} \left[ \CIRCLE_{\mathcal{E}}\right] \mathcal{T} \left[ \Circle_{\mathcal{E}}\right] \rangle_{\mathcal{E}}$ can readily be obtained  by performing Wick contractions.\footnote{In this sense, we notice that in the in-in path integral language, the $\Circle$ vertices correspond to field insertions from the $+$ branch of the path integral and the $\CIRCLE$ vertices to field insertions from the $-$ branch of the path integral (see \Refs{breuerTimelocalMasterEquations2002, Colas:2023wxa} for a more in-depth discussion of the relations between operator and path integral formalisms).} Beyond this simple case, one may wish to further connect the current representation to the standard in-in diagrammatics~\cite{Chen:2017ryl} in order to evaluate the blobs in terms of time and momenta integrals of the system and environment propagators. In this work we will not consider setups with system or environment self-interactions, although we come back to this issue in \Sec{sec:conclu}.   

\clearpage

\subsection{Other quantum information measures}
\label{sec:vN}
	
	Although the focus so far has been on computing the purity, we mention here that our formalism can be easily extended to computing other quantum information measures. Let us first mention the R\'enyi-$N$ entropy~\cite{renyi1961measures}
\begin{eqnarray}
S_{N} & \equiv & \frac{1}{1-N} \ln \left[ \mathrm{Tr}_{\mathcal{S}}\left( \widehat{\rho}_{\mathrm{red}}^N \right) \right] .
\end{eqnarray}
When $N=2$, this reduces to the R\'enyi-2 entropy given in \Eq{eq:Renyi2}, hence to the purity. The R\'enyi-$N$ entropy turns out to lend itself well to our formalism, as one must simply take $N$ product of the reduced density matrix already computed in previous subsections. In particular, using \Eq{rhored_n} one finds that for integers $N \geq 3$
\bea
\mathrm{Tr}_{\mathcal{S}}\left( \widehat{\rho}_{\mathrm{red}}^N \right) = &  \mathrm{Tr}_{\mathcal{S}}\left[ \widehat{\rho}_{\mathrm{red}}^{(0)} \right] + g N \mathrm{Tr}_{\mathcal{S}}\left[  \widehat{\rho}_{\mathrm{red}}^{(0)} \widehat{\rho}_{\mathrm{red}}^{(1)} \right] \label{Ndef} \\
& +  g^2 \left\lbrace N \mathrm{Tr}_{\mathcal{S}}\left[  \widehat{\rho}_{\mathrm{red}}^{(0)} \widehat{\rho}_{\mathrm{red}}^{(2)} \right] + N \mathrm{Tr}_{\mathcal{S}}\left[  \widehat{\rho}_{\mathrm{red}}^{(0)} \big( \widehat{\rho}_{\mathrm{red}}^{(1)} \big)^2 \right] + \frac{N(N-3)}{2} \mathrm{Tr}_{\mathcal{S}}\left[ \widehat{\rho}_{\mathrm{red}}^{(0)} \widehat{\rho}_{\mathrm{red}}^{(1)} \widehat{\rho}_{\mathrm{red}}^{(0)} \widehat{\rho}_{\mathrm{red}}^{(1)} \right] \right\rbrace \\
& + \mathcal{O}(g^3) \, ,
\eea
where cyclicity of the trace has been applied and we have used that the initial state is pure, as in \Eq{IC_start}, so that $( \widehat{\rho}^{(0)}_{\mathrm{red}})^2 = \widehat{\rho}^{(0)}_{\mathrm{red}}$. Note furthermore that $\mathrm{Tr}_{\mathcal{S}} [\widetilde{\rho}_{\mathrm{red}}^{(0)}\, \widetilde{\rho}_{\mathrm{red}}^{(1)} ]  = 0$ as found in \Eq{eq:pert10}, which also implies $\mathrm{Tr}_{\mathcal{S}}\big[ \widehat{\rho}_{\mathrm{red}}^{(0)} \widehat{\rho}_{\mathrm{red}}^{(1)} \widehat{\rho}_{\mathrm{red}}^{(0)} \widehat{\rho}_{\mathrm{red}}^{(1)} \big] = 0$. Finally one also has $\mathrm{Tr}_{\mathcal{S}} [ \widehat{\rho}_{\mathrm{red}}^{(0)} ]=1$ and from \Eq{eq:rho1:expanded} one can show explicitly that $\mathrm{Tr}_{\mathcal{S}}\big[  \widehat{\rho}_{\mathrm{red}}^{(0)} \big( \widehat{\rho}_{\mathrm{red}}^{(1)} \big)^2 \big]  = \frac{1}{2} \mathrm{Tr}_{\mathcal{S}}\big[ \big( \widehat{\rho}_{\mathrm{red}}^{(1)} \big)^2 \big] $, so the above formula reduces to
\begin{eqnarray} \label{Ndef2}
\mathrm{Tr}_{\mathcal{S}}\left( \widehat{\rho}_{\mathrm{red}}^N \right) & = & 1 + N g^2   \left\lbrace \mathrm{Tr}_{\mathcal{S}}\left[ \widehat{\rho}^{(0)}_{\mathrm{red}}  \widehat{\rho}^{(2)}_{\mathrm{red}} \right] + \frac{1}{2} \mathrm{Tr}_{\mathcal{S}}\left[ \left( \widehat{\rho}_{\mathrm{red}}^{(1)} \right)^2 \right] \right\rbrace  + \mathcal{O}(g^3)\, .
\end{eqnarray}
Since the term involving $N(N-3)$ dropped out, this is in fact valid for all integers $N \geq 2$. One recognizes the combination giving $\gamma^{(2)}$ in \Eq{eq:gammanorder}, hence we have found that
\bea
S_N = \frac{N}{N-1} \frac{1-\gamma}{2}+\mathcal{O}(g^3)\, ,
\eea
which is obviously consistent with \Eq{eq:Renyi2} when $N=2$.
	
Another quantity of interest is the von-Neumann entanglement entropy of the system, defined as 
\begin{equation}
\SEE = - \mathrm{Tr}\left( \widehat{\rho}_{\mathrm{red}} \ln \widehat{\rho}_{\mathrm{red}} \right) .
\end{equation}
This is the quantum analog of the so-called Shannon entropy in classical-information theory, and it encapsulates the information inherent to the possible outcomes of a measurement on a quantum system. It is central to the computation of many other key quantum information metrics. Perturbation theory of von-Neumann entropies has been considered in various contexts before \cite{2010ChPhB..19d0308C,Balasubramanian:2011wt,Wong:2013gua,Rosenhaus:2014woa,Rodrigues:2019cnb,Tomaras:2019sjq,Chen:2020ild,Dadras:2020xfl,Grace:2021jsf, Cheung:2023hkq, Fedida:2024dwc}, and we now connect our formalism to these works. Formally, one can show that $\SEE = \lim_{N\to 1} S_N$, hence the entanglement entropy can be obtained from analytic continuation of the R\'enyi entropy, using what is known as the replica trick~\cite{Holzhey:1994we, Calabrese:2004eu}. In the present framework, this cannot be done from \Eq{Ndef2} directly, since it is valid for $N\geq 2$ only. 

However once can expand the entanglement entropy around a pure state for which $\widehat{\rho}_{\mathrm{red}}^2\simeq \widehat{\rho}_{\mathrm{red}}$ and close to which the Mercator series can be employed
\bea
\mathrm{Tr}_{\mathcal{S}}\left[ \widehat{\rho}_{\mathrm{red}}\ln\left(\widehat{\rho}_{\mathrm{red}}\right) \right] = -
\mathrm{Tr}_{\mathcal{S}}\left[ \widehat{\rho}_{\mathrm{red}}\sum_{n=1}^\infty\frac{\left(1-\widehat{\rho}_{\mathrm{red}}\right)^n}{n} \right]\, .
\eea
At leading order, one finds $\mathrm{Tr}_{\mathcal{S}} [ \widehat{\rho}_{\mathrm{red}}\ln\left(\widehat{\rho}_{\mathrm{red}}\right)  ] \simeq \mathrm{Tr}_{\mathcal{S}} [ \widehat{\rho}_{\mathrm{red}}^2]-1$, hence
\bea
\SEE=1-\gamma + \order{g^3}.
\eea
In fact, this relation is the reason why the R\'enyi-$2$ entropy is sometimes called the ``linear entropy'':  it corresponds to a linearized version of the von-Neumann entropy where only the first term in the Mercator series is kept. At higher order, the linear entropy and the von-Neumann entropy are different, but the above framework still allows one to express the von-Neumann entropy in terms of the diagrams involved in the linear entropy.
	
	\section{Application}
	\label{sec:applic}
	
	In this section, we apply the techniques developed above to a model of phenomenological interest for primordial cosmology.
	
	\subsection{EFT-inspired model}\label{subsec:model}

        Based on the Effective Field Theory of Inflation (EFToI)  \cite{Creminelli:2006xe, Cheung:2007st, Senatore:2010wk, Piazza:2013coa} and quasi-single field inflation/the gelaton scenario \cite{Tolley:2009fg,Chen:2009zp, Noumi:2012vr, Assassi:2013gxa, Tong:2017iat, An:2017hlx, Kim:2021pbr, Pimentel:2022fsc, Wang:2022eop, Jazayeri:2022kjy, Jazayeri:2023xcj}, we illustrate our method on a subset of possible interactions between the curvature perturbation $\zeta$ (the system) and a massive scalar degree of freedom $\sigma$ (the environment). We describe the model below.
        
        Perturbing about a quasi de-Sitter background and always working to leading order in the slow-roll parameter $\epsilon_1 := - \dot{H} / H^2 \ll 1$, where $H=\dot{a}/a$ is the Hubble expansion rate and $a$ is the scale factor, the first ingredient in the considered model is the pseudo-Goldstone boson $\pi$ which is related to the curvature perturbation $\zeta$ as $\zeta = - H \pi$ at leading order in cosmological perturbation theory. The formalism describes inflation as a symmetry-breaking phenomenon in which $\pi$ non-linearly realizes the broken time-diffeomorphism invariance due to the presence of the inflaton field. Being an effective field theory, the Goldstone boson $\pi$ determines the low-energy spectrum of the theory with higher-order  effects parametrized by operators compatible with the remaining spatial diffeomorphism symmetry, and systematically organized in a derivative expansion. We work in the decoupling regime\footnote{This is valid at energies larger than $E_{\mathrm{mix}} \sim M_2^2 / \Mp $, see \Refa{Cheung:2007st}.} in which the physics of the Goldstone boson decouples from the other metric fluctuations so that $\pi$ is the only scalar degree of freedom considered. Its action takes the form \cite{Cheung:2007st}
	\begin{equation}
        \label{eq:S:pi}
	S_{\mathcal{S}} = \int \dd^4x \; \sqrt{-g}\left[ \frac{1}{2} \Mp ^2 R + \Mp ^2 H^2 \left( 2 \epsilon_1 - 3 \right) - \Mp^2 H^2 \epsilon_1 \; \delta g^{00} +\frac{M^4_2}{2!}\left(\delta g^{00}\right)^2 + \frac{M^4_3}{3!} \left(\delta g^{00}\right)^3 + \cdots\right] .
	\end{equation}
    The scales $M_{2}$ and $M_{3}$ are in general time-dependent\footnote{In setups where $H$ and $\dot{H}$ do not vary significantly over a Hubble time, it is natural to assume that $M_{2}$ and $M_{3}$ do not vary much either, and treat these coefficients as constants.}, and $\pi$ is related to the time-time component of the metric fluctuation $\delta g^{00}$ via
	\begin{align}
\label{g00_def}
	\delta g^{00} = -2 \dot{\pi} - \dot{\pi}^2 + \frac{\left(\partial_i \pi \right)^2}{a^2} 
	\end{align}
where $(\partial_i \pi)^2 = \delta^{ij} \partial_i \pi \partial_j \pi$.
	Functions of the background in \Eq{eq:S:pi} are evaluated at time $t+\pi$ such that, expanding in $\pi$, we observe a cancellation of the terms linear in $\pi$ (tadpoles). This cancellation takes place ``on-shell'', \ie when the Friedmann equations hold \cite{Cheung:2007st}. The action (\ref{eq:S:pi}) neglects terms $\mathcal{O}\big[ (\delta g^{00})^4 \big]$ and others involving the extrinsic curvature (and so neglects the two transverse modes of the graviton), which are indeed subdominant in the decoupling regime mentioned above. Higher-order derivatives of the fluctuations are also dropped.
	 
	 The second ingredient to the model considered is an additional massive field $\sigma$, minimally-coupled, with an action \cite{Assassi:2013gxa,Jazayeri:2022kjy,Jazayeri:2023xcj}
       	\begin{align}
\label{eq:S:sigma}
		S_{\mathcal{E}} &= \int \dd^4x \sqrt{-g} \left( - \frac{1}{2} \partial_\mu \sigma \partial^\mu \sigma - \frac{1}{2} m^2 \sigma^2 + \cdots \right) .
	\end{align} 
Here we neglect cubic self-interactions of $\sigma$ (6 operators in total) and higher. The interactions between $\pi$ and $\sigma$ are then fixed by the EFToI symmetries \cite{Noumi:2012vr} -- we choose to examine only a subset of the possible interactions which are linear in $\sigma$ so that
	\begin{align}
        \label{eq:S:pi:sigma}
		S_{\mathrm{int}} & = \int \dd^4x\; \sqrt{-g} \left[\widetilde{M}_1^3 \delta g^{00} \sigma + \widetilde{M}_3^3 \left(\delta g^{00}\right)^2 \sigma +\ldots \right] \; 
	\end{align}
	and neglect the other possible interactions for simplicity (which means ignoring 9 other operators like $\sim \delta g^{00} \sigma^2$). Note that phenomenological properties of this model have been studied in \Refs{Arkani-Hamed:2015bza, Pimentel:2022fsc, Wang:2022eop,Jazayeri:2022kjy,Jazayeri:2023xcj}. In this work, we treat $\pi$ as the ``system'' and $\sigma$ as the ``environment''.
 
    Expanding \Eq{eq:S:pi} in terms of $\pi$, the system free action (\ref{eq:S:pi}) simplifies to (working in terms of conformal time $\eta$ and enforcing background evolution to cancel tadpoles)
	\begin{align}
\label{eq:S_S2}
	S_{\mathcal{S}} & = \int \dd \eta\; \dd^3 \bm{x} \; a^2 \left[ \frac{1}{2} \pi^{\prime 2 }_c - \frac{1}{2} \cs ^2 \left( \partial_i \pi_c \right)^2 \right] 
	\end{align}
	where cubic self interactions of $\pi$ have been neglected, and we have introduced the canonically normalized field
	\begin{align}\label{eq:pic:o}
	\pi_c \equiv \frac{\sqrt{2\epsilon_1}H \Mp }{\cs}  \pi\, \qquad \mathrm{with} \qquad \cs ^2 \equiv \left( 1 + \frac{2M_2^4}{\Mp ^2 H^2 \epsilon_1 } \right)^{-1}.
	\end{align}
        In this expression, we used the speed of sound $\cs$, forbidding superluminal propagation, $\cs^2<1$, by enforcing $M^4_2 > 0$. 
	The environment free action (\ref{eq:S:sigma}) is simply
	\begin{align}
\label{eq:S_E2}
	S_{\mathcal{E}} & = \int \dd \eta\; \dd^3\bm{x} \; a^2 \left[ \frac{1}{2} \sigma^{\prime 2 } - \frac{1}{2} ( \partial_i \sigma )^2 - \frac{1}{2} m^2 a^2 \sigma^2 \right] 
	\end{align}
	and the interaction action, expanded to cubic order in the field operators, has the form
	\begin{align}
        \label{eq:S_int2}
	S_{\mathrm{int}} & = - \int \dd \eta\; \dd^3\bm{x} \; g \left[ a^3 \pi'_c \sigma  - \frac{1}{\Lambda_1} a^2 \left(\partial_i \pi_c\right)^2 \sigma - \frac{1}{\Lambda_2} a^2 \pi_c^{\prime 2} \sigma \right] \ ,
	\end{align}
	with the definitions
	\begin{align}\label{eq:Lambda}
	g \equiv \frac{2 \cs \widetilde{M}_1^3}{\sqrt{2 \epsilon_1} H M_\mathrm{Pl} }  \ , \qquad \frac{1}{\Lambda_1} \equiv \frac{\cs}{2\sqrt{2 \epsilon_1} H \Mp  } \qquad \mathrm{and} \qquad \frac{1}{\Lambda_2} \equiv \frac{\cs}{2\sqrt{2 \epsilon_1} H \Mp  } \left(\frac{4 \widetilde{M}_{3}^3}{\widetilde{M}_{1}^3} - 1 \right) \ .
	\end{align}
	One can see that $S_{\mathrm{int}}$ is controlled by an overall dimensionful coupling $g$ and contains three terms. On the one hand, the first two terms in \Eq{eq:S_int2} are both generated by the operator $\widetilde{M}_1^3 \delta g^{00} \sigma$ upon using \Eq{g00_def}. On the other hand, the term $\pi_c^{\prime 2} \sigma$ is generated by both $\widetilde{M}_1^3 \delta g^{00} \sigma$ and $ \widetilde{M}_3^3 (\delta g^{00})^2 \sigma$, which is why the scale $\Lambda_1$ is not always fixed in the same way as $\Lambda_2$. Note that in the case where the term $\widetilde{M}_3^3 (\delta g^{00})^2 \sigma$  in \Eq{eq:S:pi:sigma} can be neglected, the relation $\Lambda_2 = - \Lambda_1$ holds, and this will play a key role in dealing with some of the divergences we encounter below [we return to this point below \Eq{C3_text} later on].

In what follows, it will also be convenient to parametrize our results in terms of the amplitude of the reduced power spectrum of curvature perturbations~\cite{Mukhanov:1985rz, Mukhanov:1988jd, Wei:2004xx},
	\begin{eqnarray} \label{PowSpec_def}
	\Delta_{\zeta}^2 \equiv \frac{H^2}{8 \pi^2 \epsilon_1 M_\mathrm{Pl}^2 \cs} \simeq  2.2 \times 10^{-9}
	\end{eqnarray}
since this combination will appear in various expressions. The numerical value is quoted from measurements of the CMB temperature anisotropies~\cite{Planck:2018jri}.

\subsubsection{Canonical variables and Hamiltonian}

We rescale the field variables according to
    \begin{equation}
    v_{\pi}=a \pi_c\qquad \mathrm{and} \qquad v_\sigma=a \sigma\ , 
    \end{equation}
    in order for their kinetic term to be canonically normalized. By further adding the total derivative $\frac{1}{2}\frac{\dd}{\dd \eta}(\frac{a'}{a}v_\pi^2 + \frac{a'}{a} v_\sigma^2)$ to the Lagrangian density $\mathcal{L}$ in $S = \int \exd \eta\; \exd^3\bm{x}\; \mathcal{L}$ (see \Refs{Braglia:2024zsl, Sou:2024tjv} for recent discussions on this topic), the action $S = S_{\mathcal{S}} + S_{\mathcal{E}} + S_{\mathrm{int}} $ is given by
     \begin{align}
     S_{\mathcal{S}} = & \int \dd \eta\; \dd^3\bm{x} \; \frac{1}{2} \left[ (v_\pi ')^2 + \frac{a''}{a} v_\pi^2 -  \cs^2 (\partial_i v_\pi)^2 \right] ,\\
      S_{\mathcal{E}} = & \int \dd \eta\; \dd^3\bm{x} \;  \frac{1}{2}\left[(v_\sigma ')^2 + \left(\frac{a''}{a}-a^2 m^2\right) v_\sigma^2 -  (\partial_i v_\sigma)^2 \right] ,\\
      S_{\mathrm{int}} = & - g \int \dd \eta\; \dd^3\bm{x} \;   \left[a v_\sigma \left(v_\pi'-\frac{a'}{a}v_\pi\right) - \frac{v_\sigma}{a\Lambda_1} \left(\partial_i v_\pi\right)^2 - \frac{v_\sigma}{a\Lambda_2}\left(v_\pi' - \frac{a'}{a} v_\pi\right)^2\right] .
     \end{align}
The canonical momenta are determined from
\begin{align}
p_\pi & =  \frac{\partial \mathcal{L}}{\partial v_\pi'} = v_\pi' \left( 1 + \frac{2 g}{a\Lambda_2}   v_{\sigma}\right) - g a v_\sigma - \frac{2 g a'}{a^2 \Lambda_2} v_\pi v_\sigma ,\label{eq:momdef} \\
p_\sigma & = \frac{\partial \mathcal{L}}{\partial v_\sigma'} = v_\sigma'\, ,
\end{align}
and the Hamiltonian is then obtained by performing a Legendre transform
\begin{align}
H = & \int \exd^3\bm{x} \; \big[  p_{\pi} v_\pi^\prime + p_\sigma \sigma^\prime - \mathcal{L} \big] \\
 = &  \int \exd^3\bm{x}  \; \bigg\{ \frac{1}{2}\left[p_\pi^2 + \cs^2 \left(\partial_i v_\pi\right)^2 - \frac{a''}{a} v_\pi^2\right]
+\frac{1}{2}\left[p_\sigma^2 +  \left(\partial_i v_\sigma\right)^2 +\left(a^2 m^2- \frac{a''}{a}\right) v_\sigma^2\right] 
\nonumber \\ &
+ a g  v_\sigma \left(p_\pi - \frac{a'}{a}v_\pi \right)
- \frac{g v_\sigma}{a\Lambda_1} \left(\partial_i v_\pi\right)^2 - \frac{gv_\sigma }{a \Lambda_2}\left( p_\pi - \frac{a'}{a} v_\pi\right)^2
\nonumber \\ &
+ \frac{g^2 v_\sigma^2}{2 a^2 \Lambda_2^2}\frac{1}{1+\frac{2 g}{a \Lambda_2}v_\sigma} \left(a^2 \Lambda_2 - 2 p_\pi + 2 \frac{a'}{a} v_\pi\right)^2 \; \bigg\}\, .
\label{eq:Hamilt:EFT:full}
\end{align}
In this expression, the terms in the first line are quadratic and correspond to the ``free'' Hamiltonian, \ie they leave the $\pi$ and $\sigma$ sectors uncoupled. The terms in the second line belong to the interaction Hamiltonian -- they are either quadratic or cubic, and arise at order $g$. When expanding the terms in the third line at quadratic order in the fields, one finds $a^2 g^2 v_\sigma^2/2$, which is non interacting. It can thus be appended to the first line, which corresponds to renormalizing the mass of the $\sigma$ field according to $m^2\to m^2+g^2$. All other terms of the third line are interacting terms, they are cubic or higher order in the fields, and they arise at order $g^2$ and above. At leading order in $g$ they can thus be neglected (the validity of this truncation is further discussed in \Sec{sec:validity}), and the Hamiltonian $H = H_{\mathcal{S}} + H_{\mathcal{E}} + g H_{\mathrm{int}}$ becomes
\begin{align}
H_{\mathcal{S}} & = \int \exd^3\bm{x} \; \frac{1}{2}\left[p_\pi^2 + \cs^2 \left(\partial_i v_\pi\right)^2 - \frac{a''}{a} v_\pi^2\right]  , \\
H_{\mathcal{E}} & = \int \exd^3\bm{x} \; \frac{1}{2}\left\{p_\sigma^2 +  \left(\partial_i v_\sigma\right)^2 +\left[a^2 \left(m^2+g^2 \right) - \frac{a''}{a}\right] v_\sigma^2\right\}  , \\
g H_{\mathrm{int}} & = \int \exd^3\bm{x} \;  
g \left[a  v_\sigma \left(p_\pi - \frac{a'}{a}v_\pi \right)
- \frac{v_\sigma}{a\Lambda_1} \left(\partial_i v_\pi\right)^2 -\frac{v_\sigma }{a \Lambda_2}\left( p_\pi - \frac{a'}{a} v_\pi\right)^2\, \right]  .
\label{eq:Hint1resc_pert}
\end{align}

\subsubsection{Mode functions}

In the interaction picture, the field operators  evolve in the free theory and are thus most usefully expressed in momentum space by Fourier expanding
\begin{equation}
\widetilde{v}_{i}(\eta,\bm{x}) = \int \frac{\exd^3\bm{k}}{(2\pi)^{3}} \; \widetilde{v}_{i}(\bm{k},\eta) e^{i\bm{k}\cdot \bm{x}} \quad \mathrm{and} \quad \widetilde{p}_{i}(\eta,\bm{x}) = \int \frac{\exd^3\bm{k}}{(2\pi)^{3}} \; \widetilde{p}_{i}(\bm{k},\eta) e^{i\bm{k}\cdot \bm{x}}
\end{equation}
for each field $i = \pi, \sigma$, where $\widetilde{v}_{i}(\bm{k},\eta)$ and $\widetilde{p}_{i}(\bm{k},\eta)$ can be further expanded in creation and annihilation operators
	\begin{align}\label{eq:modefctdecomp} 
		\widetilde{v}_{i}(\bmk,\eta) = v_{i}(k,\eta) \widehat{a}^{i}_{\bs{k}} +  v^{*}_{i}(k,\eta) \widehat{a}^{i\dag}_{-\bs{k}} \quad \mathrm{and} \quad  \widetilde{p}_{i}(\bmk,\eta)  = p_{i}(k,\eta) \widehat{a}^{i}_{\bs{k}} +  p^{*}_{i}(k,\eta) \widehat{a}^{i\dag}_{-\bs{k}}\, .
	\end{align}
The creation and annihilation operators $\widehat{a}^{i \dagger}_{\bs{k}}$ and $\widehat{a}^{i}_{\bs{k}}$ obey the canonical commutation relations
\begin{align}
    \left[\widehat{a}^{i}_{\bs{k}}, \widehat{a}_{\bs{q}}^{j\dagger}\right] = \delta(\bs{k}-\bs{q})\delta_{i,j},
\end{align}
which leads to the Wronskian conditions
\begin{equation}
v_{i}(k,\eta) p_{i}^*(k,\eta) - v_{i}^*(k,\eta) p_{i}(k, \eta)=i
\end{equation}
for the mode functions. Heisenberg's equation yields the classical equations of motion for the mode functions, \ie
\begin{eqnarray}
v_{\pi}'&=&p_\pi\, ,\qquad p_\pi'=\left(\cs^2 k^2 - \frac{a''}{a}\right)v_\pi\, ,\\
v_\sigma' &=& p_\sigma\,  , \qquad p_\sigma' = \left(k^2+a^2 m^2-\frac{a''}{a}\right) v_\sigma\, ,
\end{eqnarray}
where the $k$ arguments are omitted for simplicity, since Fourier modes decouple in the free theory. Hamilton's equations can be combined into the so-called Mukhanov-Sasaki equations
	\begin{eqnarray}
		\label{eq:dyn1} v''_{\pi} + \left(\cs^2 k^2 -  \frac{2}{\eta^2}\right)v_{\pi} = 0
		\quad\quad\text{and}\quad\quad
		\label{eq:dyn2} v''_{\sigma} + \left( k^2 -  \frac{\nu^2_{\sigma}-\frac{1}{4}}{\eta^2}\right)v_{\sigma} = 0\, ,
	\end{eqnarray}
where at leading order in slow roll we have replaced $a=-(H \eta)^{-1}$, and where we have introduced
\begin{equation}
\label{eq:nusigma:def}
\nu_{\sigma} = \sqrt{ \frac{9}{4} - \frac{m^2 }{H^2}} \equiv i \mu_{\sigma} \ ,
\end{equation}
which is a pure imaginary if $m^2  > \frac{9}{4}H^2$ and real otherwise. By normalizing the mode functions to the Bunch-Davies vacuum~\cite{Bunch:1978yq} in the asymptotic, sub-Hubble past, one obtains\footnote{In the asymptotic past, $k/a\gg H, m, g$, hence the fields become decoupled and their mode functions evolve on an effectively quasi-static background. This allows one to set both sectors in the free vacuum state of Minkowski space-times. We stress that decoupling at small scales is necessary for the Bunch-Davies prescription to be applied~\cite{Grain:2019vnq, Colas:2021llj}.
\label{footnote:BunchDavies}
}
	\begin{align}
		\label{eq:modefctvp}	v_{\pi}(k,\eta) & = \frac{\ee^{- i \cs k \eta}}{\sqrt{2\cs k}} \left(1 - \frac{i}{\cs k \eta}\right)  , \\
		\label{eq:modefctvs}	v_{\sigma}(k,\eta) & = \frac{\sqrt{- \pi \eta}}{2} \; \ee^{-\tfrac{\pi}{2}\mu_{\sigma}+i\tfrac{\pi}{4}} H_{i\mu_{\sigma}}^{(1)}\left(- k \eta \right)  ,
	\end{align}
where $H^{(1)}_{\nu}$ is the Hankel function of the first kind and of order $\nu$. The mode functions of the momentum operators read
\begin{align}
		\label{eq:modefctpp}	p_{\pi}(k,\eta) &= - i \sqrt{\frac{\cs  k}{2}} \; \ee^{- i \cs k \eta} \left(1-\frac{i}{\cs k\eta}-\frac{1}{\cs^2 k^2 \eta^2}\right)  , \\
		\label{eq:modefctps}	p_{\sigma}(k,\eta) &= - \frac{1}{2} \sqrt{ \frac{\pi}{-\eta} }\; \ee^{-\tfrac{\pi}{2}\mu_{\sigma}+i\tfrac{\pi}{4}} \left[ \left(i\mu_{\sigma}+\frac{1}{2}\right) H_{i\mu_{\sigma}}^{(1)}(-k \eta)+ k \eta H_{i\mu_{\sigma}+1}^{(1)}(- k \eta) \right]   . 
\end{align}
One can check that \Eqs{eq:modefctvp}-\eqref{eq:modefctps} obey the Wronskian condition given above.

\subsubsection{Early-time breakdown of the $g$ expansion}
\label{sec:validity} 

Above we have restricted the Hamiltonian to terms that are linear in $g$, and it is important to assess the regime of validity of this truncation. Upon inspection of \Eq{eq:Hamilt:EFT:full} it is clear that the terms we have neglected are controlled by $g v_\sigma/(a \Lambda)$, hence our results only apply when the condition
\begin{equation} \label{bound_start}
\left| \frac{ g v_\sigma}{a \Lambda_2} \right| \ll 1 
\end{equation}
is satisfied. The validity of this condition can be qualitatively assessed as follows. In real space, $\langle \widetilde{v}_\sigma(\bm{x})\rangle =0$ and $\langle \widetilde{v}^2_\sigma(\bm{x})\rangle =\int \dd\ln k \,  \frac{k^3}{2\pi^2} v_\sigma^2(k)$. The contribution from a given scale $k$ to $v_\sigma$ can thus be estimated as $k^{3/2}\vert v_\sigma(k) \vert$, which is of order $k$ on sub-Hubble scales and of order $a H$ on super-Hubble scales if $\sigma$ is massless (if $\sigma$ is massive, the super-Hubble behavior is further suppressed, hence the following estimate is conservative). At super-Hubble scales, the condition~\eqref{bound_start} thus requires that
        \begin{align}
\label{eq:rho:bound:superH}
            g \ll \frac{\sqrt{\epsilon_1} \Mp }{\cs} 
        \end{align}
        upon using \Eq{eq:Lambda}, where we further assume $\widetilde{M}_3 \lesssim \widetilde{M}_1$. This implies that $g$ needs to be parametrically suppressed compared to the Planck mass, and that the theory breaks down when $\epsilon_1 \to 0$, which is expected from the EFToI perspective.  At sub-Hubble scales, the condition~\eqref{bound_start} reduces to
        \begin{align}\label{eq:rho:cond2ini}
            \frac{a H}{k} \gg \frac{ g \cs }{\sqrt{\epsilon_1}  \Mp } \; .
        \end{align}
This implies that our expansion breaks down at small distances, and becomes valid only once the scales of interest cross  $g\cs /(\sqrt{\epsilon_1}  \Mp)$ times the Hubble radius. It is interesting to notice that \Eq{eq:rho:bound:superH} requires that the scale below which the expansion breaks down lies well within the Hubble radius. Moreover, if one works with
        \begin{align}\label{eq:rho:cond2}
            g \lesssim \frac{\sqrt{\epsilon_1} H}{\cs },
        \end{align}
which is more restrictive than \Eq{eq:rho:bound:superH}, then that critical scale falls below the Planck length. In this case, the theory becomes valid at all super-Planckian distances, to which it should be restricted anyway. The main conclusion is that the effective theory we are working with is expected to break down at early time, hence below we employ it to investigate late-time behavior only. 
		
	\subsection{Perturbative entropies}
	\label{subsec:ppapp}

Let us now determine the in-in evolution of the purity of the system. We use the quantum interaction Hamiltonian (\ref{eq:Hint1resc_pert}) recast in momentum space as
\begin{equation}
g \widetilde{H}_{\mathrm{int}}(\eta) = g \int \frac{\exd^3\bm{k}}{(2\pi)^3} \; \left[ \OSlin(\bm{k},\eta) + \OSnon(\bm{k},\eta)  \right] \otimes \widetilde{v}_{\sigma}(-\bm{k},\eta) \label{Ham_kspace}
\end{equation}
where
\begin{align}
\OSlin(\bm{k},\eta)  \equiv & \; a(\eta)  \widetilde{p}_\pi(\bm{k},\eta) - a'(\eta) \widetilde{v}_{\pi}(\bm{k},\eta) \label{OSlin_def} \\
\OSnon(\bm{k},\eta)  \equiv & \;  \int \frac{\exd^3 \bm{p}_1}{(2\pi)^3} \int \frac{\exd^3 \bm{p}_2}{(2\pi)^3} \; \delta(\bm{p}_1 + \bm{p}_2 - \bm{k}) \bigg\{ \frac{\bm{p}_1 . \bm{p}_2}{a(\eta)\Lambda_1} \widetilde{v}_\pi(\bm{p}_1,\eta) \widetilde{v}_{\pi}(\bm{p}_2,\eta) \label{OSnon_def} \\
& \; - \frac{1}{a(\eta)\Lambda_2} \left[  \widetilde{p}_\pi(\bm{p}_1,\eta) - \frac{a'(\eta)}{a(\eta)} \widetilde{v}_{\pi}(\bm{p}_1,\eta) \right] \left[ \widetilde{p}_\pi(\bm{p}_2,\eta) - \frac{a'(\eta)}{a(\eta)} \widetilde{v}_{\pi}(\bm{p}_2,\eta) \right]  \bigg\} \ . \notag
\end{align}
We have written $g \widetilde{H}_{\mathrm{int}}$ in the form of \Eq{eq:HintFourier} so that the machinery from \Sec{sec:methods} can be employed directly. Recall that, according to \Eq{eq:S2:KS:KE}, at leading order in the coupling one has to compute
	\begin{align} \label{sec3S2}
		S_2(k) \simeq \frac{ 4 g^2 }{ (2\pi)^{6} } \int_{-\infty}^\eta \dd \eta_1  \int_{-\infty}^{\eta} \dd \eta_2  \; 
		\bar{\mathcal{K}}_{\mathcal{S}}^{(2)}(k, \eta_1,\eta_2) \bar{\mathcal{K}}_{\mathcal{E}}^{(2)}(k, \eta_1,\eta_2)  + \mathcal{O}(g^3) \; .
	\end{align}

\begin{figure}[tbp]
		\centering
		\fbox{\includegraphics[height=0.236\textheight]{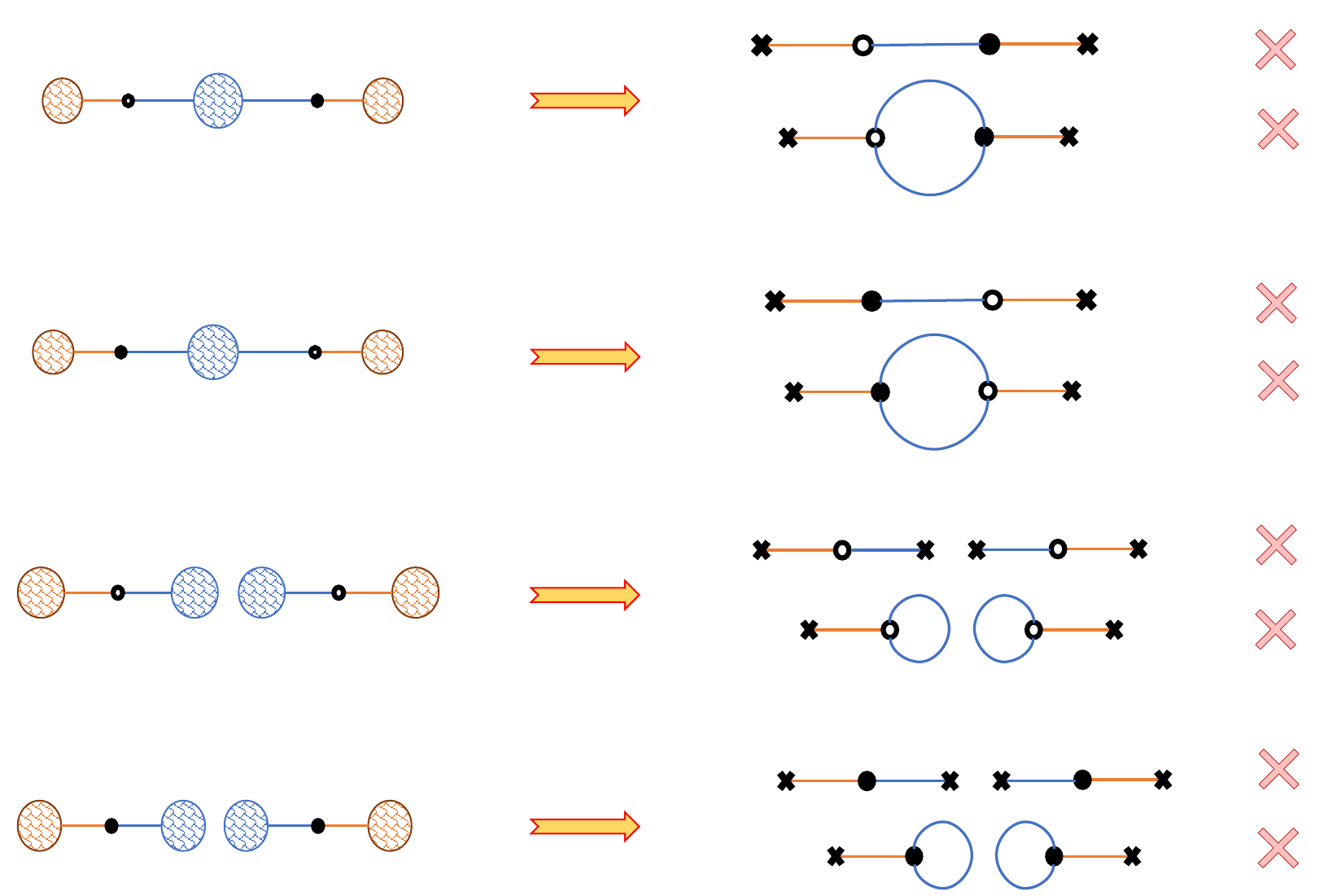}}
	    \fbox{\includegraphics[height=0.236\textheight]{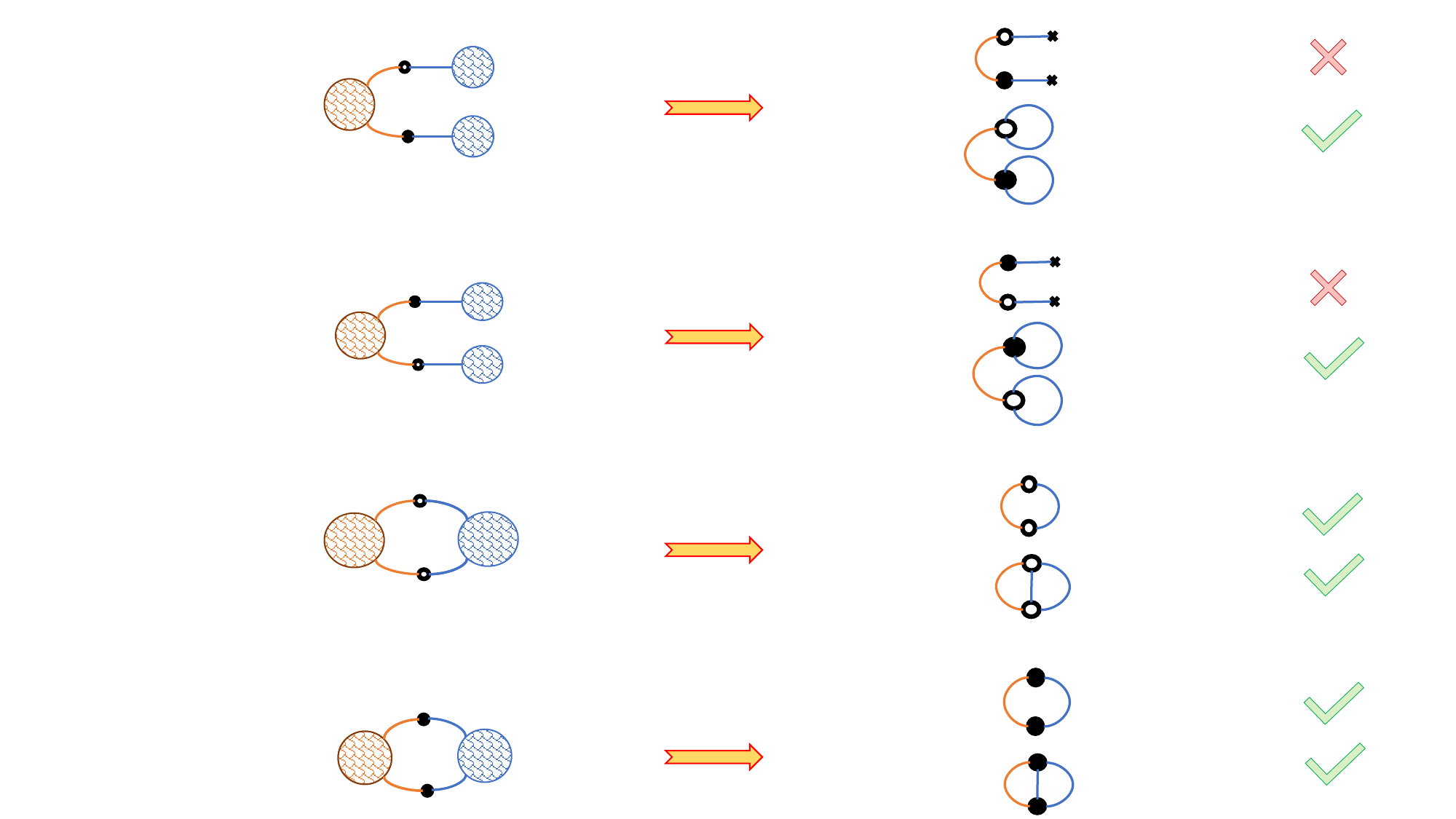}}
		\caption{Diagrams involved in the calculation of $\gamma^{(1,1)}$ (left panel) and $\gamma^{(2,0)}$ (right panel), with quadratic and cubic vertices of the form $\pi \sigma$ and $\pi^2 \sigma$. Contributions with a cross represent vanishing one-point functions. One can see that $\gamma^{(1,1)} = 0$, all contributions being proportional to the environment one-point function. For $\gamma^{(2,0)}$, both the quadratic and cubic vertices lead to non-vanishing second-order contributions, which induce departures from a pure state.}
		\label{fig:modeldiagram}
	\end{figure}

\paragraph{Memory kernels} 
Using the definition~\eqref{eq:def:Kreduced} together with \Eqs{eq:K2:def} and (\ref{centred}), the two-point function in the environment takes the simple form
	\begin{align}
		\bar{\mathcal{K}}_{\mathcal{E}}^{(2)}(k, \eta_1,\eta_2) = v_\sigma(k,\eta_1) v^*_\sigma(k,\eta_2) \ ,  \label{eq:Wightman:env:gen}
	\end{align} 
whose mass-dependence we take various limits of in the proceeding subsections. For the system, the memory kernel is made of two pieces,
	\begin{align} \label{KSsplit}
\bar{\mathcal{K}}_{\mathcal{S}}^{(2)}(k, \eta_1,\eta_2) = \KSlin(k,\eta_1,\eta_2) + \KSnon(k,\eta_1,\eta_2) \, ,
	\end{align} 
which correspond to the two-point functions of $\OSlin$ and $\OSnon$ respectively. The cross-term between $\OSlin$ and $\OSnon$ vanishes since it involves cubic powers of field operators, the expectation value of which vanishes on Gaussian states. A detailed calculation is presented in \App{App:2ptfunc}, where it is shown that
\begin{equation} \label{KSlin_ans}
\KSlin(k,\eta_1,\eta_2)  = \frac{\cs k}{2 H^2 \eta_1 \eta_2} e^{- i \cs k ( \eta_1 - \eta_2)}
\end{equation}
and
\bea
 \label{KSnon_ans}
\KSnon(k, \eta_1,\eta_2) =& \frac{H^2 \eta_1 \eta_2}{(2\pi)^{12}} \frac{\pi}{32 k} \int_{k}^\infty \exd P \; \int_0^k \exd Q\; e^{- i \cs P (\eta_1 - \eta_2)}  \\
	& \times \left\{ \frac{P^2+Q^2-2 k^2 }{\cs \Lambda _1}  \left[1-\frac{2 i}{\cs (P+Q)\eta _1}\right] \left[1-\frac{2 i}{\cs (P-Q)\eta _1 }\right] - \frac{\cs (P^2-Q^2)}{\Lambda _2} \right\}  \\
	& \times \left\{ \frac{P^2+Q^2-2 k^2}{\cs \Lambda_1} \left[1+\frac{2 i}{\cs (P-Q) \eta _2}\right] \left[1+\frac{2 i}{\cs (P+Q) \eta _2 }\right] - \frac{\cs (P^2-Q^2)}{\Lambda _2}  \right\} .
\eea
In practice, after Wick contractions of the field operators, and once conservation of momentum is imposed, the system correlator can be expressed as an integral over two momenta (these are the two momenta flowing through the double loops in the right panel of \Fig{fig:modeldiagram}) . The expression~\eqref{KSnon_ans} is what remains after the angular part these two momenta has been integrated away [and after an additional rotation is performed -- see \Eq{PQ_transf}].

It is of course possible to perform the two remaining integrals over $P$ and $Q$ in \Eq{KSnon_ans} before inserting the result in \Eq{sec3S2}. In particular, by examining the $P \to \infty$ limit of the integrand, one can show that the dominant contribution in the coincident limit $\eta_1 \to \eta_2$ reads
\bea
 \label{KSnon_singular}
\KSnon(k, \eta_1,\eta_2) =& \ \frac{H^2 \eta_1 \eta_2}{(2\pi)^{12}} \frac{\pi}{32 k} \int_{k}^\infty \exd P \; \int_0^k \exd Q\; e^{- i \cs P (\eta_1 - \eta_2)} \left[ \left(\frac{1}{\cs\Lambda_1} - \frac{\cs}{\Lambda_2} \right) P^4 + \mathcal{O}(P^3) \right]  \\
\simeq & \ \frac{H^2 \eta_1 \eta_2}{(2\pi)^{12}} \frac{\pi}{32 k} \left\{ 24 i \frac{   \frac{\cs}{\Lambda_2} - \frac{1}{\cs\Lambda_1}  }{\cs^5 (\eta_1-\eta_2)^5} + \mathcal{O}\left[ \left(\eta_1-\eta_2 \right)^{-4} \right]  \right\} \ .
\eea
This singular coincident limit can be understood as contributing to the UV divergences that later arise, which may be regulated using a short-time cutoff. A more desirable regularization scheme however can be used if one instead keeps the correlator represented in terms of the momentum integrals as in \Eq{KSnon_ans}, and then performs the $P$ and $Q$ integrals only {\it after} any time integrals occurring in the linear entropy formula (\ref{sec3S2}). This is a useful tactic because the $P$ and $Q$ variables cleanly separate the UV and IR behavior of the integrand and therefore, more standard methods of UV regularization from QFT (\eg~dimensional regularization) can be straightforwardly applied (this would also apply to IR divergences if other interactions with less derivatives were studied).

\paragraph{Spectral purity} The total linear entropy at second-order in the coupling can be represented by the diagrams depicted in \Fig{fig:modeldiagram}. They are obtained by applying the techniques of \Sec{subsubsec:diagram} to the model considered here. With notations analogous to \Eq{KSsplit},  the linear entropy receives two contributions,
\begin{eqnarray} \label{S2_split}
S_2(k)  \simeq \Slin (k) + \Snon(k) + \mathcal{O}(g^3)
\end{eqnarray}
with
	\begin{align} \label{S2_j_def}
		S^i_2(k) \equiv \frac{ 4 g^2 }{ (2\pi)^{6} } \int_{-\infty}^\eta \dd \eta_1  \int_{-\infty}^{\eta} \dd \eta_2  \; 
		\bar{\mathcal{K}}_{\mathcal{S}, i }^{(2)}(k, \eta_1,\eta_2) \bar{\mathcal{K}}_{\mathcal{E}}^{(2)}(k, \eta_1,\eta_2)  \qquad \text{for\ $i= $lin, NL} \ .
	\end{align}
Using \Eqs{eq:Wightman:env:gen} and~\eqref{KSlin_ans}, the linear contribution reduces to~\cite{Colas:2024xjy,Burgess:2024eng}
\begin{equation} \label{Slin_Ldef}
\Slin(k)  =  \frac{g^2 \cs}{(2 \pi)^6 H^2} | \mathfrak{L}_k(\eta) |^2 
\quad \mathrm{with} \quad 
\mathfrak{L}(\eta) \equiv \int_{-\infty}^{\eta} \exd \eta' \; e^{- i \cs k \eta'} \sqrt{ \frac{\pi k}{- 2 \eta'} } \; e^{- \tfrac{\pi}{2} \mu_\sigma + \tfrac{i\pi}{4} } H_{i\mu_{\sigma}}^{(1)}(- k \eta') \ .
\end{equation}
In this form it is obvious that the linear entropy is always non-negative, hence that purity remains below one and that the map is CPTP, see the discussion below \Eq{eq:deriv2}. Similarly, the non-linear contribution can be expressed as
\begin{eqnarray} 
\Snon(k) = \frac{g^2 H^2}{(2 \pi)^{18}} \frac{\pi }{16 k^4} \int_{k}^\infty \exd P \int_0^k \exd Q \; \left| \mathfrak{N}_{k}(P,Q,\eta) \right|^2 \label{Snon_Ndef}
\end{eqnarray}
with
\begin{align} \label{Ndef_integral}
\mathfrak{N}_{k}(P,Q,\eta)  \equiv & \ \sqrt{\frac{\pi}{2}} \; e^{ - \tfrac{\pi}{2} \mu_\sigma + \tfrac{i \pi}{4} }  \int_{-\infty}^{\eta} \exd \eta ' \; (-k\eta')^{\tfrac{3}{2}} H_{i \mu}^{(1)}(- k \eta') e^{- i \cs P \eta'} \\
&\ \times \left\{ \frac{P^2+Q^2-2 k^2}{\cs \Lambda _1}  \left[1-\frac{2 i}{\cs (P+Q)\eta'}\right] \left[1-\frac{2 i}{\cs (P-Q)\eta' }\right]  - \frac{\cs (P^2-Q^2)}{\Lambda _2} \right\}  . \notag
\end{align}
In the analysis that follows, we first perform the time integral  in \Eq{Ndef_integral} before performing the one over $P$ and $Q$ in \Eq{Snon_Ndef}, for the reasons mentioned above. 

Finally, a quantity of physical interest below is the decoherence \emph{rate}, obtained by differentiating \Eq{eq:S2:KS:KE} with respect to time,
	\begin{align} \label{eq:entropyrateeg}
		\frac{\exd S_2(k)}{\exd \eta} \simeq \frac{ 8 g^2 }{ (2\pi)^{6} } \int_{-\infty}^\eta \dd \eta'  \; 
		\Rea\left[ \bar{\mathcal{K}}_{\mathcal{S}}^{(2)}(k, \eta,\eta') \bar{\mathcal{K}}_{\mathcal{E}}^{(2)}(k, \eta,\eta') \right] + \mathcal{O}(g^3) \; .
	\end{align}
This has the advantage of involving a single integral over time, and it can be divided into a linear and a non-linear contribution by time differentiating \Eqs{Slin_Ldef} and~\eqref{Snon_Ndef} respectively. The other advantage of working with the decoherence rate is that, as explained in \Sec{sec:validity}, the EFT treatment of the model we consider is expected to break down at early time. As a consequence, finite contributions to the linear entropy coming from early-time interactions cannot be properly accounted for. Such contributions drop out from the late-time decoherence rate, which may thus be argued to be the proper quantity to compute in the EFT.

 In what follows, we compute the spectral purity in three cases for the environment mass: massless ($m=0$), conformal ($m=\sqrt{2}H$) and heavy ($m\gg H$).

\subsubsection{Massless environment}
	
Let us first consider the case where the environment is massless, $m=0$, such that $i \mu_\sigma = 3/2$ and \Eq{eq:modefctvs} reduces to
\bea
\label{eq:vsigma:massless}
v_\sigma(k,\eta) = \frac{\ee^{-ik\eta}}{\sqrt{2k}}\left(1-\frac{i}{k\eta}\right) .
\eea
		
\subsubsection*{Linear contribution}
	
\begin{figure}[tbp]
		\centering
		\includegraphics[width=0.8\textwidth]{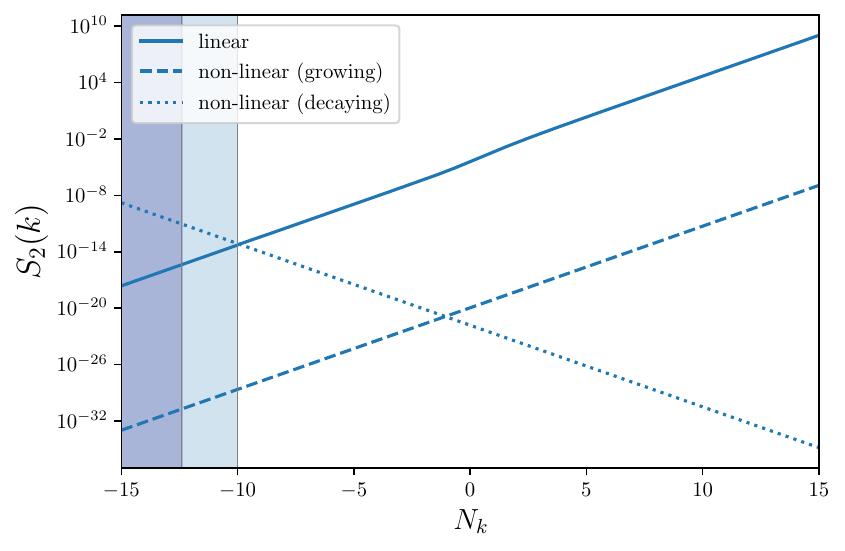}
		\caption{Spectral purity $S_2(k)$ as a function of the number of $e$-folds $N_k$ from Hubble-exit time, in the case of a massless environment with $g=0.01\, H$ and $\cs=1$. Loop corrections overtake the tree-order result at early times given by \Eq{eq:SNL:gt:Slin:massless}, which is displayed with the light-blue shaded region. The EFT also breaks down at early time, see \Eq{eq:rho:cond2ini:Deltazeta}, which is displayed in dark blue. At late time, both the linear and non-linear contributions grow like $a^2$, the latter being suppressed with respect to the former by $\Delta^2_\zeta /(4\pi^2 \cs)^4$.}
		\label{fig:S2}
	\end{figure}

The linear contribution to the entropy depends on the integral $\mathfrak{L}_k(\eta)$ defined in \Eq{Slin_Ldef}, which is evaluated for the massless case in \App{App:Lin_Int} and one finds
\begin{equation} 
\label{eq:Slin:massless:full}
\Slin(k) = \frac{g^2 \cs}{(2\pi)^6 H^2} \left| \frac{i e^{- i (1+\cs) k \eta}}{- k \eta} + \cs \left\{ \mathrm{Ei}\left[-i(1+\cs)k\eta\right] - i \pi \right\} \right|^2 \ .
\end{equation}
Here, $\mathrm{Ei}$ is the exponential integral function defined in (\ref{EiInt}).
It is convenient to express the above in terms of the number of $e$-folds $N=\ln(a)$ spent by the mode $k$ outside the Hubble radius. In a de-Sitter universe it is given by 
\begin{eqnarray} \label{efold_def}
N_{k} = - \log ( - k \eta)\, ,
\end{eqnarray}
such that $N_k > 0$ and $N_k< 0$ correspond to before and after Hubble-crossing, respectively. The early ($N_k\ll -1$) and late-time ($N_k\gg 1$) behavior of the entropy can be obtained by expanding \Eq{eq:Slin:massless:full} and one finds
\bea
\label{eq:Slin:massless:asymptotics}
\Slin(k) \simeq \begin{cases} \dfrac{ \cs g^2 }{(2\pi)^6 H^2}  \dfrac{e^{ 2 N_k}}{(1+\cs)^2}   \qquad &\text{(sub-Hubble)} \\
 \dfrac{\cs g^2 }{(2\pi)^6 H^2}   e^{ 2 N_k} \qquad & \text{(super-Hubble)} 
\end{cases}\, .
\eea
The entropy is thus monotonously increasing, at a rate which is the same in both asymptotic regimes, \ie $\Slin \propto a^2$, with only a slight increase in the overall amplitude at around Hubble crossing. This can be further checked in \Fig{fig:S2} where \Eq{eq:Slin:massless:full} is displayed. 

	\subsubsection*{Non-linear contribution}	
	
Evaluating \Eq{Ndef_integral} with \Eq{eq:vsigma:massless}, in \App{App:non_int} we show that
\bea 
\label{Nans_massless}
\mathfrak{N}_{k}(P,Q,\eta) = & e^{-i (k + \cs P)\eta} \left\{ -\frac{4 i k (P^2+Q^2 -2 k^2 )}{\cs^3 (P^2-Q^2) \Lambda_1 }  \frac{1}{-k\eta} + \frac{\cs (P^2-Q^2 ) ( 2 k +\cs P)}{ (k+\cs P)^2\Lambda_2 } \right. \\
& -\frac{(P^2+Q^2-2 k^2)}{\cs (k+\cs P) \Lambda_1}  \left[\frac{4 k P}{\cs (P^2-Q^2)}+\frac{2k + \cs P}{ k+ \cs P}\right]  \\
& \left. + \frac{i}{\cs ( k+ \cs P )} \left[ \frac{P^2+Q^2-2 k^2}{\Lambda_1}-\frac{\cs^2(P^2-Q^2)}{\Lambda_2} \right] (- k \eta) \right\rbrace .
\eea
Inserting this expression into the formula (\ref{Snon_Ndef}) one obtains UV divergences of the form $\Snon \propto \int_{k}^\infty \exd P\; P^{n} $ for $P \to \infty$ with  $n=-1,0,1,2$. These are reminiscent of the singular equal-time limit of the environment correlator \cite{Burgess:2024eng} as discussed below \Eq{KSnon_singular}. Here they appear as large-momentum divergences, which makes it possible to employ a variant of dimensional regularization where we take for example\footnote{Strictly speaking, this regularization scheme is only tangentially related to dimensional regularization, which would require the use of the mode functions in $(4 + \varepsilon)$-dimensional de Sitter space.}
\begin{eqnarray} \label{main_dimreg}
\int_{k}^\infty  \frac{\exd P}{k + \cs P} \to \int_{k}^\infty \frac{\exd P}{k + \cs P} \left( \frac{P}{\mu} \right)^\varepsilon \simeq - \frac{1}{\cs} \left\{ \frac{1}{\varepsilon} - \log \left[ \frac{\cs \mu}{(1 + \cs) k} \right] \right\} + \mathcal{O}(\varepsilon)\, .
\end{eqnarray}
Here we have expanded near $\varepsilon \simeq 0$ and taken $\mu > 0$ to be an arbitrary mass scale. Performing the $P$- and then $Q$-integrals in \Eq{Snon_Ndef} leads to
\begin{eqnarray}  \label{Snon_Cans}
\Snon(k) = \frac{g^2 H^2}{(2 \pi)^{18}} \frac{\pi }{16}  \left( \mathcal{C}_1\; e^{2 N_k} + \mathcal{C}_2 + \mathcal{C}_3\; e^{- 2 N_k}\right) ,
\end{eqnarray}
written in terms of the number of $e$-folds defined in \Eq{efold_def} and with time-independent coefficients given by
\begin{eqnarray} 
\mathcal{C}_1 & = & - \frac{48}{\cs^6 \Lambda_1^2}\, , \label{C1_text} \\
\mathcal{C}_2 & = & \frac{4}{\cs^3} \left(\frac{9}{\cs^4 \Lambda _1^2}-\frac{1}{\Lambda _2^2}\right) \left\lbrace\frac{1}{\varepsilon}-\log \left[\frac{\cs \mu }{(\cs+1) k} \right]\right\rbrace  +\tfrac{264 \cs^6+792 \cs^5-1064 \cs^4-3975 \cs^3-1625 \cs^2+1935 \cs+1185}{45 \cs^7 (\cs+1)^3 \Lambda _1^2} \notag \\
&\ & -\tfrac{2 \left(96 \cs^6+288 \cs^5+544 \cs^4+855 \cs^3+885 \cs^2+585 \cs+195\right)}{45 \cs^5 (\cs+1)^3 \Lambda _1 \Lambda _2} + \tfrac{24 \cs^6+72 \cs^5+216 \cs^4+405 \cs^3+275 \cs^2-45 \cs-75}{45 \cs^3 (\cs+1)^3 \Lambda _2^2} \label{C2_text} \\
\mathcal{C}_3 & = & \frac{4 }{\cs^3}\left(\frac{1}{\cs^2 \Lambda _1}-\frac{1}{\Lambda _2}\right) \left(\frac{\cs^2-3}{3 \Lambda _2}+\frac{3-5 \cs^2}{3 \cs^2 \Lambda _1}\right) \left\lbrace\frac{1}{\varepsilon}-\log \left[\frac{\cs \mu }{(\cs+1) k}\right]\right\rbrace \notag \\
&& +\tfrac{88 \cs^4+60 \cs^3-80 \cs^2-45 \cs+15}{15 \cs^7 (\cs+1) \Lambda _1^2}-\tfrac{2 \left(32 \cs^4+40 \cs^3-60 \cs^2-45 \cs+15\right)}{15 \cs^5 (\cs+1) \Lambda _1 \Lambda _2}+\tfrac{8 \cs^4+20 \cs^3-40 \cs^2-45 \cs+15}{15 \cs^3 (\cs+1) \Lambda _2^2} \label{C3_text}
\end{eqnarray}
with further details of the derivation given in Appendix \ref{App:non_int}.

\paragraph{UV logarithmic divergences:} The coefficients $\mathcal{C}_2$ and $\mathcal{C}_3$ feature UV-divergent terms that we now discuss. As argued below \Eq{eq:entropyrateeg}, a quantity that is immune to finite non-EFT effects at small scales is the entropy production rate,
\begin{eqnarray} \label{SnonRATE_Cans}
\frac{\exd \Snon(k)}{\exd N_k} = \frac{g^2 H^2}{(2 \pi)^{18}} \frac{\pi }{16} \left(  \mathcal{C}_1 e^{2 N_k} - \mathcal{C}_3 e^{- 2 N_k} \right) ,
\end{eqnarray}
from which $\mathcal{C}_2$ absent. At late time indeed, the constant term $\mathcal{C}_2$ may receive contributions from effects occurring before the EFT became valid, which we cannot properly account for and that may cancel out the divergences in \Eq{C2_text}. By considering the decoherence rate we are focusing on late-time effects, which are within the reach of the EFT. 

For $\mathcal{C}_3$, the divergent term is controlled by $(\cs^2-3)/\Lambda_2 + (3-5\cs^2)/(\cs^2\Lambda_1)$. As mentioned below \Eq{eq:Lambda}, in the case where the term controlled by $\widetilde{M}_3 $ can be neglected in the interaction Lagrangian~\eqref{eq:S:pi:sigma}, the relation $\Lambda_2 = - \Lambda_1$ holds. Moreover, if the term controlled by ${M}_2$ in the system's Lagrangian~\eqref{eq:S:pi} can be neglected too, then $\cs^2=1$, see \Eq{eq:pic:o}. Under these conditions the divergent term in $\mathcal{C}_3$ thus cancels out. Both ${M}_2 $ and $\widetilde{M}_3 $ control terms involving $(\delta g^{00})^2 \supset \dot{\pi}^3, \dot{\pi}(\partial_i \pi)^2$, which are expected to contribute to spectral purity (by coupling a given Fourier mode to all other Fourier modes) but which we have neglected. It is therefore reasonable to assume that these contributions cancel out the divergent term in $\mathcal{C}_3$. In other words, only when $\cs  = 1$ and $\Lambda_2 = - \Lambda_1$ can one consistently discard terms of order $(\delta g^{00})^2$ in the Lagrangian and we will therefore assume that this is the case in what follows. 

Under these conditions $\mathcal{C}_3$ is finite and given by 
	\begin{align}
		\mathcal{C}_3 =  -\frac{4}{3\Lambda_1^2} \ .
	\end{align}
The term it controls in \Eq{Snon_Cans} is displayed with the dotted line in \Fig{fig:S2}, while the term controlled by $\mathcal{C}_1$ is displayed with the dashed line. Let us further discuss the contributions of these terms at early and late time.

	\paragraph{Early-time perturbative breakdown:} 
At sub-Hubble scales, the non-linear contribution takes over the linear contribution, since
\begin{eqnarray}
\label{eq:SNL:gt:Slin:massless}
\left. \frac{\Snon(k)}{\Slin(k)} \right\vert_{\text{sub-Hubble}}= - \frac{\Delta_\zeta^2}{24 (2\pi)^9}\ee^{-4 N_k},
\end{eqnarray}
where we have used \Eqs{eq:Lambda} and~\eqref{PowSpec_def}. This becomes larger than unity (in absolute value) at early time, when 
\bea
\label{eq:rho:cond2ini:Deltazeta}
\frac{aH}{k}<\left[\frac{\Delta_\zeta^2}{24 (2\pi)^9}\right]^{1/4}\simeq 5 \times 10^{-5}\, ,
\eea
\ie $\simeq 10$ \efolds before Hubble crossing. One should recall nonetheless that the EFT treatment breaks down when
\bea
\frac{aH}{k}<\frac{g}{H}\sqrt{8\pi^2 \Delta_\zeta^2},
\eea
see \Eq{eq:rho:cond2ini}. The two breaking scales coincide when $g/H=0.1$. Therefore, when $g$ is parametrically smaller than $H$, the EFT becomes valid before the loop corrections we have computed become sub-dominant. In any case, the above discussion confirms that our results should be trusted at late time only, \ie around and after Hubble crossing.

	\paragraph{Late-time decoherence:} At super-Hubble scales, both the linear and the non-linear contributions to spectral purity grow like $a^2$ and adding them together leads to
\begin{align}
\label{eq:massless_final}
	\left.S_2(k)\right\vert_{\mathrm{super-Hubble}} = \dfrac{\cs}{(2\pi)^6} \frac{g^2}{H^2} \left[1 - \frac{1}{(2\pi)^{9}} \frac{3}{8}  \frac{\Delta_{\zeta}^2}{\cs^4} \right] e^{ 2 N_k} \, .
\end{align}	
At late time, the entropy thus grows as the universe's ``area'' $a^2$. The fact that it grows confirms the CPTP nature of the non-unitary evolution, see the discussion below \Eq{eq:deriv2}. Loop contributions only provide a small correction (as long as $\cs>10^{-4}$), and decoherence, which we may define as $S_2(k)> 1$, occurs when
\bea
\frac{aH}{k}>(2\pi)^3 \frac{H}{g\sqrt{\cs}}\, .
\eea

The link between purity and linear entropy is given by \Eq{eq:Renyi2}, which allows one to define an effective purity per Fourier mode
\bea
\label{eq:gammak:S2}
\gamma_k = \ee^{-S_2(k)}\, .
\eea
This quantity is displayed in \Fig{fig:gamma_full}, which confirms that decoherence takes place soon after Hubble crossing. One might be concerned that, since $S_2$ has been computed to order $g^2$ in perturbation theory, only a linearized version of \Eq{eq:gammak:S2} should be used, namely $\gamma_k\simeq 1-S_2(k)$, for consistency. However, in \Refa{Colas:2024xjy} it is shown that \Eq{eq:gammak:S2} is able to efficiently resum secular effects due to the late-time growth of the linear entropy. This was shown by comparison with an exact, transport-equation based solution as presented in \Refa{Colas:2022kfu}. This is why we trust \Eq{eq:gammak:S2} to provide a reliable assessment of decoherence even in the strongly decohered regime.

		\subsubsection{Conformal environment}
	
Let us consider the case where the environmental field $\sigma$ is conformally coupled to gravity, \ie $i \mu_\sigma = \frac{1}{2}$. In this case \Eq{eq:modefctvs} reduces to $v_\sigma(k,\eta)=\ee^{-ik\eta}/\sqrt{2\pi}$ and in \App{app:Integration:details} we show that the linear contribution to the entropy is given by 
	\begin{equation}
	\label{eq:Slin:asymptotics:conformal}
	\Slin = \frac{g^2 \cs}{(2\pi)^6 H^2} \left| \text{Ei}\left[ - i (\cs+1) k \eta \right] -  i \pi  \right|^2
	 \simeq \begin{cases} \dfrac{ \cs g^2 }{(2\pi)^6 H^2}  \dfrac{e^{ 2 N_k}}{(1+\cs)^2}   \qquad &\text{(sub-Hubble)} \\
 \dfrac{\cs g^2 }{(2\pi)^6 H^2}   N_k^2 \qquad & \text{(super-Hubble)} 
\end{cases}\, .
	\end{equation}
We have checked that the agreement between these expressions and the result of the transport-equation method (exact numerical treatment) presented in \Refs{Colas:2022kfu, Colas:2024xjy} is excellent. It is interesting to notice that, at sub-Hubble scale, we get the exact same behavior as in the massless case, see \Eq{eq:Slin:massless:asymptotics}. However, at super-Hubble scales, entropy production is much slower, since it proceeds as $S_2 \propto \ln^2(a)$ instead of $S_2\propto a^2$.

The non-linear contribution can be computed following similar steps as in the massless case.  At early time, before the Fourier mode crosses the Compton length, \ie when $k/a>m$, the same result as in the massless-environment case are recovered. At late time, \Eq{Ndef_integral} is evaluated by first integrating over $\eta'$, then expanding the result in $-k\eta \ll 1$, and finally integrate over $P$ and $Q$. This leads to
	\begin{align}\label{eq:oneloopconf}
		\left. S_2(k)\right\vert_{\text{super-Hubble}} \simeq \frac{ \cs }{(2\pi)^6} \frac{g^2}{H^2}\left[1 - \frac{1}{(2\pi)^9} \frac{3}{8} \frac{\Delta_{\zeta}^2}{\cs^4} \right] N_k^2 \, .
	\end{align}    
As before, the non-linear contribution increases with time at the same rate as the linear one, and is suppressed by $\Delta_\zeta^2/\cs^4$. The case of a conformal environment is displayed in \Fig{fig:gamma_full} with orange curves. 
	
\subsubsection{Heavy environment}
	
	Let us finally consider the case where the environment is heavy ($m \gg H$), where for simplicity we additionally assume that $\cs = 1$ and focus on the entropy production rate. Inserting \Eq{eq:modefctvs} into \Eq{Slin_Ldef} leads to~\cite{Colas:2022kfu, Colas:2024xjy}
	\begin{align}\label{eq:heavypurpert}
		\frac{\dd \Slin(k)}{\dd \eta} \bigg|_{\cs=1} = - \frac{k g^2}{32\pi^5 H^2} \; \Rea\left[ \ee^{-\pi \mu} H^{(1)}_{i\mu}(-k\eta) \ee^{-ik\eta} F_\mu(-k\eta) \right] 
	\end{align}
for the linear contribution, where
	\begin{align}\label{eq:FI1}
		F_\mu(z) &\equiv  \gamma^{*}_{\mu_{\mathcal{F}}}(z) g_{\mu_{\mathcal{F}}}( z) + \delta^{*}_{\mu_{\mathcal{F}}}(z) 	g_{-\mu_{\mathcal{F}}}(z)
	\end{align} 
	and we have introduced the notations
	\begin{align}
		\gamma_{\mu_{\mathcal{F}}}(z) &\equiv \frac{1+  \coth \pi \mu_{\mathcal{F}}}{\Gamma(1+i\mu_{\mathcal{F}})}\left(\frac{z}{2}\right)^{i\mu_{\mathcal{F}}},\quad\quad~~\,
		\delta_{\mu_{\mathcal{F}}}(z) \equiv  \frac{-1}{\sinh \pi\mu_{\mathcal{F}}} \frac{1}{\Gamma(1-i\mu_{\mathcal{F}})}\left(\frac{z}{2}\right)^{ -i\mu_{\mathcal{F}} }
	\end{align}
	and
	\begin{align}
		g_{\mu_{\mathcal{F}}}( z)= \frac{1}{ 1 - 2 i \mu_{\mathcal{F}}}{{}_2F_2}^{\frac{1}{2} - i \mu_{\mathcal{F}},\frac{1}{2} - i \mu_{\mathcal{F}}}_{\frac{3}{2} - i \mu_{\mathcal{F}}, 1 - 2 i \mu_{\mathcal{F}}}(- 2iz) \ ,
	\end{align}
	with ${}_2 F_2 $ being the $(2,2)$-generalized hypergeometric function. We have checked that the agreement of these expressions with the transport-equation method is again excellent. The above expressions can be expanded at early and late time and one finds
	\begin{align}\label{eq:heavypurpertef}
		\left.\frac{\dd \Slin(k)}{\dd N_k} \right|_{\cs=1} =\begin{dcases}
			 \frac{g^2}{2(2\pi)^6 H^2}  \ee^{2 N_k}  & \quad \text{(sub-Hubble)}\\
			-\frac{g^2}{(2\pi)^6 H^2}  \frac{4}{\mu(1+4\mu^2)} \ee^{-N_k}  & \quad \text{(super-Hubble)}
		\end{dcases} \, .
	\end{align}
At sub-Hubble scales, one recovers again the same result as in the massless and the conformal case, see \Eqs{eq:Slin:massless:asymptotics} and~\eqref{eq:Slin:asymptotics:conformal} respectively. At super-Hubble scales however, the situation is quite different since the entropy production rate becomes negative, which signals the presence of recoherence~\cite{Colas:2022kfu}. In the heavy mass limit $m\gg H$, $\mu\simeq m/H$, see \Eq{eq:nusigma:def}, and at late time one obtains
	\begin{align}\label{eq:heavypurpertresum}
		\left. \gamma_k\right\vert_{\text{super-Hubble}} = \gamma_\infty - \frac{g^2}{16\pi^6 H^2} \frac{H^3}{m^3} \ee^{-N_k} \, ,
	\end{align}
where $\gamma_\infty$ has been computed in \Refa{Colas:2024xjy} and is given by
        \begin{align}
            \gamma_\infty \simeq 1 -  \frac{g^2}{8\pi^4 H^2} \ee^{- 2 \pi \frac{m}{H}}. 
        \end{align}

\begin{figure}[tbp]
                \begin{minipage}{6in}
        		\centering
        		\raisebox{-0.5\height}{\includegraphics[width=.48\textwidth]{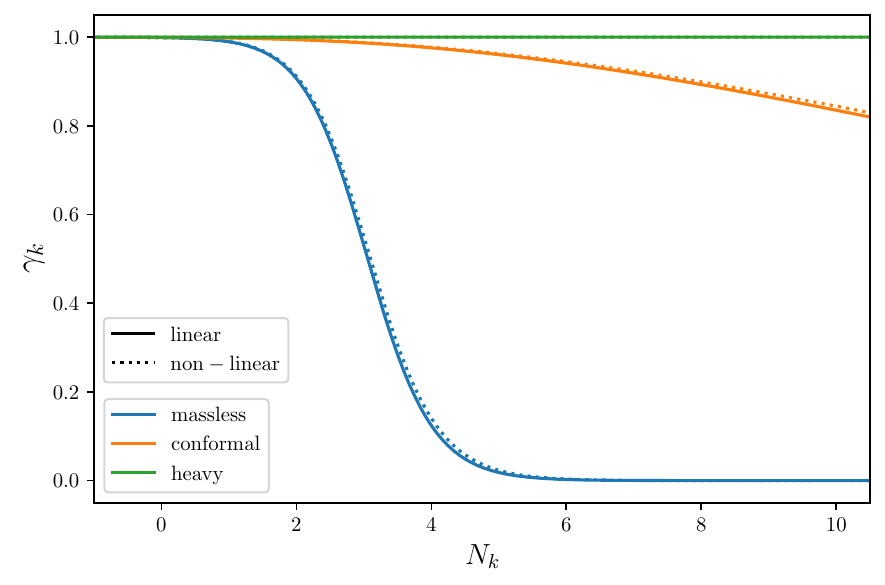}}
        		%\hspace*{.2in}
        		\raisebox{-0.5\height}{\includegraphics[width=.48\textwidth]{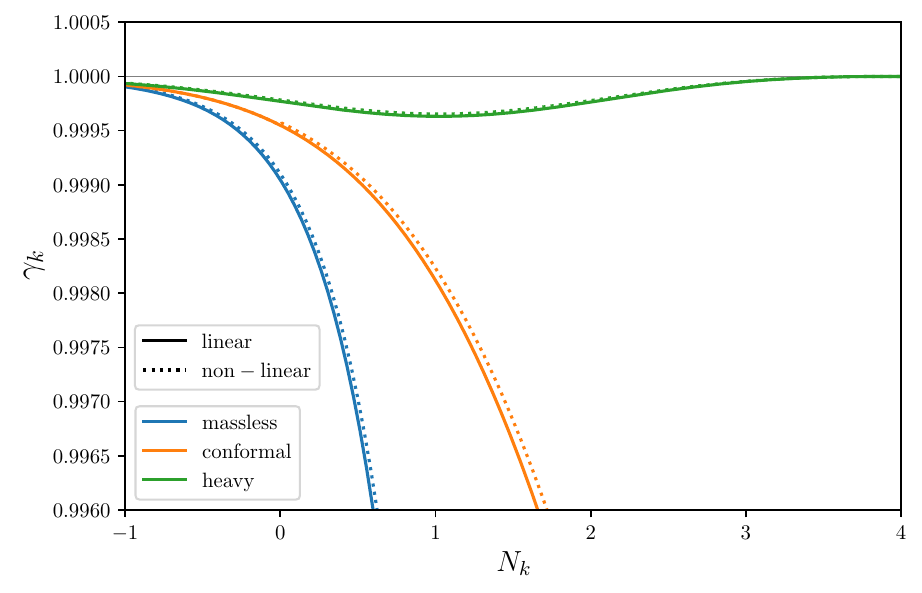}}
        	\end{minipage}
     		\caption{Per-mode purity $\gamma_k$, see \Eq{eq:gammak:S2}, as a function of the number of \efolds $N_k$ spent by a given scale outside the Hubble radius, for $g=0.05 H$, $\cs=1$ and $\Delta^2_\zeta/ (4\pi^2 \cs )^4=1$. The value of $\Delta^2_\zeta$ is unrealistically large, but this is to amplify non-linear corrections that would otherwise not be visible by eye. The solid line stand for the linear contributions, and the dotted lines correspond to the sum of the linear and non-linear contributions. If the environment is massless (blue curves) or conformally coupled (orange curves) decoherence takes place at late time, while recoherence occurs after a transient phase of slight decoherence if the environment is heavy (green curves). The right panel zooms in on the region $\gamma_k\simeq 1$.}
     		\label{fig:gamma_full}
     	\end{figure}

For the non-linear contribution, closed-forms expressions cannot be reached, but the relevant integrals can be expanded in the sub- and super-Hubble regimes. At early time, before the Compton length is crossed out, the same results as in the massless and conformal cases are recovered. At late time, we find
	\begin{align}\label{eq:oneloopheavy}
		\left. \frac{\dd S_2(k)}{\dd N_k} \right\vert_{\text{super-Hubble}} \simeq \frac{g^2}{(2\pi)^6 H^2}   \frac{4}{\mu(1+4\mu^2)}  \left[ 1 -\frac{1}{(2\pi)^9} \frac{3}{8} \frac{\Delta_{\zeta}^2}{\cs^4}  \right] \ee^{-N_k} \, ,
	\end{align}
where we have reinstated $\cs$. Again, the non-linear contribution has the same time dependence as the linear one, and is suppressed by the exact same amount, see \Eqs{eq:massless_final} and~\eqref{eq:oneloopconf}. It is thus interesting to notice that the recoherence mechanism identified in \Refa{Colas:2022kfu} seem robust to loop corrections. Since non-linear interactions mix all Fourier modes, hence substantially enlarge the size of the effective environment compared to linear interactions that only mix Fourier modes of the same wavevector, this may not be obvious a priori and constitutes an important verification. The heavy-environment case is displayed in green in \Fig{fig:gamma_full}.
 
	\section{Outlook}\label{sec:outlook}
	
We have shown how quantum purity (also known as linear entropy), although not an observable, can be computed perturbatively using in-in methods. We have applied our formalism to a model of cosmological interest and found that under conditions that we identified, the perturbative calculation is indeed under control, and provides a relevant framework to investigate quantum decoherence of cosmological fluctuations. Let us now mention a few prospects our work opens, both conceptual and phenomenological, before presenting concluding remarks.

	\subsection{Scaling dimensions} 
	
Beyond the specific model we have considered, a generic question of interest consists in identifying the interactions leading to late-time decoherence of the curvature perturbation. Let us consider \Eq{eq:entropyrateeg} again, which we reproduce here in the case where the interaction strength can depend on time explicitly
\bea
		\frac{\dd S_2(k)}{\dd \eta} \simeq  \frac{8}{(2\pi)^6} g(\eta) \int_{-\infty}^{\eta} \dd \eta' \,  g(\eta') \Rea\left[ \bar{\mathcal{K}}_{\mathcal{S}}^{(2)}(k, \eta,\eta') \bar{\mathcal{K}}_{\mathcal{E}}^{(2)}(k, \eta,\eta') \right].
\eea
Let us assume that $g$ evolves in time as $g\propto a^{\Delta_g}$, where $\Delta_g$ is the scaling dimension of the coupling. Our goal is to determine the scaling dimension of the decoherence rate, \ie the quantity $\Delta_s$ such that
	\begin{align}
		\frac{\dd S_2}{\dd \eta} \propto a^{\Delta_s}\quad\text{at late time.}
	\end{align}
If $\Delta_s < 0$, the purity asymptotes a constant at late time, while if $\Delta_s \geq 0$, the linear entropy keeps increasing at late time and decoherence takes place.
	
We parametrize the late-time (or ``near-boundary'' in the language of \Refs{Mirbabayi:2015hva, Cohen:2020php}) scaling behavior of the operators appearing in the interaction Hamiltonian $\widetilde{H}_{\mathrm{int}} (\eta)$ with
	\begin{align}\label{eq:scaling}
		\widetilde{\mathcal{O}}^{\mathcal{S/E}}_{\bmk}(\eta) \rightarrow a^{\Delta_+^{\mathcal{S/E}}} \widetilde{\mathcal{O}}_+^{\mathcal{S/E}} 
	\end{align}
	where $\widetilde{\mathcal{O}}_+^{\mathcal{S/E}}$ are time-independent up to $\mathcal{O}(a^{-1})$ corrections. Assuming that the late-time behavior of the entropy production is controlled by the late-time scaling of the two-point functions of the system and the environment,
	\begin{align}
		\bar{\mathcal{K}}_{\mathcal{S}}^{(2)}(\bmk_1,\bmk_2, \eta_1,\eta_2)  &\rightarrow \left[a(\eta_1) a( \eta_2)\right]^{\Delta_+^{\mathcal{S}} }\, , \\
		\bar{\mathcal{K}}_{\mathcal{E}}^{(2)}(\bmk_1,\bmk_2, \eta_1,\eta_2)  & \rightarrow \left[a(\eta_1)  a(\eta_2)\right]^{\Delta_+^{\mathcal{E}} }\, ,
	\end{align}
	we obtain 
	\begin{tcolorbox}[%
		enhanced, 
		breakable,
		skin first=enhanced,
		skin middle=enhanced,
		skin last=enhanced
		]
		\paragraph{Scaling dimension of entropy production:}
		\begin{align}\label{eq:Sentscaling}
			\Delta_s = 2\Delta_+^{\mathcal{S}} + 2\Delta_+^{\mathcal{E}}+2\Delta_g-1.
		\end{align}
	\end{tcolorbox}	

Let us consider a few example of interests. For the model discussed in \Sec{sec:applic}, the system is described by a massless field, and by expanding \Eq{eq:modefctvp} at late time one finds $\pi \propto \mathrm{constant}+\eta^2$, hence $\Delta^{\pi}=0$ and $\Delta^{\pi'}=-1$. The environment consists of a test field with mass $m_\sigma$, for which by expanding \Eq{eq:modefctvs} at late time one finds~\cite{Mirbabayi:2015hva, Cohen:2020php} 
	\begin{align}
		\Delta_+^{\mathcal{E}} = \Rea \left( \sqrt{\frac{9}{4} - \frac{m^2}{H^2}}  \right)  - \frac{3}{2}.
	\end{align}
There are three interaction terms in \Eq{eq:S_int2}, which come with coupling scaling-dimensions $\Delta_g^{\pi'\sigma} = 3$, $\Delta_g^{(\partial_i \pi)^2 \sigma} = 2$ and $\Delta_g^{\pi^{\prime2} \sigma} = 2$ respectively. With \Eq{eq:Sentscaling} one thus finds
\bea
\Delta_s^{\pi'\sigma}= &\Delta_s^{(\partial_i \pi)^2 \sigma}\equiv \Delta_s=3+ 2\Delta_+^{\mathcal{E}}\, ,\\
\Delta_s^{\pi^{\prime2} \sigma}= &-1+ 2\Delta_+^{\mathcal{E}}\, .
\eea
Several comments are in order. First, the $\pi^{\prime2} \sigma$ interaction becomes negligible at late time since its scaling dimension is systematically smaller than for the other interaction terms. This is consistent with the results of \Sec{sec:applic} where $\mathcal{C}_1$ depends on $\Lambda_1$ but not on $\Lambda_2$, see \Eq{C1_text}. Second, the scaling dimension of the non-linear contributions coincide with the one of the linear term, which is again consistent with the results of \Sec{sec:applic}, see \Eqs{eq:massless_final}, \eqref{eq:oneloopconf} and \eqref{eq:oneloopheavy}. Third, for $m=0$ one obtains $\Delta_s=3$ which is compatible with \Eq{eq:massless_final}, for $m=\sqrt{2} H$ (conformal environment) one obtains $\Delta_s=1$ which is compatible with \Eq{eq:oneloopconf}, and for $m>3H/2$ one finds $\Delta_s=0$, which is again compatible with \Eq{eq:oneloopheavy}, recalling that $\partial_\eta = a \partial_N$.

More than a mere consistency check of the \Sec{sec:applic}, this shows that the rate at which decoherence or recoherence proceeds at late time can be readily anticipated from scaling arguments, which proves particularly convenient.\footnote{Note that there are cases where the late-time behavior of the purity is not only controlled by the late-time scaling~\eqref{eq:scaling} and the scaling argument fails. For instance, in \Refa{Burgess:2024eng}, only light environments satisfy the scaling argument but heavy environments do not.} Another example of the scaling argument is the case where, instead of letting the curvature perturbation interact with a scalar hidden sector, one considers environments made of higher-spin particles, like for instance the tensor modes $\gamma_{ij}$ of General Relativity \cite{Burgess:2022nwu}. Applying the above procedure to Maldacena's seminal cubic action derived in \Refa{Maldacena:2002vr}, we observe that among the tensor-scalar-scalar interaction terms, only $\gamma_{ij}(\partial_i \zeta) (\partial_j \zeta)$ leads to late-time decoherence of the curvature perturbations with $\Delta_s = 3$. Similarly, in the scalar-tensor-tensor sector, the only late-time decohering operator is $\zeta (\partial_\ell  \gamma_{ij}) (\partial_\ell  \gamma_{ij})$ with $\Delta_s = 3$.  Both of these contributions are $\epsilon_1^2$ suppressed and lead to decoherence as found in \Refa{Burgess:2022nwu}.\footnote{Resumming secular effects can sometimes change the scaling and this is why \Refa{Burgess:2022nwu} rather finds $\Delta_s=4$ at late times using a Markovian master equation.} Importantly, we note that interactions generating late-time decoherence are not so common.\footnote{If instead of considering a bipartition in terms of quantum numbers (mass, spin), one considers a bipartition in terms of a momentum cutoff $k_{\mathrm{UV}}$ \cite{Brahma:2020zpk, Burgess:2022nwu}, one could also investigate EFToI self-interactions $\pi^{\prime3}$ and $(\partial_i \pi)^2 \pi'$, which would naively lead to scaling dimensions $\Delta_s = -6$ and $\Delta_s = -2$ respectively, hence no late-time decoherence of the long-wavelength modes.}  One could also apply similar techniques to spin $1/2$  (see \eg \Refa{Tong:2023krn}) and spin $1$ (see \eg \Refs{Anber:2009ua, Bordin:2018pca, Lee:2016vti, Peloso:2022ovc}) environments, which we leave for future work.  

Finally, let us stress that the scaling dimension may not only allow one to determine whether decoherence takes place at late time or not, it may also determine whether quantum features get erased or not. Indeed, in \Refs{Martin:2021znx, Martin:2022kph}, it was found that due to the competition between decoherence and quantum squeezing at large scales, quantum discord still grows at late time if decoherence proceeds slow enough, \ie $\gamma_k \propto a^{-p}$ with $p<4$. This corresponds to a scaling dimension $\Delta_s<5$. Three cases can therefore be encountered:
	\begin{tcolorbox}[%
		enhanced, 
		breakable,
		skin first=enhanced,
		skin middle=enhanced,
		skin last=enhanced
		]
		\paragraph{Scaling dimension and quantum-to-classical transition.}
\begin{itemize}
\item $\Delta_s<0$: purity asymptotes a constant (possibly with recoherence) and decoherence does not take place.
\item $0\leq \Delta_s<5$: decoherence takes place but quantum signatures remain (quantum discord still increases).
\item $\Delta_s\geq 5$: decoherence takes place and quantum signatures are erased.
\end{itemize}
	\end{tcolorbox}	
	
	\subsection{Kraus measurement}\label{subsec:Kraus_measurement}
	
	We have seen in \Sec{subsubsec:NLO} that the $n^{\mathrm{th}}$-order correction to the total density matrix $\widetilde{\rho}^{(n)}$ can be written under the form~\eqref{eq:Kraus}. This brings our perturbative framework to a language that is close to the one of \textit{quantum operations} or \textit{dynamical maps}~\cite{breuerTheoryOpenQuantum2002}, where the evolution of the system is written as
	\begin{align}
		\widetilde{\rho}^{(n)}(\eta) = \sum_{m \in \mathcal{M}}  \Phi_m \left(\ket{\mathrm{BD}}\bra{\mathrm{BD}} \right) .
	\end{align}
Here, $\mathcal{M}$ is the set of possible outcomes or a generalized measurement. The maps $\Phi_m$ are completely positive and trace preserving if and only if there exists a countable set of operators $\Omega_{mk}$ fulfilling
	\begin{align}\label{eq:CPTP}
		\sum_k \Omega_{mk}^\dag \Omega_{mk}  \leq 1\, , 
	\end{align}
such that the maps act on the initial vacuum state according to~\cite{1983LNP...190.....K}
	\begin{align}
		\Phi_m \left(\ket{\mathrm{BD}}\bra{\mathrm{BD}} \right) = \sum_k \Omega_{mk} \ket{\mathrm{BD}}\bra{\mathrm{BD}}   \Omega_{mk}^\dag\, .
	\end{align}

From \Eq{eq:Kraus}, it should be possible to relate the $\Omega_{mk}$ operators to the ones appearing in the Kraus representation of \Sec{sec:Kraus}, \ie to the system and environment unequal-time correlators. In this way, the CPTP constraint~\eqref{eq:CPTP} should give rise to a set of conditions on these correlators. Those might be interpretable in terms of statistical inequalities that would guarantee the existence of underlying well-defined distributions, hence the CPTP nature of the dynamical map, along the lines of \Refs{Green:2023ids, Aoude:2024xpx}. This would be similar in spirit to the cosmological bootstrap program (see \Refa{Baumann:2022jpr} for a recent review) where unitarity is used to constrain the structure of the correlators~\cite{Goodhew:2020hob, Goodhew:2021oqg, Cespedes:2020xqq, Albayrak:2023hie}. In the absence of unitarity, other symmetries can be used to constrain the effective theory, see \Refs{LopezNacir:2011kk, Salcedo:2024smn}, and this picture would certainly benefit from CPTP constraints, the exploration of which is left for future work.

	\section{Conclusions}\label{sec:conclu}
	
The extent to which cosmological fluctuations decohere in the early universe is key to characterizing the strength of possible signatures of their quantum origin, as well as investigating the imprint of environmental degrees of freedom in general. However, the use of standard in-in techniques to compute decoherence measures, such as quantum purity, is not straightforward for two reasons. First, in-in methods are typically developed for unitary theories, whereas cosmological fluctuations evolve non-unitarily when coupled to an environment and should therefore be treated as an open quantum system. Second, purity and entanglement entropy cannot be simply expressed as the expectation value of observable operators. Consequently, standard perturbative in-in techniques must be adapted to compute these quantum information measures.

The goal of this work is to develop such a framework. We have shown how, at leading order in a perturbative expansion in the interaction strength, purity can be expressed in terms of the convolution of unequal time correlators of the system and the environment. For linear interactions, the Fourier modes of the system decouple and they remain in a Gaussian state. The amount by which each Fourier sector decoheres is decided by the determinant of its covariance matrix, hence the two-point functions of the system entirely determine its quantum state, as expected.

In the presence of non-linear interactions, the Fourier modes cannot be treated separately but we have introduced the concept of \emph{spectral purity}, in order to quantify the contribution of each scale to the overall entropy of the field. We have also found that, when the interaction is quadratic in the system's field variables, the spectral purity is still given by the determinant of the covariance matrix (with an additional factor of $1/2$). Technically, this is because the perturbations of the covariance matrix determinant and the particle number expectation value are related to one another. In practice, this means that even when the system is in a non-Gaussian state, its purity is still determined by its Gaussian moments at leading order, provided the interaction is quadratic in its field variables. In the case where both linear and quadratic interactions are present, non-linearities appear through loop corrections to the power spectra at higher order in perturbation theory.

We have cast our perturbative framework in terms of a diagrammatic expansion presented in \Sec{subsubsec:diagram}, which involves unequal time correlators represented by \textit{blobs}. One has to combine this diagrammatic approach to the usual in-in techniques \cite{Chen:2017ryl} to compute the blobs. The benefit of this formulation is however to single out the contributions that are effectively contributing to a change in the purity. If one had used an in-in approach from the get-go, a much larger number of diagrams would have had to be considered. This highlights how our formalism can be used in synergy with the usual in-in techniques. 

We then applied our formalism to a model of cosmological interest, where adiabatic perturbations couple to entropic degrees of freedom during inflation. By considering a subset of interactions within the Effective Field Theory of Inflation (EFToI), we were able to explore a non-linear extension of the results presented in \Refa{Colas:2022kfu}. 

We have found that the perturbative expansion breaks down at early time, but this is also where the EFT is expected to become invalid. At late times however, our framework can be trusted and it indicates that loop corrections, \ie the effects of the considered non-linear interaction terms, are suppressed by $\Delta_\zeta^2/\cs^4$, where $\Delta_\zeta^2$ is the amplitude of the curvature power spectrum at the scale of interest, and $\cs$ is the effective speed of sound. When the environment is massless, we show that linear entropy grows as $a^2$, leading to rapid decoherence after Hubble crossing. For an environment that is conformally coupled, linear entropy grows as $\ln^2a$, resulting in slower decoherence. Finally when the environment is heavy compared to the Hubble scale, the system experiences a transient phase of limited decoherence before recohering. In this case, the purity asymptotes to a constant at late times, remaining exponentially close to unity.

Note that this phenomenon of recoherence was observed in \Refa{Colas:2022kfu}, but doubts arose in the community whether this was merely a result of considering a bilinear mixing $\pi' \sigma$ between system and environment, raising concerns about the persistence of recoherence in the presence of non-linear interactions. The heuristic reason is the following. Recoherence is a clear symptom of non-Markovianity, while Markovian behavior is usually expected when the environment is ``large enough''. In the case of linear interactions, the system (\ie a given Fourier mode of the curvature perturbation) couples to a single degree of freedom (\ie the same Fourier mode of the isocurvature field), and this is why observing non-Markovian signatures is maybe not so surprising. However, in the presence of non-linear interactions, the system couples to \emph{all} Fourier modes of the environment, which might lead one to expect primarily Markovian evolution. Nevertheless, we have demonstrated that recoherence persists for the class of non-linear operators considered in this work. This indicates a robustness to loop effects, and our discussion of scaling dimensions reveals that far from being an academic curiosity, recoherence is in fact expected in a large number of cases. \\

The line of research presented in this work could be extended in various directions. First, the formalism we developed is generic and one may want to apply it to other cosmological and non-cosmological setups (see \eg \Refs{Danielson:2022tdw, Danielson:2022sga, Danielson:2024yru} in the context of black holes and \Refa{Belfiglio:2024qsa} for holography). When doing so, one might encounter situations in which non-linear self-interactions of the system and/or the environment are at play. In this case, three perturbative expansions should be implemented simultaneously: one in the self-interaction strength of the system $g_{\mathcal{S}}$, one in the self-interaction strength of the environment $g_{\mathcal{E}}$, and one in the system-environment coupling $g$. If $g_{\mathcal{S}}, g_{\mathcal{E}} \ll g$ (in the right units), the $g_{\mathcal{S}}$ and $g_{\mathcal{E}}$ expansions should be performed first, and the framework presented in this work can still be employed, where the self-interaction terms only give rise to corrections in the system and environment unequal-time correlators. In the opposite regime however, a different expansion scheme might be required. 

It may also happen that the relevant system and environment bipartition is not defined in terms of fields but rather in terms of different regions in physical space or different mode intervals of a given field. In that case, although the perturbative expansion should follow a similar structure as the one developed in this work, it might need to be generalised if the interaction Hamiltonian is not of the local form~\eqref{eq:HintFourier}. 

Second, one could consider alternative measures of decoherence and quantum entanglement, such as entanglement entropy or logarithmic negativity. At perturbative order, these measures can be simply interconnected, as discussed in \Sec{sec:vN}, but they may deliver different outcomes upon resummation.

Third, as divergences often appear in perturbative computations~\cite{Baidya:2017eho, Avinash:2019qga}, the renormalization of open quantum systems in the field theoretic context might also requires further investigations (see \eg \Refs{Agon:2014uxa, Agon:2017oia}). In particular, we have found that a divergent contribution to the entropy arises from early-time interactions in the model we considered [this is the term proportional to $\mathcal{C}_2$ in \Eq{Snon_Cans}]. When considering the entropy production rate this term drops out, hence at late time where the EFT become valid it plays no role, but the entropy rate still involves an integral over time, see \Eq{eq:entropyrateeg}. How early-time interactions decouple from late-time dynamics is therefore not obvious in general. A better understanding of divergences and renormalization in open quantum systems -- potentially informed by recent studies of finite-dimensional or harmonic oscillator open systems, such as those in \Refs{Winczewski:2021bpr,Correa:2023pwg,Crowder:2023zwq} -- may be necessary to determine more clearly whether and how a theory decouples at small scales.

Fourth, one might speculate that just as two-point functions characterize purity even for non-linear quadratic interactions at leading order, higher-order statistics like the bispectrum or trispectrum could potentially characterize purity at higher orders in the coupling or in the presence of higher-order interactions.

Fifth and finally, now that the problem is framed in terms of (equal or unequal-time) correlators, further investigation into the role of symmetries, UV unitarity, and locality in the evolution of the system's quantum information properties would be valuable. We leave this exploration for future work.
	
	\section*{Acknowledgements:} We thank C. Burgess, C. Duaso Pueyo, R. Holman and E. Pajer for useful discussions. This work is supported by STFC consolidated grant ST/X001113/1, ST/T000791/1, ST/T000694/1, ST/X000664/1 and EP/V048422/1.
	
	\appendix
	
	\section{Perturbative Gaussianity}\label{app:proof}
	
In the case of linear interactions, the system remains in a Gaussian state, hence it is fully characterized by its covariance matrix
	\begin{align}\label{eq:covmat:def}
		\delta(\bm{0})\; \bs{\mathrm{Cov}}_{ij}(k, \eta) = \frac{1}{2}\left\langle  \widetilde{\bm{z}}_i(\bm{k}, \eta) \widetilde{\bm{z}}_j^\dagger(\bm{k}, \eta) +  \widetilde{\bm{z}}_j(\bm{k}, \eta) \widetilde{\bm{z}}_i^\dagger(\bm{k}, \eta) \right\rangle\, ,
	\end{align}
see \Eq{eq:cov:def:matrix}. Here, $\widetilde{\bm{z}}(\bm{k}, \eta)$ is a phase-space vector gathering the phase-space variables describing the $\bm{k}$ Fourier mode of the system. For explicitness, we will consider the case where the system is made of a single scalar degree of freedom denoted $\pi$, hence $\widetilde{\bm{z}} = [\widetilde{v}_\pi(\bm{k}, \eta), \widetilde{p}_\pi(\bm{k}, \eta) ]^{\mathrm{T}}$ contains a position operator, $\widetilde{v}_\pi(\bm{k}, \eta)$, and its conjugate momentum, $\widetilde{p}_\pi(\bm{k},\eta)$.
The quantum state for the system factorizes between different Fourier modes, $\widetilde{\rho}_{\mathrm{red}}=\bigotimes _{\bmk\in\mathbb{R}^{3+}} \widetilde{\rho}_{\mathrm{red}} (\bmk)$, see \Sec{sec:Spectral:Purity}, and the purity for each Fourier mode is fully determined by the determinant of the covariance matrix, 
\begin{align}
S_2(k) = \ln\left[4 \det \bs{\mathrm{Cov}}(k,\eta) \right]\, ,
\end{align}
see \Sec{sec:Gaussian:Purity}. At order $g^2$, this gives rise to
\begin{align}
\label{eq:S2:Gauss}
S_2^{(2)}(k) =4  \mathrm{det}^{(2)}  \bs{\mathrm{Cov}}(k,\eta) \, .
\end{align}
When interactions are non linear, these considerations do not apply, and the purity receives contribution from all $n$-point correlation functions. In this appendix, our goal is to show that at order $g^2$, for interactions that are quadratic in the system's phase-space variables, only the two-point functions contribute, hence \Eq{eq:S2:Gauss} remains valid (with an additional factor $2$) even though the system is not in a Gaussian state anymore. \\

Without loss of generality, we consider the case where the operators appearing in the interaction Hamiltonian are centred, see the discussion below \Eq{eq:pert10}, hence the system is not modified at order $g$ and the first environmental effects appear at order $g^2$. At this order, one has
	\begin{align}\label{eq:covpert}
		\mathrm{det}^{(2)} \bs{\mathrm{Cov}} = \bs{\mathrm{Cov}}^{(2)}_{11} \bs{\mathrm{Cov}}^{(0)}_{22} + \bs{\mathrm{Cov}}^{(2)}_{22} \bs{\mathrm{Cov}}^{(0)}_{11} - 2 \bs{\mathrm{Cov}}^{(2)}_{12} \bs{\mathrm{Cov}}^{(0)}_{12}\, .
	\end{align}
In the free theory, the phase-space operators can be expanded onto creation and annihilation operators by means of the free, Bunch-Davies normalized mode functions $\bs{z}(k, \eta)$, according to
\bea
\label{eq:mode:function:expansion}
\widetilde{\bm{z}}(\bm{k},\eta) = \bm{z}(k,\eta) \widehat{c}_{\bm{k}} + \bm{z}^*(k,\eta) \widehat{c}_{-\bm{k}}^\dagger,
\eea
where $\widehat{c}_{\bm{k}}$ and $\widehat{c}_{\bm{k}}^\dagger$ are the annihilation and creation operators and $\bm{z}(k,\eta)=[v_\pi(k,\eta), p_\pi(k,\eta)]^{\mathrm{T}}$ gathers the mode functions. This implies that $\bs{\mathrm{Cov}}^{(0)}(k, \eta)$ contains the free power spectra and can be expressed in terms of the mode functions as 
	\begin{align}\label{eq:covmat}
		\bs{\mathrm{Cov}}^{(0)}_{ij}(k, \eta) = \frac{1}{2} \left[\bs{z}_i(k, \eta) \bs{z}_j^*(k, \eta) +  \bs{z}_i^*(k, \eta) \bs{z}_j(k, \eta)  \right] .
	\end{align}
The entries of $\bs{\mathrm{Cov}}^{(2)}$ correspond to expectation values computed against $\widetilde{\rho}_{\mathrm{red}}^{(2)}$. For a generic quantum operator $\widehat{Q}$, one has
	\begin{align}
		\left< \widehat{Q}(\eta) \right>^{(2)} \equiv \mathrm{Tr} \left[ \widetilde{Q}(\eta)\widetilde{\rho}_{\mathrm{red}}^{(2)} (\eta) \right]
	\end{align}
	where $\widetilde{\rho}_{\mathrm{red}}^{(2)}(\eta)$ is given in \Eq{eq:rho:red:2} that we reproduce here for clarity
	\begin{align}
		\widetilde{\rho}^{(2)}_{\mathrm{red}}(\eta) &= - \int_{-\infty}^\eta \dd \eta_1  \int_{-\infty}^{\eta_1} \dd \eta_2 \int  \frac{\dd^3\bmp}{(2\pi)^3} \int  \frac{\dd^3\bmq}{(2\pi)^3}   \\
		\bigg\{& \left[\widetilde{\mathcal{O}}^{\mathcal{S}}_{\bmp}(\eta_1) \widetilde{\mathcal{O}}^{\mathcal{S}}_{\bmq}(\eta_2) \ket{\mathrm{BD}}\bra{\mathrm{BD}}_{\mathcal{S}} -  \widetilde{\mathcal{O}}^{\mathcal{S}}_{\bmq}(\eta_2) \ket{\mathrm{BD}}\bra{\mathrm{BD}}_{\mathcal{S}} \widetilde{\mathcal{O}}^{\mathcal{S}}_{\bmp}(\eta_1) \right]  \mathcal{K}_{\mathcal{E}}^{(2)}(\bmp,\bmq, \eta_1,\eta_2) \nonumber \\
		-&\left[ \widetilde{\mathcal{O}}^{\mathcal{S}}_{\bmp}(\eta_1)\ket{\mathrm{BD}}\bra{\mathrm{BD}}_{\mathcal{S}} \widetilde{\mathcal{O}}^{\mathcal{S}}_{\bmq}(\eta_2) - \ket{\mathrm{BD}}\bra{\mathrm{BD}}_{\mathcal{S}} \widetilde{\mathcal{O}}^{\mathcal{S}}_{\bmq}(\eta_2)\widetilde{\mathcal{O}}^{\mathcal{S}}_{\bmp}(\eta_1)  \right] \mathcal{K}_{\mathcal{E}}^{(2)}(\bmq,\bmp, \eta_2,\eta_1) \bigg\}.\nonumber
	\end{align}
	It leads to 
	\begin{align}\label{eq:secondorder}
		\left< \widehat{Q}(\eta) \right>^{(2)} = - & \int_{-\infty}^\eta \dd \eta_1  \int_{-\infty}^{\eta_1} \dd \eta_2   \int \frac{\dd^3 \bmp}{(2\pi)^3} \int \frac{\dd^3 \bmq}{(2\pi)^3} \nonumber \\
		\bigg\{ &\mathcal{K}_{\mathcal{E}}^{(2)}(\bmp,\bmq, \eta_1,\eta_2)
		\left< \left[\widetilde{Q}(\eta) , \widetilde{\mathcal{O}}^{\mathcal{S}}_{\bmp}(\eta_1) \right]  \widetilde{\mathcal{O}}^{\mathcal{S}}_{\bmq}(\eta_2)\right>  \\
		-&\mathcal{K}_{\mathcal{E}}^{(2)}(\bmq,\bmp, \eta_2,\eta_1) \left<\widetilde{\mathcal{O}}^{\mathcal{S}}_{\bmq}(\eta_2) \left[\widetilde{Q}(\eta),\widetilde{\mathcal{O}}^{\mathcal{S}}_{\bmp}(\eta_1) \right]  \right>   \bigg\}\nonumber
	\end{align}
	where brackets represent vacuum expectation values in the initial vacuum state. 	

In practice, the determinant in \Eq{eq:covpert} can be obtained from \Eq{eq:secondorder} by replacing
	\begin{align}
		\widetilde{Q}(\eta)  = \bs{\mathrm{Cov}}^{(0)}_{22}(k, \eta)  &\widetilde{v}_\pi(\bmk, \eta)  \widetilde{v}_\pi(-\bmk, \eta) + \bs{\mathrm{Cov}}^{(0)}_{11}(k, \eta)  \widetilde{p}_\pi(\bmk, \eta)  \widetilde{p}_\pi(-\bmk, \eta) \nonumber\\
		-&  \bs{\mathrm{Cov}}^{(0)}_{12}(k, \eta) \left[\widetilde{v}_\pi(\bmk, \eta) \widetilde{p}_\pi(-\bmk, \eta) + \widetilde{p}_\pi(\bmk, \eta)\widetilde{v}_\pi(-\bmk, \eta)\right] .
	\end{align}
Making use of the canonical commutation relations, 
\begin{align}
\label{eq:com:1}
[\widetilde{v}_\pi(\bmk, \eta), \widetilde{v}_\pi^\dagger(\bmk', \eta)] =& [\widetilde{p}_\pi(\bmk, \eta), \widetilde{p}_\pi^\dagger(\bmk', \eta)] =0\, ,\\
[\widetilde{v}_\pi(\bmk, \eta), \widetilde{p}_\pi^\dagger(\bmk', \eta)] =& i\delta(\bmk-\bmk')\, ,
\label{eq:com:2}
\end{align}
and recalling that $\widetilde{v}_\pi(-\bmk, \eta)=\widetilde{v}_\pi^\dagger(\bmk, \eta)$ and $\widetilde{p}_\pi(-\bmk, \eta)=\widetilde{p}_\pi^\dagger(\bmk, \eta)$, the above can be rewritten as
	\begin{align}
		\widetilde{Q}(\eta)  = \bs{\mathrm{Cov}}^{(0)}_{22}(k, \eta)  & \widetilde{v}_\pi^\dagger(\bmk, \eta) \widetilde{v}_\pi(\bmk, \eta)  + \bs{\mathrm{Cov}}^{(0)}_{11}(k, \eta)    \widetilde{p}_\pi^\dagger(\bmk, \eta) \widetilde{p}_\pi(\bmk, \eta) \nonumber\\
		-&  \bs{\mathrm{Cov}}^{(0)}_{12}(k, \eta) \left[ \widetilde{p}_\pi^\dagger(\bmk, \eta) \widetilde{v}_\pi(\bmk, \eta)+ \widetilde{v}_\pi^\dagger(\bmk, \eta) \widetilde{p}_\pi(\bmk, \eta)\right] .
	\end{align}
	
Let us cast this relationship in matricial form. First, one has
	\begin{align}
	\widetilde{Q}(\eta)  = \frac{1}{4}\widetilde{\bs{z}}^{\dag}(\bmk, \eta) \left[\bs{\mathrm{Cov}}^{(0)}(k, \eta)\right]^{-1} \widetilde{\bs{z}}(\bmk, \eta) \, ,
		\end{align}
	where we have used that the inverse of the free covariance matrix can be written as
	\begin{align}
		\left[\bs{\mathrm{Cov}}^{(0)}(k, \eta)\right]^{-1} = \frac{1}{\det \bs{\mathrm{Cov}}^{(0)}}  \begin{pmatrix}
			\bs{\mathrm{Cov}}^{(0)}_{22}(k, \eta)& -\bs{\mathrm{Cov}}^{(0)}_{12}(k, \eta) \\
			-\bs{\mathrm{Cov}}^{(0)}_{12}(k, \eta) & \bs{\mathrm{Cov}}^{(0)}_{11}(k, \eta)
		\end{pmatrix}
	\end{align}
where  $\det \bs{\mathrm{Cov}}^{(0)} = 1/4$ since in the free theory the system is placed in a pure Gaussian state. Then, we rewrite \Eq{eq:mode:function:expansion} as
\begin{align}
\widetilde{\bm{z}}(\bm{k},\eta) = \bm{M}(k,\eta) \cdot \widehat{\bm{c}}_{\bm{k}}
\quad\text{where}\quad 
\bm{M}(k,\eta) = 
\begin{pmatrix}
v_\pi(k,\eta) & v^*_\pi(k,\eta)\\
p_\pi(k,\eta) & p^*_\pi(k,\eta)
\end{pmatrix}
\quad\text{and}\quad 
\widehat{\bm{c}}_{\bm{k}} = 
\begin{pmatrix}
 \widehat{c}_{\bm{k}} \\
\widehat{c}_{-\bm{k}}^\dagger
\end{pmatrix}\, .
\end{align}
The creation and annihilation operators obey the commutation relations
\begin{align}
\label{eq:com:c}
[\widehat{c}_{\bmk},\widehat{c}^\dagger_{\bmk'}]=\delta(\bmk-\bmk')
\quad\text{and}\quad
[\widehat{c}_{\bmk},\widehat{c}_{\bmk}']=[\widehat{c}_{\bmk}^\dagger,\widehat{c}_{\bmk '}^\dagger]=0
\end{align}
hence for \Eqs{eq:com:1} and~\eqref{eq:com:2} to be satisfied the free mode functions must be such that 
\begin{align}
v_\pi(k,\eta) p_\pi^*(k,\eta)-v_\pi^*(k,\eta) p_\pi(k,\eta) =i \, .
\end{align}
Using this relation repeatedly, one can readily show that
\begin{align}
\bm{M}^{\mathrm{T}*}(k,\eta) \cdot \left[\bs{\mathrm{Cov}}^{(0)}(k, \eta)\right]^{-1}  \cdot \bm{M}(k,\eta) =2\, \bm{\mathrm{Id}}_{2}\, ,
\end{align}
where $\bm{\mathrm{Id}}_{2}$ is the $2\times 2$ identity matrix. As a consequence,
\begin{align}
\label{eq:Q:eta:simp}
\widetilde{Q}(\eta)  = \frac{1}{2} \widehat{\bm{c}}_{\bmk}^\dagger \widehat{\bm{c}}_{\bmk} = \frac{1}{2}\left(\widehat{c}_{\bmk}^\dagger \widehat{c}_{\bmk} + \widehat{c}_{-\bmk}\widehat{c}_{-\bmk}^\dagger  \right)\, ,
\end{align}
hence $\widetilde{Q}(\eta)$ is in fact independent of $\eta$ and is directly related to the particle-number operator. Inserting \Eq{eq:Q:eta:simp} into \Eq{eq:secondorder}, one thus obtains
	\begin{align}
		\mathrm{det}^{(2)} \bs{\mathrm{Cov}}=  & - \frac{1}{2} \int_{-\infty}^\eta \dd \eta_1  \int_{-\infty}^{\eta_1} \dd \eta_2   \int \frac{\dd^3 \bmp}{(2\pi)^3} \int \frac{\dd^3 \bmq}{(2\pi)^3} \nonumber \\
		\bigg\{ &\mathcal{K}_{\mathcal{E}}^{(2)}(\bmp,\bmq, \eta_1,\eta_2)
		\left< \left[\widehat{c}_{\bmk}^\dagger \widehat{c}_{\bmk} + \widehat{c}_{-\bmk}\widehat{c}_{-\bmk}^\dagger , \widetilde{\mathcal{O}}^{\mathcal{S}}_{\bmp}(\eta_1) \right]  \widetilde{\mathcal{O}}^{\mathcal{S}}_{\bmq}(\eta_2)\right>  \\
		-&\mathcal{K}_{\mathcal{E}}^{(2)}(\bmq,\bmp, \eta_2,\eta_1) \left<\widetilde{\mathcal{O}}^{\mathcal{S}}_{\bmq}(\eta_2) \left[\widehat{c}_{\bmk}^\dagger \widehat{c}_{\bmk} + \widehat{c}_{-\bmk}\widehat{c}_{-\bmk}^\dagger,\widetilde{\mathcal{O}}^{\mathcal{S}}_{\bmp}(\eta_1) \right]  \right>   \bigg\}\nonumber\, .
	\end{align}
Recalling that $\mathcal{K}_{\mathcal{S}/\mathcal{E}}^{(2)}(-\bmk_2,-\bmk_1, \eta_2,\eta_1) = [\mathcal{K}_{\mathcal{S}/\mathcal{E}}^{(2)}(\bmk_1,\bmk_2, \eta_1,\eta_2)]^*$, and using the fact that the $\widetilde{O}^{\mathcal{S}}(\bmx)$ operators are Hermitian, hence $\widetilde{O}_{-\bmk}^{\mathcal{S}}=(\widetilde{O}_{\bmk}^{\mathcal{S}})^\dagger$, the two terms appearing in the above expression are complex conjugate of each other, hence one can write
	\begin{align}
		\mathrm{det}^{(2)} \bs{\mathrm{Cov}}=  & - \int_{-\infty}^\eta \dd \eta_1  \int_{-\infty}^{\eta_1} \dd \eta_2   \int \frac{\dd^3 \bmp}{(2\pi)^3} \int \frac{\dd^3 \bmq}{(2\pi)^3} \nonumber \\ &
		\Rea\bigg\{ \mathcal{K}_{\mathcal{E}}^{(2)}(\bmp,\bmq, \eta_1,\eta_2)
		\left< \left[\widehat{c}_{\bmk}^\dagger \widehat{c}_{\bmk} + \widehat{c}_{-\bmk}\widehat{c}_{-\bmk}^\dagger , \widetilde{\mathcal{O}}^{\mathcal{S}}_{\bmp}(\eta_1) \right]  \widetilde{\mathcal{O}}^{\mathcal{S}}_{\bmq}(\eta_2)\right> \bigg\}\, .
	\end{align}
Upon introducing the particle number operator
\begin{align}
\widehat{N}=\int\frac{\dd^3\bmk}{(2\pi)^3} \widehat{c}_{\bmk} \widehat{c}^\dagger_{\bmk}\, ,
\end{align}
the above leads to
\begin{align}
\int \frac{\dd^3\bmk}{(2\pi)^3}  \mathrm{det}^{(2)} \bs{\mathrm{Cov}}=  & - 2 \int_{-\infty}^\eta \dd \eta_1  \int_{-\infty}^{\eta_1} \dd \eta_2   \int \frac{\dd^3 \bmp}{(2\pi)^3} \int \frac{\dd^3 \bmq}{(2\pi)^3} \nonumber \\ &
		\Rea\left\{ \mathcal{K}_{\mathcal{E}}^{(2)}(\bmp,\bmq, \eta_1,\eta_2)
		\left< \left[\widehat{N}, \widetilde{\mathcal{O}}^{\mathcal{S}}_{\bmp}(\eta_1) \right]  \widetilde{\mathcal{O}}^{\mathcal{S}}_{\bmq}(\eta_2)\right> \right\}\, .
\end{align}
Since $\widehat{N}$ vanishes when acted on the vacuum, this finally gives
\begin{align}
\int \frac{\dd^3\bmk}{(2\pi)^3} \mathrm{det}^{(2)} \bs{\mathrm{Cov}}=  & 2 \int_{-\infty}^\eta \dd \eta_1  \int_{-\infty}^{\eta_1} \dd \eta_2   \int \frac{\dd^3 \bmp}{(2\pi)^3} \int \frac{\dd^3 \bmq}{(2\pi)^3}
 \nonumber \\ &
		\Rea\bigg\{ \mathcal{K}_{\mathcal{E}}^{(2)}(\bmp,\bmq, \eta_1,\eta_2)
		\left<  \widetilde{\mathcal{O}}^{\mathcal{S}}_{\bmp}(\eta_1) \widehat{N}  \widetilde{\mathcal{O}}^{\mathcal{S}}_{\bmq}(\eta_2)\right> \bigg\}\, .
		\label{eq:int:detCov}
\end{align}
	
Our next task is to compute the correlator appearing in the above expression. For explicitness, let us assume that $\widetilde{\mathcal{O}}^{\mathcal{S}}$ is of order $n$ in the phase-space operators, 
\begin{align}
\widetilde{\mathcal{O}}^{\mathcal{S}}(\bmx,\eta) = \mathcal{C}_{\alpha_1 \cdots \alpha_n}(\eta) \widetilde{\bs{z}}_{\alpha_1}(\bmx, \eta) \cdots  \widetilde{\bs{z}}_{\alpha_n}(\bmx, \eta) ,
\end{align}
where $\alpha_i\in\{1,2\}$ and $\mathcal{C}_{\alpha_1 \cdots \alpha_n}$ are free coefficients, so when Fourier expanding the above expression $\widetilde{\mathcal{O}}^{\mathcal{S}}_{\bmp}$ is of the form
\begin{align}\label{eq:Jope}
		\widetilde{\mathcal{O}}^{\mathcal{S}}_{\bmp}(\eta_1)  = (2\pi)^3 \int \frac{\dd^3 \bmp_1}{(2\pi)^3} \cdots \int \frac{\dd^3 \bmp_n}{(2\pi)^3} \mathcal{C}_{\alpha_1 \cdots \alpha_n} \left(\eta_1\right)  %\nonumber \\
		  \widetilde{\bs{z}}_{\alpha_1}(\bmp_1, \eta_1) \cdots  \widetilde{\bs{z}}_{\alpha_n}(\bmp_n, \eta_1) \delta\left(  \bmp_1 + \cdots + \bmp_n-\bmp \right). 
	\end{align}
As mentioned above, we consider the case where $\langle \widetilde{\mathcal{O}}^{\mathcal{S}}_{\bmk}(\eta)\rangle = 0$. This implies that, when $n$ is even, the mean value is implicitly subtracted from the above expression. Let us now investigate the first values of nn to gain more intuition on the above results. 
\begin{itemize}
\item $n=1$
\end{itemize}
When $n=1$, one has
\begin{align}
\left<  \widetilde{\mathcal{O}}^{\mathcal{S}}_{\bmp}(\eta_1) \widehat{N}  \widetilde{\mathcal{O}}^{\mathcal{S}}_{\bmq}(\eta_2)\right> = &
\int \frac{\dd^3\bmk}{(2\pi)^3}
C_{\alpha_1}(\eta_1) C_{\alpha_2}(\eta_2)
\nonumber \\ &
\left\langle\left[z_{\alpha_1}(p,\eta_1) \widehat{c}_{\bmp} + z_{\alpha_1}^*(p,\eta_1) \widehat{c}_{-\bmp}^\dagger \right]\widehat{N}_{\bmk} \left[z_{\alpha_2}(q,\eta_2) \widehat{c}_{\bmq} + z_{\alpha_2}^*(q,\eta_2) \widehat{c}_{-\bmq}^\dagger \right] \right\rangle ,
\end{align}
where $\widehat{N}_{\bmk} = \widehat{c}_\bmk c^\dagger_{\bmk}$. The only non-vanishing contribution arises from creating a particle with $\widehat{c}_{-\bmq}^\dagger$ and destroying it with $\widehat{c}_{\bmp}$, so one finds
 \begin{align}
\left<  \widetilde{\mathcal{O}}^{\mathcal{S}}_{\bmp}(\eta_1) \widehat{N}  \widetilde{\mathcal{O}}^{\mathcal{S}}_{\bmq}(\eta_2)\right> = &
\int \frac{\dd^3\bmk}{(2\pi)^3}
C_{\alpha_1}(\eta_1) C_{\alpha_2}(\eta_2)z_{\alpha_1}(p,\eta_1)z_{\alpha_2}^*(q,\eta_2)
\left\langle \widehat{c}_{\bmp} \widehat{N}_{\bmk}  \widehat{c}_{-\bmq}^\dagger  \right\rangle 
\nonumber \\ = &
\int \frac{\dd^3\bmk}{(2\pi)^3}
C_{\alpha_1}(\eta_1) C_{\alpha_2}(\eta_2)z_{\alpha_1}(p,\eta_1)z_{\alpha_2}^*(q,\eta_2)
\left\langle \widehat{c}_{\bmp}   \widehat{c}_{-\bmq}^\dagger  \right\rangle \delta\left(\bmk+\bmq\right)
\nonumber \\ = &
\int \frac{\dd^3\bmk}{(2\pi)^3}
C_{\alpha_1}(\eta_1) C_{\alpha_2}(\eta_2)z_{\alpha_1}(p,\eta_1)z_{\alpha_2}^*(q,\eta_2)
\delta\left(\bmp+\bmq\right)\delta\left(\bmk+\bmq\right)
\nonumber \\ = &
C_{\alpha_1}(\eta_1) C_{\alpha_2}(\eta_2)z_{\alpha_1}(p,\eta_1)z_{\alpha_2}^*(q,\eta_2)
\frac{\delta\left(\bmp+\bmq\right)}{(2\pi)^3}\, .
\end{align}
One can readily see that a similar calculation gives the same result for $\left<  \widetilde{\mathcal{O}}^{\mathcal{S}}_{\bmp}(\eta_1)  \widetilde{\mathcal{O}}^{\mathcal{S}}_{\bmq}(\eta_2)\right>$, up to the factor $(2\pi)^{-3}$, hence
\bea
\left<  \widetilde{\mathcal{O}}^{\mathcal{S}}_{\bmp}(\eta_1) \widehat{N}  \widetilde{\mathcal{O}}^{\mathcal{S}}_{\bmq}(\eta_2)\right>=
(2\pi)^{-3}\left<  \widetilde{\mathcal{O}}^{\mathcal{S}}_{\bmp}(\eta_1) \widetilde{\mathcal{O}}^{\mathcal{S}}_{\bmq}(\eta_2)\right> = 
(2\pi)^{-3} \bar{K}_{\mathcal{S}}^{(2)}(p,\eta_1,\eta_2)\delta\left(\bmp+\bmq\right)\, ,
\eea
see \Eqs{eq:K2:def} and~\eqref{eq:def:Kreduced}. Inserting this formula into \Eq{eq:int:detCov}, one has
\bea
\int  \dd^3 \bmk\,  \mathrm{det}^{(2)} \bs{\mathrm{Cov}}=  & 2 (2\pi)^{-6}
 \int_{-\infty}^\eta \dd \eta_1  \int_{-\infty}^{\eta_1} \dd \eta_2   \int  \dd^3 \bmp\,   \Rea\left[\bar{K}_{\mathcal{E}}^{(2)}(p,\eta_1,\eta_2)\bar{K}_{\mathcal{S}}^{(2)}(p,\eta_1,\eta_2)\right] ,
 \eea
 where we have used \Eq{eq:def:Kreduced} again. By comparison with \Eq{eq:S2:KS:KE}, this can be written as
 \bea
 \int  \dd^3\bmk\,  \mathrm{det}^{(2)} \bs{\mathrm{Cov}}=  & \frac{1}{4} \int \dd^3\bm{k}\, S_2^{(2)}(k)\, ,
 \eea
 hence
 \bea
 \label{eq:purity:cov:n_eq_1}
 S_2^{(2)}(k) = 4 \mathrm{det}^{(2)} \bs{\mathrm{Cov}}\, ,
 \eea
 which is the result announced, see \Eq{eq:S2:Gauss}. When $n=1$, as explained in \Sec{sec:Gaussian:Purity} the system remains in a Gaussian state and this result is already known, it also holds at all orders in $g$. The above derivation is however a consistency check of the validity of the present formalism.
 
\begin{itemize}
\item $n=2$
\end{itemize}
When $n=2$, one has to compute
\begin{align}
\left<  \widetilde{\mathcal{O}}^{\mathcal{S}}_{\bmp}(\eta_1) \widehat{N}  \widetilde{\mathcal{O}}^{\mathcal{S}}_{\bmq}(\eta_2)\right> = &
\int \frac{\dd^3\bmk}{(2\pi)^3}\int \frac{\dd^3\bmp_1}{(2\pi)^3}\int \frac{\dd^3\bmp_2}{(2\pi)^3}\int \frac{\dd^3\bmq_1}{(2\pi)^3}\int \frac{\dd^3\bmq_2}{(2\pi)^3}
\nonumber \\ & 
\delta(\bmp_1+\bmp_2-\bmp)
\delta(\bmq_1+\bmq_2-\bmq)
C_{\alpha_1 \alpha_2}(\eta_1) C_{\beta_2 \beta_2}(\eta_2)
\nonumber \\ &
\left\langle\left[z_{\alpha_1}(p_1,\eta_1) \widehat{c}_{\bmp_1} + z_{\alpha_1}^*(p_1,\eta_1) \widehat{c}_{-\bmp_1}^\dagger \right]
\left[z_{\alpha_2}(p_2,\eta_1) \widehat{c}_{\bmp_2} + z_{\alpha_2}^*(p_2,\eta_1) \widehat{c}_{-\bmp_2}^\dagger \right]
\widehat{N}_{\bmk} 
\right. \nonumber \\ & \left.
\left[z_{\beta_1}(q_1,\eta_1) \widehat{c}_{\bmq_1} + z_{\beta_1}^*(q_1,\eta_2) \widehat{c}_{-\bmq_1}^\dagger \right]
\left[z_{\beta_2}(q_2,\eta_1) \widehat{c}_{\bmq_2} + z_{\beta_2}^*(q_2,\eta_2) \widehat{c}_{-\bmq_2}^\dagger \right]
 \right\rangle  .
\end{align}
Amongst the several terms arising from developing the square brackets, the only one that is not annihilated by the particle-number operator is the one involving $\widehat{c}_{\bmp_1} \widehat{c}_{\bmp_2} \widehat{N}_{\bm k}\widehat{c}_{-\bmq_1}^\dagger \widehat{c}_{-\bmq_2}^\dagger$. Making use of the commutation relations~\eqref{eq:com:c}, one has
\bea
\label{eq:Nk:comm:cdag:cdag}
\left[\widehat{N}_{\bmk},\widehat{c}_{-\bmq_1}^\dagger \widehat{c}_{-\bmq_2}^\dagger\right]=\widehat{c}_{-\bmq_1}^\dagger \widehat{c}_{-\bmq_2}^\dagger \left[\delta\left(\bmk+\bmq_1\right)+\delta\left(\bmk+\bmq_2\right)\right]\, .
\eea
Upon integrating over $\bmk$, one thus finds
\begin{align}
\left<  \widetilde{\mathcal{O}}^{\mathcal{S}}_{\bmp}(\eta_1) \widehat{N}  \widetilde{\mathcal{O}}^{\mathcal{S}}_{\bmq}(\eta_2)\right> = &
2 (2\pi)^{-3}\int \frac{\dd^3\bmp_1}{(2\pi)^3}\int \frac{\dd^3\bmp_2}{(2\pi)^3}\int \frac{\dd^3\bmq_1}{(2\pi)^3}\int \frac{\dd^3\bmq_2}{(2\pi)^3}
\nonumber \\ & 
\delta(\bmp_1+\bmp_2-\bmp)
\delta(\bmq_1+\bmq_2-\bmq)
C_{\alpha_1 \alpha_2}(\eta_1) C_{\beta_2 \beta_2}(\eta_2)
\nonumber \\ &
z_{\alpha_1}(p_1,\eta_1) z_{\alpha_2}(p_2,\eta_1)z_{\beta_1}^*(q_1,\eta_2) z_{\beta_2}^*(q_2,\eta_2)
\left\langle \widehat{c}_{\bmp_1} \widehat{c}_{\bmp_2} \widehat{c}_{-\bmq_1}^\dagger\widehat{c}_{-\bmq_2}^\dagger \right\rangle .
\label{eq:Op:N:Oq:quad}
\end{align}
This coincides with the expression one obtains for $\langle  \widetilde{\mathcal{O}}^{\mathcal{S}}_{\bmp}(\eta_1)   \widetilde{\mathcal{O}}^{\mathcal{S}}_{\bmq}(\eta_2)\rangle$, up to two differences. First, \Eq{eq:Op:N:Oq:quad} contains an additional factor $2$, which is coming from the two terms appearing in the right-hand side of \Eq{eq:Nk:comm:cdag:cdag}. In other words, since two particles are created and annihilated, the presence of the particle-number operator yields a factor $2$. Second, in $\langle  \widetilde{\mathcal{O}}^{\mathcal{S}}_{\bmp}(\eta_1)   \widetilde{\mathcal{O}}^{\mathcal{S}}_{\bmq}(\eta_2)\rangle$ there are additional terms, for instance the one involving $ \widehat{c}_{\bmp_1} \widehat{c}_{-\bmp_2}^\dagger \widehat{c}_{\bmq_1}\widehat{c}_{-\bmq_2}^\dagger$, which do not appear in \Eq{eq:Op:N:Oq:quad} since they are annihilated by the particle-number operator. However, these terms contribute to $\langle  \widetilde{\mathcal{O}}^{\mathcal{S}}_{\bmp}(\eta_1)\rangle \langle   \widetilde{\mathcal{O}}^{\mathcal{S}}_{\bmq}(\eta_2)\rangle$, which is subtracted from the final result, see the remark below \Eq{eq:Jope}. From these considerations one concludes that $\left<  \widetilde{\mathcal{O}}^{\mathcal{S}}_{\bmp}(\eta_1) \widehat{N}  \widetilde{\mathcal{O}}^{\mathcal{S}}_{\bmq}(\eta_2)\right>=2 (2\pi)^{-3}\left<  \widetilde{\mathcal{O}}^{\mathcal{S}}_{\bmp}(\eta_1) \widetilde{\mathcal{O}}^{\mathcal{S}}_{\bmq}(\eta_2)\right> $.	 The rest of the calculation is identical to the case $n=1$, the factor $2$ being the only difference, and one thus finds
 \bea
2 S_2^{(2)}(k) = 4 \mathrm{det}^{(2)} \bs{\mathrm{Cov}}\, ,
 \eea
 where the factor $2$ has been singled out in the left-hand side. Compared to \Eq{eq:purity:cov:n_eq_1}, spectral purity is thus half of what it would be if the state was Gaussian, with the same covariance. 

\begin{itemize}
\item $n\geq 3$
\end{itemize}
When $n=3$, one has to compute
\begin{align}
\left<  \widetilde{\mathcal{O}}^{\mathcal{S}}_{\bmp}(\eta_1) \widehat{N}  \widetilde{\mathcal{O}}^{\mathcal{S}}_{\bmq}(\eta_2)\right> = &
\int \frac{\dd^3\bmk}{(2\pi)^3}\int \frac{\dd^3\bmp_1}{(2\pi)^3}\int \frac{\dd^3\bmp_2}{(2\pi)^3} \int \frac{\dd^3\bmp_3}{(2\pi)^3}\int \frac{\dd^3\bmq_1}{(2\pi)^3}\int \frac{\dd^3\bmq_2}{(2\pi)^3} \int \frac{\dd^3\bmq_3}{(2\pi)^3}
\nonumber \\ & 
\delta(\bmp_1+\bmp_2+\bmp_3-\bmp)
\delta(\bmq_1+\bmq_2+\bmq_3-\bmq)
C_{\alpha_1 \alpha_2 \alpha_3}(\eta_1) C_{\beta_2 \beta_2\beta_3}(\eta_2)
\nonumber \\ &
\left\langle\left[z_{\alpha_1}(p_1,\eta_1) \widehat{c}_{\bmp_1} + z_{\alpha_1}^*(p_1,\eta_1) \widehat{c}_{-\bmp_1}^\dagger \right]
\left[z_{\alpha_2}(p_2,\eta_1) \widehat{c}_{\bmp_2} + z_{\alpha_2}^*(p_2,\eta_1) \widehat{c}_{-\bmp_2}^\dagger \right]
\right. \nonumber \\ & \left.
\left[z_{\alpha_3}(p_3,\eta_1) \widehat{c}_{\bmp_3} + z_{\alpha_3}^*(p_3,\eta_1) \widehat{c}_{-\bmp_3}^\dagger \right]
\widehat{N}_{\bmk} 
\left[z_{\beta_1}(q_1,\eta_1) \widehat{c}_{\bmq_1} + z_{\beta_1}^*(q_1,\eta_2) \widehat{c}_{-\bmq_1}^\dagger \right]
\right. \nonumber \\ & \left.
\left[z_{\beta_2}(q_2,\eta_1) \widehat{c}_{\bmq_2} + z_{\beta_2}^*(q_2,\eta_2) \widehat{c}_{-\bmq_2}^\dagger \right]
\left[z_{\beta_3}(q_3,\eta_1) \widehat{c}_{\bmq_3} + z_{\beta_3}^*(q_3,\eta_2) \widehat{c}_{-\bmq_3}^\dagger \right]
 \right\rangle .
\end{align}
Several terms need to be distinguished. A first term arises from selecting creation operators only in the right-hand side of the particle-number operator. Making use of the commutation relations~\eqref{eq:com:c}, one has
\bea
\left[\widehat{N}_{\bmk} ,\widehat{c}_{-\bmq_1}^\dagger\widehat{c}_{-\bmq_2}^\dagger\widehat{c}_{-\bmq_3}^\dagger \right] = \widehat{c}_{-\bmq_1}^\dagger\widehat{c}_{-\bmq_2}^\dagger\widehat{c}_{-\bmq_3}^\dagger \left[\delta\left(\bmk + \bmq_1\right) + \delta\left(\bmk + \bmq_2\right) + \delta\left(\bmk + \bmq_3\right) \right]\, .
\eea
When integrated over $\bmk$, this thus leaves out a factor $3$. This is because $\widehat{c}_{-\bmq_1}^\dagger\widehat{c}_{-\bmq_2}^\dagger\widehat{c}_{-\bmq_3}^\dagger$ creates a net number of particles equal to $3$. A second category of terms arises from selecting two creation operators and one annihilation operator  in the right-hand side of the particle-number operator. There are three such terms: $\widehat{c}_{\bmq_1}\widehat{c}_{-\bmq_2}^\dagger\widehat{c}_{-\bmq_3}^\dagger$, $\widehat{c}_{-\bmq_1}^\dagger  \widehat{c}_{\bmq_2}\widehat{c}_{-\bmq_3}^\dagger$ and $\widehat{c}_{-\bmq_1}^\dagger \widehat{c}_{-\bmq_2}^\dagger\widehat{c}_{\bmq_3}$ (the latter vanishes when acted on the vacuum). For the first term for instance, using the commutation relations~\eqref{eq:com:c}, one finds
\bea
\left[\widehat{N}_{\bmk} ,\widehat{c}_{\bmq_1}\widehat{c}_{-\bmq_2}^\dagger\widehat{c}_{-\bmq_3}^\dagger \right] = \widehat{c}_{\bmq_1}\widehat{c}_{-\bmq_2}^\dagger\widehat{c}_{-\bmq_3}^\dagger
\left[\delta\left(\bmk+\bmq_2\right)+\delta\left(\bmk+\bmq_3\right)-\delta\left(\bmk+\bmq_1\right)\right]\, .
\eea
When integrated over $\bm{k}$, this leaves out a factor $1$, in agreement again with the fact that $\widehat{c}_{\bmq_1}\widehat{c}_{-\bmq_2}^\dagger\widehat{c}_{-\bmq_3}^\dagger$ creates a net number of particles equal to $1$. Therefore, these terms come with multiplicities that differ from their counterpart in the correlator $\langle  \widetilde{\mathcal{O}}^{\mathcal{S}}_{\bmp}(\eta_1)   \widetilde{\mathcal{O}}^{\mathcal{S}}_{\bmq}(\eta_2)\rangle$, hence there is no simple relationship between $\langle  \widetilde{\mathcal{O}}^{\mathcal{S}}_{\bmp}(\eta_1) \widehat{N}  \widetilde{\mathcal{O}}^{\mathcal{S}}_{\bmq}(\eta_2)\rangle$ and $\langle  \widetilde{\mathcal{O}}^{\mathcal{S}}_{\bmp}(\eta_1)   \widetilde{\mathcal{O}}^{\mathcal{S}}_{\bmq}(\eta_2)\rangle$. As a consequence, for $n\geq 3$ the leading-order purity is not only given by $\mathrm{det}^{(2)}\bm{\mathrm{Cov}}$.

        \section{Unnesting commutators by induction}\label{sec:unnest}

        In this Appendix, we demonstrate how the commutators appearing in \Eq{eq:rho:iterative:SPT} can be un-nested as in \Eq{eq:Krauss:n}, which we reproduce here for convenience
        \bea
        \label{eq:Krauss:nn}
        \left[\widetilde{H}_{\mathrm{int}}(\eta_1),\left[ \widetilde{H}_{\mathrm{int}}(\eta_2),\cdots \left[\widetilde{H}_{\mathrm{int}}(\eta_n),\ket{\mathrm{BD}}\bra{\mathrm{BD}} \right] \cdots \right] \right] = 
        \sum_{i=0}^n (-1)^{n-i} \sum_{j=1}^{\binom{n}{i}} \widetilde{\bs{\mathcal{K}}}^{(i)}_j \ket{\mathrm{BD}}\bra{\mathrm{BD}}\left( \widetilde{\bs{\mathcal{K}}}^{(n-i)}_{\binom{n}{i}-j+1}\right)^\dagger  .
        \eea
This identity can be shown by induction as follows. At orders $n=0$ and $n=1$, it is trivially satisfied. Let us now assume that it holds at order $n$. The commutator between $\widetilde{H}_{\mathrm{int}}(\eta_0)$ and \Eq{eq:Krauss:nn} can be written as
        \begin{align}
        &\left[\widetilde{H}_{\mathrm{int}}(\eta_0),\left[ \widetilde{H}_{\mathrm{int}}(\eta_1),\cdots \left[\widetilde{H}_{\mathrm{int}}(\eta_n),\ket{\mathrm{BD}}\bra{\mathrm{BD}} \right] \cdots \right] \right] =
        \sum_{i=0}^n (-1)^{n-i} \sum_{j=1}^{\binom{n}{i}}
         \nonumber \\ &
        \left[ 
        \widetilde{H}_{\mathrm{int}}(\eta_0){}^{(n)}\widetilde{\bs{\mathcal{K}}}^{(i)}_j \ket{\mathrm{BD}}\bra{\mathrm{BD}} \left({}^{(n)}\widetilde{\bs{\mathcal{K}}}^{(n-i)}_{\binom{n}{i}-j+1}\right)^\dagger
        -
        {}^{(n)}\widetilde{\bs{\mathcal{K}}}^{(i)}_j \ket{\mathrm{BD}}\bra{\mathrm{BD}}\left( {}^{(n)}\widetilde{\bs{\mathcal{K}}}^{(n-i)}_{\binom{n}{i}-j+1}\right)^\dagger \widetilde{H}_{\mathrm{int}}(\eta_0)
        \right]   .
        \label{eq:Krauss:interm}
        \end{align}
        In this expression, we have added an upper left index $n$ to stress that the $\widetilde{\bs{\mathcal{K}}}^{(i)}_j$ vectors are constructed out of $\widetilde{H}_{\mathrm{int}}(\eta_1),\, \cdots,\, \widetilde{H}_{\mathrm{int}}(\eta_n)$. If ${}^{(n+1)}\widetilde{\bs{\mathcal{K}}}^{(i)}_j$ denote the $\widetilde{\bs{\mathcal{K}}}^{(i)}_j$ vectors constructed out of $\widetilde{H}_{\mathrm{int}}(\eta_0),\, \cdots\, \widetilde{H}_{\mathrm{int}}(\eta_n)$, from \Eqs{eq:Kij:def:in}-\eqref{eq:Kij:def:end}, the $\binom{n}{i-1}$ first elements of the ${}^{(n+1)}\widetilde{\bs{\mathcal{K}}}^{(i)}$ vector contain $\widetilde{H}_{\mathrm{int}}(\eta_0)$ as the first term, and they can thus be obtained by multiplying $\widetilde{H}_{\mathrm{int}}(\eta_0)$ with the $\binom{n}{i-1}$ elements of the $ {}^{(n)}\widetilde{\bs{\mathcal{K}}}^{(i-1)}$ vector. In other words,
        \bea
        {}^{(n+1)}\widetilde{\bs{\mathcal{K}}}^{(i)}_j =\widetilde{H}_{\mathrm{int}}(\eta_0) {}^{(n)}\widetilde{\bs{\mathcal{K}}}^{(i-1)}_j
        \quad\text{for}\quad j\leq \binom{n}{i-1}\, .
        \eea
        The subsequent elements of the ${}^{(n+1)}\widetilde{\bs{\mathcal{K}}}^{(i)}$ vector do not contain $\widetilde{H}_{\mathrm{int}}(\eta_0)$, and they coincide with the $\binom{n}{i}$ elements of the ${}^{(n)}\widetilde{\bs{\mathcal{K}}}^{(i)}$ vector, hence
        \bea
        {}^{(n+1)}\widetilde{\bs{\mathcal{K}}}^{(i)}_{\binom{n}{i-1}+j} = {}^{(n)}\widetilde{\bs{\mathcal{K}}}^{(i)}_j \quad\text{for}\quad j\leq \binom{n}{i}\, .
        \eea
        Since $\binom{n}{i-1}+\binom{n}{i}=\binom{n+1}{i}$, this indeed determines all elements of the ${}^{(n+1)}\widetilde{\bs{\mathcal{K}}}^{(i)}$ vector, and the two above relations fix the element ordering in \Eqs{eq:Kij:def:in}-\eqref{eq:Kij:def:end}. Inserting them into \Eq{eq:Krauss:interm} leads to
        \begin{align}
        &\left[\widetilde{H}_{\mathrm{int}}(\eta_0),\left[ \widetilde{H}_{\mathrm{int}}(\eta_1),\cdots \left[\widetilde{H}_{\mathrm{int}}(\eta_n),\ket{\mathrm{BD}}\bra{\mathrm{BD}} \right] \cdots \right] \right] =
        \sum_{i=0}^n (-1)^{n-i} \sum_{j=1}^{\binom{n}{i}}
         \nonumber \\ &
        \left[ 
        {}^{(n+1)}\widetilde{\bs{\mathcal{K}}}^{(i+1)}_j \ket{\mathrm{BD}}\bra{\mathrm{BD}} \left({}^{(n+1)}\widetilde{\bs{\mathcal{K}}}^{(n-i)}_{\binom{n}{n-i-1}+\binom{n}{i}-j+1}\right)^\dagger
        -
        {}^{(n+1)}\widetilde{\bs{\mathcal{K}}}^{(i)}_{\binom{n}{i-1}+j} \ket{\mathrm{BD}}\bra{\mathrm{BD}}\left( {}^{(n+1)}\widetilde{\bs{\mathcal{K}}}^{(n-i+1)}_{\binom{n}{i}-j+1}\right)^\dagger
        \right]  ,
        \end{align}
        and noticing that $\binom{n}{n-i-1}+\binom{n}{i}=\binom{n+1}{i+1}$, one finds
        \begin{align}
        &\left[\widetilde{H}_{\mathrm{int}}(\eta_0),\left[ \widetilde{H}_{\mathrm{int}}(\eta_1),\cdots \left[\widetilde{H}_{\mathrm{int}}(\eta_n),\ket{\mathrm{BD}}\bra{\mathrm{BD}} \right] \cdots \right] \right] =
        \sum_{i=0}^n (-1)^{n-i} \sum_{j=1}^{\binom{n}{i}}
         \nonumber \\ &
        \left[ 
        {}^{(n+1)}\widetilde{\bs{\mathcal{K}}}^{(i+1)}_j \ket{\mathrm{BD}}\bra{\mathrm{BD}} \left({}^{(n+1)}\widetilde{\bs{\mathcal{K}}}^{(n-i)}_{\binom{n+1}{i+1}-j+1}\right)^\dagger
        -
        {}^{(n+1)}\widetilde{\bs{\mathcal{K}}}^{(i)}_{\binom{n}{i-1}+j} \ket{\mathrm{BD}}\bra{\mathrm{BD}}\left( {}^{(n+1)}\widetilde{\bs{\mathcal{K}}}^{(n-i+1)}_{\binom{n}{i}-j+1}\right)^\dagger
        \right]   .
        \end{align}
        In the first term, let us relabel $i+1\to i$, and in the second term, let us relabel $\binom{n}{i-1}+j\to j$. This gives
        \begin{align}
        &\kern-2cm \left[\widetilde{H}_{\mathrm{int}}(\eta_0),\left[ \widetilde{H}_{\mathrm{int}}(\eta_1),\cdots \left[\widetilde{H}_{\mathrm{int}}(\eta_n),\ket{\mathrm{BD}}\bra{\mathrm{BD}} \right] \cdots \right] \right] = \nonumber \\ &
        \sum_{i=1}^{n+1} (-1)^{n+1-i} \sum_{j=1}^{\binom{n}{i-1}}
        {}^{(n+1)}\widetilde{\bs{\mathcal{K}}}^{(i)}_j \ket{\mathrm{BD}}\bra{\mathrm{BD}} \left({}^{(n+1)}\widetilde{\bs{\mathcal{K}}}^{(n+1-i)}_{\binom{n+1}{i}-j+1}\right)^\dagger
         \nonumber \\ &
        + \sum_{i=0}^n (-1)^{n+1-i} \sum_{j=\binom{n}{i-1}+1}^{\binom{n}{i-1}+\binom{n}{i}}
         {}^{(n+1)}\widetilde{\bs{\mathcal{K}}}^{(i)}_{j} \ket{\mathrm{BD}}\bra{\mathrm{BD}}\left( {}^{(n+1)}\widetilde{\bs{\mathcal{K}}}^{(n-i+1)}_{\binom{n}{i}+\binom{n}{i-1}-j+1}\right)^\dagger .
        \end{align}
        Using again that $\binom{n}{i-1}+\binom{n}{i}=\binom{n+1}{i}$, this reduces to	
        \begin{align}
        &\kern-2cm \left[\widetilde{H}_{\mathrm{int}}(\eta_0),\left[ \widetilde{H}_{\mathrm{int}}(\eta_1),\cdots \left[\widetilde{H}_{\mathrm{int}}(\eta_n),\ket{\mathrm{BD}}\bra{\mathrm{BD}} \right] \cdots \right] \right] = \nonumber \\ &
        \sum_{i=1}^{n+1} (-1)^{n+1-i} \sum_{j=1}^{\binom{n}{i-1}}
        {}^{(n+1)}\widetilde{\bs{\mathcal{K}}}^{(i)}_j \ket{\mathrm{BD}}\bra{\mathrm{BD}} \left({}^{(n+1)}\widetilde{\bs{\mathcal{K}}}^{(n+1-i)}_{\binom{n+1}{i}-j+1}\right)^\dagger
         \nonumber \\ &
         +\sum_{i=0}^n (-1)^{n+1-i} \sum_{j=\binom{n}{i-1}+1}^{\binom{n+1}{i}}
         {}^{(n+1)}\widetilde{\bs{\mathcal{K}}}^{(i)}_{j} \ket{\mathrm{BD}}\bra{\mathrm{BD}}\left( {}^{(n+1)}\widetilde{\bs{\mathcal{K}}}^{(n-i+1)}_{\binom{n+1}{i}-j+1}\right)^\dagger .
        \end{align}
        The first sum does not have a term with $i=0$, but its argument vanishes when evaluated at $i=0$, hence the sum can be extended to $i=0\cdots n+1$. Likewise, the argument of the second sum vanishes when evaluated at $i=n+1$, so the two sums can be extended to $i=0\cdots n+1$, and one finds
        \begin{align}
        &\kern-2cm \left[\widetilde{H}_{\mathrm{int}}(\eta_0),\left[ \widetilde{H}_{\mathrm{int}}(\eta_1),\cdots \left[\widetilde{H}_{\mathrm{int}}(\eta_n),\ket{\mathrm{BD}}\bra{\mathrm{BD}} \right] \cdots \right] \right] = \nonumber \\ &
        \sum_{i=0}^{n+1} (-1)^{n+1-i} \sum_{j=1}^{\binom{n+1}{i}}
        {}^{(n+1)}\widetilde{\bs{\mathcal{K}}}^{(i)}_j \ket{\mathrm{BD}}\bra{\mathrm{BD}} \left({}^{(n+1)}\widetilde{\bs{\mathcal{K}}}^{(n+1-i)}_{\binom{n+1}{i}-j+1}\right)^\dagger
         \, ,\end{align}
        which concludes the proof of \Eq{eq:Krauss:nn} by induction.
	
	\section{System memory kernel}
	\label{App:2ptfunc}
	
	This appendix aims at computing the memory kernel of the system corresponding to the interaction Hamiltonian~\eqref{Ham_kspace}, and derive the expressions~\eqref{KSlin_ans} and \eqref{KSnon_ans}.
		
	Using the definition~\eqref{centred} together with \Eqs{eq:K1:def} and~\eqref{eq:K2:def}, one finds
\bea\label{KS_App}
	\bar{\mathcal{K}}_{\mathcal{S}}^{(2)}(\bmk_1,\bmk_2, \eta_1,\eta_2) = &\bra{\mathrm{BD}} \OSlin(\bm{k}_1,\eta_1) \OSlin(\bm{k}_2,\eta_2) \ket{\mathrm{BD}}_{\mathcal{E}} + \bra{\mathrm{BD}} \OSlin(\bm{k}_1,\eta_1) \OSnon(\bm{k}_2,\eta_2) \ket{\mathrm{BD}}_{\mathcal{E}}  \Big.\\
	& + \bra{\mathrm{BD}} \OSnon(\bm{k}_1,\eta_1) \OSlin(\bm{k}_2,\eta_2) \ket{\mathrm{BD}}_{\mathcal{E}} + \bra{\mathrm{BD}} \OSnon(\bm{k}_1,\eta_1) \OSnon(\bm{k}_2,\eta_2) \ket{\mathrm{BD}}_{\mathcal{E}}  \Big. \\
	& -\bra{\mathrm{BD}} \OSlin(\bm{k}_1,\eta_1) \ket{\mathrm{BD}}_{\mathcal{E}} \bra{\mathrm{BD}} \OSlin(\bm{k}_2,\eta_2) \ket{\mathrm{BD}}_{\mathcal{E}} \\
	&  - \bra{\mathrm{BD}} \OSlin(\bm{k}_1,\eta_1) \ket{\mathrm{BD}}_{\mathcal{E}} \bra{\mathrm{BD}} \OSnon(\bm{k}_2,\eta_2) \ket{\mathrm{BD}}_{\mathcal{E}}  \Big. \\
	& - \bra{\mathrm{BD}} \OSnon(\bm{k}_1,\eta_1) \ket{\mathrm{BD}}_{\mathcal{E}} \bra{\mathrm{BD}} \OSlin(\bm{k}_2,\eta_2) \ket{\mathrm{BD}}_{\mathcal{E}} \\
	& -\bra{\mathrm{BD}} \OSnon(\bm{k}_1,\eta_1) \ket{\mathrm{BD}}_{\mathcal{E}} \bra{\mathrm{BD}} \OSnon(\bm{k}_2,\eta_2) \ket{\mathrm{BD}}_{\mathcal{E}}. \Big.
\eea
	The operators $\OSlin$ and $\OSnon$ were introduced in \Eqs{OSlin_def} and \eqref{OSnon_def}, which we repeat here for convenience
\begin{align}
\OSlin(\bm{k},\eta) = & \; a(\eta) \widetilde{\Pi}_\pi(\bm{k},\eta)  \label{OSlin_def_App} \\
\OSnon(\bm{k},\eta)  \equiv & \;  \int \frac{\exd^3 \bm{p}_1}{(2\pi)^3} \int \frac{\exd^3 \bm{p}_2}{(2\pi)^3} \; \frac{ \delta(\bm{p}_1 + \bm{p}_2 - \bm{k}) }{ a(\eta) } \left[ \frac{\bm{p}_1 . \bm{p}_2\; \widetilde{v}_\pi(\bm{p}_1,\eta) \widetilde{v}_{\pi}(\bm{p}_2,\eta)}{\Lambda_1} - \frac{\widetilde{\Pi}_\pi(\bm{p}_1,\eta) \widetilde{\Pi}_\pi(\bm{p}_2,\eta) }{\Lambda_2}  \right]\, . \label{OSnon_def_App} 
\end{align}
Here we use the shorthand notation
\begin{equation}
\widetilde{\Pi}_\pi(\bm{k},\eta) \equiv \widetilde{p}_\pi(\bm{k},\eta) - \frac{a'(\eta)}{a(\eta)} \widetilde{v}_\pi(\bm{k},\eta) \, ,
\end{equation}
whose mode functions are given by
\begin{equation}
\Pi_\pi(k,\eta) \equiv p_\pi(k,\eta) - \frac{a'(\eta)}{a(\eta)} \widetilde{v}_\pi(\bm{k},\eta) = - i \sqrt{ \frac{\cs k}{2} }\; e^{- i \cs k \eta} \, , \label{Pi_mode_functions}
\end{equation}
which follow from \Eqs{eq:modefctvp} and~\eqref{eq:modefctpp}.
Since $\OSlin$ is linear and $\OSnon$ is quadratic in the field variables, all vacuum expectation values involving an odd power of fields in \Eq{KS_App} vanish, leaving
	\begin{align} 
	\bar{\mathcal{K}}_{\mathcal{S}}^{(2)}(\bmk_1,\bmk_2, \eta_1,\eta_2)  = \KSlin
	(\bmk_1,\bmk_2, \eta_1,\eta_2)  + \KSnon(\bmk_1,\bmk_2, \eta_1,\eta_2)
	\end{align} 
where
\begin{equation} 
\KSlin(\bmk_1,\bmk_2, \eta_1,\eta_2) = \bra{\mathrm{BD}} \OSlin(\bm{k}_1,\eta_1) \OSlin(\bm{k}_2,\eta_2) \ket{\mathrm{BD}}_{\mathcal{E}} \label{KSlin_App_1}
\end{equation}
and 
\bea 
\KSnon(\bmk_1,\bmk_2, \eta_1,\eta_2) = & \bra{\mathrm{BD}} \OSnon(\bm{k}_1,\eta_1) \OSnon(\bm{k}_2,\eta_2) \ket{\mathrm{BD}}_{\mathcal{E}} \label{KSnon_App_1} \\
&- \bra{\mathrm{BD}} \OSnon(\bm{k}_1,\eta_1) \ket{\mathrm{BD}}_{\mathcal{E}} \bra{\mathrm{BD}} \OSnon(\bm{k}_2,\eta_2) \ket{\mathrm{BD}}_{\mathcal{E}} \, .
\eea
This also means that all cross-terms between the operators $\OSlin$ and $\OSnon$ in the system two-point functions vanish, as claimed in the main text. 

\paragraph{Linear contribution} Upon introducing \Eqs{OSlin_def_App} and \eqref{Pi_mode_functions} into \Eq{KSlin_App_1}, one finds that 
\bea
	\KSlin(\bmk_1,\bmk_2, \eta_1,\eta_2) = & a(\eta_1) a(\eta_2) \bra{\mathrm{BD}} \widetilde{\Pi}_\pi(\bm{k}_1,\eta_1) \widetilde{\Pi}_\pi(\bm{k}_2,\eta_2) \ket{\mathrm{BD}}_{\mathcal{E}}  \\
	= & a(\eta_1) a(\eta_2) \Pi_\pi(k_1,\eta_1) \Pi^{\ast}_\pi(k_2,\eta_2) \delta(\bm{k}_1 + \bm{k}_2) \\
	= & \frac{\cs k_1 }{2 H^2\eta_1\eta_2} e^{- i\cs k_1 (\eta_1 - \eta_2)} \delta(\bm{k}_1 + \bm{k}_2) \, .
\eea
When combined with \Eq{eq:def:Kreduced}, this results in \Eq{KSlin_ans} in the main text.

\paragraph{Non-linear contribution} The computation of $\KSnon$ is more involved due to the larger number of field operators contained within the expectation values. Inserting \Eq{OSnon_def_App} into \Eq{KSnon_App_1} ones finds
\begin{eqnarray} 
	&& \KSnon(\bmk_1,\bmk_2, \eta_1,\eta_2) = \prod_{i,j=1}^2  \int \frac{\dd^3 \bmp_i}{(2\pi)^3} \int \frac{\dd^3\bmq_j}{(2\pi)^3} \;  \frac{\delta(\bmp_1 + \bmp_2 - \bmk_1 ) \delta(\bmq_1 + \bmq_2 - \bmk_2 )}{a(\eta_1) a(\eta_2)}\qquad  \\
	&& \times \;\bigg\{ \; \frac{ (\bmp_1.\bmp_2)(\bmq_1.\bmq_2) }{\Lambda_1^2 }\left[ \Scale[0.85]{  \left<\widetilde{v}_{\pi}(\bmp_1,\eta_1) \widetilde{v}_{\pi}(\bmp_2,\eta_1) \widetilde{v}_{\pi}(\bmq_1,\eta_2) \widetilde{v}_{\pi}(\bmq_2,\eta_2)  \right>  -\left<\widetilde{v}_{\pi}(\bmp_1,\eta) \widetilde{v}_{\pi}(\bmp_2,\eta) \right> \left< \widetilde{v}_{\pi}(\bmq_1,\eta') \widetilde{v}_{\pi}(\bmq_2,\eta') \right> } \right] \notag \\
	&& \quad - \frac{ \bmq_1.\bmq_2 }{\Lambda_1\Lambda_2}  \left[  \Scale[0.85]{ \left<\widetilde{\Pi}_{\pi}(\bmp_1,\eta_1) \widetilde{\Pi}_{\pi}(\bmp_2,\eta_1) \widetilde{v}_{\pi}(\bmq_1,\eta_2) \widetilde{v}_{\pi}(\bmq_2,\eta_2)  \right> -\left<\widetilde{\Pi}_{\pi}(\bmp_1,\eta_1) \widetilde{\Pi}_{\pi}(\bmp_2,\eta_1) \right> \left< \widetilde{v}_{\pi}(\bmq_1,\eta_2) \widetilde{v}_{\pi}(\bmq_2,\eta_2) \right> } \right] \nonumber \\  
	&& \quad - \frac{\bmp_1.\bmp_2}{\Lambda_1 \Lambda_2}  \left[  \Scale[0.85]{  \left<\widetilde{v}_{\pi}(\bmp_1,\eta_1) \widetilde{v}_{\pi}(\bmp_2,\eta_1) \widetilde{\Pi}_{\pi}(\bmq_1,\eta_2) \widetilde{\Pi}_{\pi}(\bmq_2,\eta_2)  \right> -\left<\widetilde{v}_{\pi}(\bmp_1,\eta_1) \widetilde{v}_{\pi}(\bmp_2,\eta_1) \right> \left< \widetilde{\Pi}_{\pi}(\bmq_1,\eta_2) \widetilde{\Pi}_{\pi}(\bmq_2,\eta_2) \right> } \right] \notag \\
	&& \quad + \frac{1}{\Lambda_2^2} \left[ \Scale[0.85]{ \left<\widetilde{\Pi}_{\pi}(\bmp_1,\eta_1) \widetilde{\Pi}_{\pi}(\bmp_2,\eta_1) \widetilde{\Pi}_{\pi}(\bmq_1,\eta_2) \widetilde{\Pi}_{\pi}(\bmq_2,\eta_2)  \right> -\left<\widetilde{\Pi}_{\pi}(\bmp_1,\eta_1) \widetilde{\Pi}_{\pi}(\bmp_2,\eta_1) \right> \left< \widetilde{\Pi}_{\pi}(\bmq_1,\eta_2) \widetilde{\Pi}_{\pi}(\bmq_2,\eta_2) \right> }\right] \bigg\} \notag ,
\end{eqnarray} 
where we have used the shorthand $\langle \widetilde{\mathcal{O}} \rangle = \langle \mathrm{BD} | \widetilde{\mathcal{O}} | \mathrm{BD} \rangle_{\mathcal{E}}$. Performing the Wick contractions, and extracting a factor of $\delta(\bm{k}_1+\bm{k}_2)$ via \Eq{eq:def:Kreduced}, leaves
\begin{eqnarray} 
	&& \KSnon(k, \eta_1,\eta_2) = \frac{2}{(2\pi)^6} \int \frac{\dd^3 \bmp}{(2\pi)^3} \int \frac{\dd^3 \bmq}{(2\pi)^3} \; \delta(\bmp + \bmq - \bmk) \label{corr_In} \\
	&& \times \; \bigg[ \; (\bmp.\bmq)^2 \; \frac{ v_\pi(p, \eta_1)  v_\pi(q, \eta_1)  v^*_\pi(p, \eta_2) v^*_\pi(q, \eta_2) }{ a(\eta_1) a(\eta_2) \Lambda_1^2 }  \; + \; \frac{\Pi_\pi(p, \eta_1)  \Pi_\pi(q, \eta_1) \Pi^*_\pi(p, \eta_2) \Pi^*_\pi(q, \eta_2)}{ a(\eta_1) a(\eta_2) \Lambda_2^2} \notag  \\
	&& \ \ - \; \bmp.\bmq\;  \frac{ \Pi_\pi(p, \eta_1)  \Pi_\pi(q, \eta_1)  v^*_\pi(p, \eta_2) v^*_\pi(q, \eta_2)  +  v_\pi(p, \eta_1)  v_\pi(q, \eta_1)  \Pi^*_\pi(p, \eta_2) \Pi^*_\pi(q, \eta_2)}{ a(\eta_1) a(\eta_2) \Lambda_1\Lambda_2}  \bigg] \notag  .
\end{eqnarray} 
To proceed we note that the above double-momentum integral consists of terms of the form
\begin{eqnarray} 
	\mathcal{I}_{n}[f] :=  \frac{2}{(2\pi)^6} \int \frac{\dd^3 \bmp}{(2\pi)^3} \int \frac{\dd^3 \bmq}{(2\pi)^3} \; \delta(\bmp + \bmq - \bmk)\;  (\bm{p} \cdot \bm{q} )^n \; f(p,q) \label{In_def}
\end{eqnarray} 
for three choices of functions $f = f(p,q)$ depending on $p$ and $q$ (and not the angles). Using this compact notation, the correlator can be written in terms of the above integrals for $n=0,1,2$ such that
\bea \label{KSlin_In}
	\KSnon(k, \eta_1,\eta_2)  = & \mathcal{I}_{2}\left[ \tfrac{ v_\pi(p, \eta_1)  v_\pi(q, \eta_1)  v^*_\pi(p, \eta_2) v^*_\pi(q, \eta_2) }{ a(\eta_1) a(\eta_2) \Lambda_1^2 } \right] + \mathcal{I}_{0}\left[ \tfrac{\Pi_\pi(p, \eta_1)  \Pi_\pi(q, \eta_1) \Pi^*_\pi(p, \eta_2) \Pi^*_\pi(q, \eta_2)}{ a(\eta_1) a(\eta_2) \Lambda_2^2} \right] \qquad \\
	& - \mathcal{I}_1\left[ \tfrac{ \Pi_\pi(p, \eta_1) \Pi_\pi(q, \eta_1)  v^*_\pi(p, \eta_2) v^*_\pi(q, \eta_2)  +  v_\pi(p, \eta_1)  v_\pi(q, \eta_1)  \Pi^*_\pi(p, \eta_2) \Pi^*_\pi(q, \eta_2)}{ a(\eta_1) a(\eta_2) \Lambda_1\Lambda_2}  \right]  .
\eea
The next step is to integrate the angles away in \Eq{In_def}, which is why we now convert to spherical coordinates $\bm{q} = (q,\theta_{q},\varphi_{q})$ and $\bm{p} = (p,\theta_{p}, \varphi_{p})$. With the aim of integrating away $\bm{q}$ in the $\delta$-function we note that
	\begin{equation}
		\delta (\bm{q} - \bm{\ell}) = \frac{\delta(q - \ell)
			\delta(\theta_q - \theta_\ell) \delta(\varphi_q - \varphi_\ell)}
		{q^2 | \sin \theta_{q}| } 
	\end{equation}
	for arbitrary $\bm{\ell}$. Expressing the vector $\bm{\ell} \equiv \bm{k} - \bm{p}$ in spherical coordinates,
	\begin{align}
		\ell & = \sqrt{ p^2 + k^2  - 2 k p \cos\theta_p }\, , \quad 
		\theta_{\ell} = \cos^{-1}\left( \frac{k-p \cos \theta_p}
		{\sqrt{p^2 + k^2  - 2 k p \cos \theta_p}} \right) \, , \quad 
		\varphi_{\ell} =  \varphi_p + \pi \, ,
	\end{align}
turns \Eq{In_def} into
	\begin{align}
		I_{n}[f] =&\ \frac{2}{(2\pi)^{12}}
		\int_{0}^{\infty} \dd q \, q^2 \int_{0}^{\pi} \dd \theta_{q} \; \sin \theta _q
		\int_0^{2\pi} \dd \varphi_{q} \int_{0}^{\infty} \dd p \, p^2
		\int_{0}^{\pi} \dd \theta_{p} \; \sin \theta_p \int_0^{2\pi} \dd \varphi_{p}
		\nonumber \\ & \times
		\frac{1}{q^2 \sin \theta_q}
		\delta\left(q - \sqrt{ p^2 + k^2  - 2 k p \cos\theta_p } \right)
		\delta\left[\theta_{q} -
		\cos^{-1}\left(\frac{k-p \cos\theta_p}
		{\sqrt{p^2 + k^2  - 2 k p \cos\theta_p }} \right)  \right] 
		\nonumber \\ & \times
		\delta\left[ \varphi_{q} - (\varphi_p + \pi ) \right] \Big\{qp \left[ \sin \theta_{q} \sin \theta_{p}
		\cos(\varphi_{q}  - \varphi_{p}) + \cos \theta_{q} 
		\cos\theta_{p} \right]\Big\}^n \; f(p,q) \, .
	\end{align}
	The integration over $\varphi_q$ simply leads to the replacement of $\varphi_q$ by $\varphi_p + \pi$ and the integration over $\varphi_p$ just yields a factor $2\pi$, which leads to 
\bea
		I_n[f]  = &\ \frac{2}{(2\pi)^{11}}
		\int_{0}^{\infty} \dd q \,  \int_{0}^{\pi} \dd \theta_{q} 
		\int_{0}^{\infty} \dd p \, p^2
		\int_{0}^{\pi} \dd \theta_{p} \;  \sin \theta_p \; \delta\left(q - \sqrt{ p^2 + k^2  - 2 k p \cos\theta_p } \right) \\
		& \times \delta\left[\theta_{q} -\cos^{-1}\left(\frac{k-p \cos\theta_p}
		{\sqrt{p^2 + k^2  - 2 k p \cos\theta_p }} \right)  \right] \; \left[ qp\left(- \sin \theta_{q} \sin \theta_{p} + \cos \theta_{q} \cos\theta_{p} \right)\right]^n \; f(p,q)\, . 
\eea
	The next step consists in integrating over $\theta_q$. Using the identities
	\begin{align}
		\cos\left[\cos^{-1}\left(\frac{k-p \cos\theta_p}
		{\sqrt{ p^2 + k^2  - 2 k p \cos\theta_p}}  \right) \right]
		=  & \frac{k-p \cos\theta_p}{\sqrt{ p^2 + k^2  - 2 k p \cos\theta_p}}\ , \\
		\sin\left[\cos^{-1}\left(\frac{k-p \cos\theta_p}
		{\sqrt{ p^2 + k^2  - 2 k p \cos\theta_p}}  \right) \right]
		=& \frac{p \sin\theta_p}{\sqrt{ p^2 + k^2  - 2 k p \cos\theta_p}} \ ,
	\end{align}
	leads to
	\begin{align}
		I_n[f] = & \ \frac{2}{(2\pi)^{11}}
		\int_{0}^{\infty} \dd q \,  
		\int_{0}^{\infty} \dd p \, p^2
		\int_{0}^{\pi} \dd \theta_{p} \; \sin \theta_p\; 
		\delta\left(q - \sqrt{ p^2 + k^2  - 2 k p \cos\theta_p } \right)
		\nonumber \\
		&\times \left\{qp \frac{ \left[ \left(k - p \cos \theta_{p} \right) \cos \theta_{p}
			- p \sin^2 \theta_{p}\right]}{\sqrt{p^2 + k^2  - 2 k p \cos \theta_p} } \right\}^n f(p,q) \ . 
	\end{align}
	We then switch to $\mu = \cos  \theta_p$ and use the fact that 
	\begin{equation}
		\delta\left( q - \sqrt{ p^2 + k^2 - 2 p k \mu } \right)
		= \frac{q}{kp} \, \delta\left( \mu - \frac{p^2 + k^2 - q^2}{2 p k} \right) 
	\end{equation}
	to rewrite
	\begin{align}
		I_n[f] = & \ \frac{2}{(2\pi)^{11}k}
		\int_{0}^{\infty} \dd q \,  
		\int_{0}^{\infty} \dd p \,
		\int_{-1}^{1} \dd \mu\; \delta\left( \mu - \frac{p^2 + k^2 - q^2}{2 p k} \right)  \frac{  (qp)^{n+1} ( k\mu - p)^n}{( p^2 + k^2  - 2 k p \mu )^{n/2} } \;  f(p,q) \ . 
	\end{align}
	After integrating over $\mu$, the remaining $\delta$-function restricts the integration region to
	\begin{align}
		U = \big\{ \; |p-k| < q < p+k \; \big\}
	\end{align}
	which gives
	\begin{align}
	\label{eq:In[f]:interm:1}
		I_n[f] = & \frac{2}{(2\pi)^{11}k} \int_{U} \exd p \; \exd q\; p q \left( \frac{k^2-p^2-q^2}{2} \right)^n \; f(p,q) \, . 
	\end{align}
	The region $U$ is inconvenient to integrate over in these variables, which is why we now introduce~\cite{Burgess:2022nwu}
	\begin{eqnarray} \label{PQ_transf}
	P = p+q \qquad \mathrm{and} \qquad Q = p-q\, .
	\end{eqnarray}
This turns \Eq{eq:In[f]:interm:1} into 
	\begin{align}
		I_n[f] = & \ \frac{1}{4(2\pi)^{11}k} \int_{k}^\infty \exd P \; \int_0^k \exd Q\; (P^2 -Q^2) \left( \frac{2k^2-P^2-Q^2}{4} \right)^n \; f(p,q) \, . 
	\end{align}
Using this identity allows one to rewrite \Eq{KSlin_In} as
\bea
	 \KSnon(k, \eta_1,\eta_2) = & \frac{1}{(2\pi)^{11}} \frac{1}{64k} \int_{k}^\infty \exd P \; \int_0^k \exd Q\; (P^2 -Q^2) \\
	&  \left[  \frac{(P^2+Q^2-2k^2)v_\pi\left(\frac{P+Q}{2}, \eta_1\right)  v_\pi\left(\frac{P-Q}{2}, \eta_1\right)}{a(\eta_1) \Lambda_1} + \frac{4 \Pi_\pi\left(\frac{P+Q}{2}, \eta_1\right) \Pi_\pi\left(\frac{P-Q}{2}, \eta_1\right)}{a(\eta_1) \Lambda_2} \right] \\
	&  \left[  \frac{(P^2+Q^2-2k^2) v^{\ast}_\pi\left(\frac{P+Q}{2}, \eta_2\right)  v^{\ast}_\pi\left(\frac{P-Q}{2}, \eta_2\right)}{a(\eta_2) \Lambda_1} + \frac{4 \Pi^{\ast}_\pi\left(\frac{P+Q}{2}, \eta_2\right) \Pi^{\ast}_\pi\left(\frac{P-Q}{2}, \eta_2\right)}{a(\eta_2) \Lambda_2} \right] .
\eea
With the mode functions~\eqref{eq:modefctvp} and~\eqref{Pi_mode_functions} this reads
\bea
 \label{KSnon_ansApp}
\KSnon(k, \eta_1,\eta_2) =& \frac{H^2 \eta_1 \eta_2}{(2\pi)^{12}} \frac{\pi}{32 k} \int_{k}^\infty \exd P \; \int_0^k \exd Q\; e^{- i \cs P (\eta_1 - \eta_2)}  \\
	& \times \left\{ \frac{P^2+Q^2-2 k^2 }{\cs \Lambda _1}  \left[1-\frac{2 i}{\cs (P+Q)\eta _1}\right] \left[1-\frac{2 i}{\cs (P-Q)\eta _1 }\right] - \frac{\cs (P^2-Q^2)}{\Lambda _2} \right\}  \\
	& \times \left\{ \frac{P^2+Q^2-2 k^2}{\cs \Lambda_1} \left[1+\frac{2 i}{\cs (P-Q) \eta _2}\right] \left[1+\frac{2 i}{\cs (P+Q) \eta _2 }\right] - \frac{\cs (P^2-Q^2)}{\Lambda _2}  \right\} .
\eea
which coincides with \Eq{KSnon_ans} stated the main text.

\section{Integrals $\mathfrak{L}_k$ and $\mathfrak{N}_k$}
\label{app:Integration:details}

In this appendix we flesh out the mathematical details of the integrals computed in \Sec{subsec:ppapp}, using similar methods to those employed in \Refa{Melville:2021lst}.

\subsection*{Linear contribution}
\label{App:Lin_Int}

In \Eq{Slin_Ldef}, $\Slin$ is expressed in terms of a single time integral, and switching to the dimensionless integration variable
\begin{equation}
z' = - k \eta \ ,
\end{equation}
this integral takes the form
\begin{equation}
\mathfrak{L}_k(\eta) = \sqrt{ \frac{\pi}{2} } \; e^{- \tfrac{\pi}{2} \mu_\sigma + \tfrac{i\pi}{4}}  \int_{- k \eta}^\infty \exd z' \; \frac{ e^{i \cs z'} H_{i\mu_\sigma}^{(1)}(z') }{ \sqrt{z' } } \, .
\end{equation}
For the case of a massless environment with $i \mu_\sigma = \frac{3}{2}$ the integrand simplifies and one finds:
\begin{align}
\mathfrak{L}_k(\eta) \big|_{\text{massless}} = & \ \int_{- k \eta}^\infty \exd z' \;  e^{i ( 1 + \cs ) z'} \left( \frac{i}{z^{\prime 2}} + \frac{1}{z'} \right) \\
= & \ \frac{i e^{- i (1+\cs) k \eta}}{- k \eta} + \cs \left\{ \mathrm{Ei}[-i(1+\cs)k\eta] - i \pi \right\} \\
\simeq & \  \begin{cases} \ \frac{i}{-k\eta} + \ldots  \qquad & , \ -k\eta \ll 1 \\
\ e^{- i (1+\cs) k \eta} \left[ \frac{i}{(1+\cs)(-k\eta)} + \ldots \right] \qquad & , \ -k\eta \gg 1 \end{cases} \, ,
\end{align}
where $\mathrm{Ei}(z)$ is the exponential integral function defined for $z \in \mathbb{C} \setminus (-\infty,0]$ (\ie with a branch cut along the negative real axis) as
\begin{equation} \label{EiInt}
\mathrm{Ei}(z) \equiv -\int_{-z}^\infty \exd t \; \frac{e^{-t}}{t} \ ,
\end{equation}
where the principal value of the integral is taken. In the case of a conformal environment with $i \mu_\sigma = \frac{1}{2}$ we have:
\begin{align} \label{Lconformal_app}
\mathfrak{L}_k(\eta) \big|_{\text{conformal}} = & \ \int_{- k \eta}^\infty \exd z' \;  \frac{ e^{i ( 1 + \cs ) z'} }{z'}  \\
= & \ - \text{Ei}\big[ - i (\cs+1) k \eta \big] +  i \pi \\
\simeq & \  \begin{cases} \ - \log \left[ - e^{\gamma_{\mathrm{E}}} (\cs+1) k\eta\right] + \frac{i \pi}{2} + \ldots \qquad & , \ -k\eta \ll 1 \\
\  e^{- i (1+\cs) k \eta} \left[ \frac{i}{(1+\cs)(-k\eta)} + \ldots \right] \qquad & , \ -k\eta \gg 1 \end{cases} \, .
\end{align}

\subsection*{Non-linear contribution}
\label{App:non_int}

In order to determine the non-linear contribution to the entropy, we must first compute the integral $\mathfrak{N}_{k}(P,Q,\eta) $ defined in \Eq{Ndef_integral}, and then integrate over $P$ and $Q$ using \Eq{Snon_Ndef} to find $\Snon$. 

We focus on the case of a massless environment, since the calculations for conformal and heavy environments proceed along similar lines. 
In this case, we use $i \mu_\sigma = \frac{3}{2}$ and switch the integration variable to $z' = - k \eta$ as above for simplicity, so that \Eq{Ndef_integral} becomes
\bea
\mathfrak{N}_{k}(P,Q,\eta) = &\ \frac{1}{k} \int_{-k\eta}^{\infty} \exd z'\; e^{i (k + \cs P) \tfrac{z'}{k}} (i + z') \\
&\ \times \; \left\{ \frac{P^2+Q^2-2 k^2}{\cs \Lambda _1}   \left[1+ \frac{2 i k}{\cs (P+Q) z'}\right] \left[1-\frac{2 i}{\cs (P-Q) z' }\right] - \frac{\cs (P^2-Q^2)}{\Lambda _2} \right\} .
\eea
The terms proportional to $(z')^{-1}$ and $(z')^{-2}$ converge, meanwhile those proportional to $(z')^{0}$ and $z'$ formally diverge. Understood as a distribution however, such that \cite{Burgess:2022nwu}
\begin{eqnarray}
\int_{-k\eta}^\infty \exd z' \; e^{i (k + \cs P) \tfrac{z'}{k}} &  = & e^{ - i (k + \cs P) \eta}  \; \frac{ i k }{k+ \cs P} \ , \\
\int_{-k \eta}^\infty \exd z' \; z' \; e^{i (k + \cs P) \tfrac{z'}{k}} &  = & e^{ - i (k + \cs P) \eta}\left[ -\frac{k^2}{(k+\cs P)^2}+\frac{i k}{k+\cs P} (- k\eta) \right]\ , 
\end{eqnarray}
we find that $\mathfrak{N}_{k}$ evaluates to \Eq{Nans_massless} in the main text. At this stage we use \Eq{Snon_Ndef} to evaluate the entropy, however we make the change to dimensionless variables 
\begin{equation}
a = Q/k \qquad \mathrm{and} \qquad b = P/k 
\end{equation}
so that when used in tandem with (\ref{Nans_massless}) we find that
\begin{eqnarray} 
\Snon(k) =   \frac{g^2 H^2}{(2 \pi)^{18}} \frac{\pi }{16}\left[ \frac{\mathcal{C}_1}{(-k\eta)^2} + \mathcal{C}_2 + \mathcal{C}_3 (-k\eta)^{2}  \right] .
\end{eqnarray}
This is organised in powers of $-k\eta$, with time-independent coefficients given by
\begin{eqnarray} 
\mathcal{C}_1 & = & \int_{1}^\infty \exd b \int_0^1 \exd a \; \frac{16 (a^2+b^2-2)^2}{ (b^2-a^2 )^2 \cs^6 \Lambda _1^2} \\
\mathcal{C}_2 & = & \int_{1}^\infty \exd b \int_0^1 \exd a \; \bigg\{ \frac{(a^2+b^2-2)^2 }{\left(a^2-b^2\right)^2 (b \cs+1)^4 \cs^4 \Lambda _1^2 } \bigg[ b \cs^3 (b^2-a^2 )^2 (b \cs+4)+8 b \cs (a^2+3 b^2)\notag \\
&& +8 (a^2+b^2 )+4 \cs^2 (a^4-2 a^2 b^2+5 b^4)  \bigg] \\
&& + \frac{2 (a^2+b^2-2) \left[4 (b \cs+1)-\cs^2(b^2-a^2) (b \cs+2)^2\right]}{ (b \cs+1)^4 \cs^2 \Lambda _1 \Lambda _2 } + \frac{\cs^2(b^2-a^2)^2 (b \cs+2)^2}{(b \cs+1)^4 \Lambda_2^2 } \bigg\} \notag \\
\mathcal{C}_3 & = & \int_{1}^\infty \exd b \int_0^1 \exd a  \left[ \frac{\left(a^2+b^2-2\right)^2}{(b \cs+1)^2\cs^2 \Lambda _1^2}-\frac{2 \left(b^2-a^2\right) \left(a^2+b^2-2\right)}{ (b \cs+1)^2\Lambda _1 \Lambda _2} +\frac{\cs^2 \left(b^2-a^2\right)^2}{ (b \cs+1)^2\Lambda _2^2} \right] .
\end{eqnarray}
We first perform the $b$ integrals in the above, which contain the UV divergences described in the main text. For example, for $\mathcal{C}_1$, taking a partial fraction expansion in $b$ gives
\begin{equation} 
\mathcal{C}_1 = \frac{16}{\cs^6 \Lambda_1^2} \int_{1}^\infty \exd b \int_0^1 \exd a \left\{1 + \frac{a^4-1}{a^3} \left(\frac{1}{b-a}-\frac{1}{b+a}\right) + \frac{(a^2-1)^2 }{a^2} \left[\frac{1}{(b+a)^2}+\frac{1}{(b-a)^2}\right] \right\} .
\end{equation}
The integrand in the above behaves as $1 + 4(a^2 - 1) b^{-2} + \mathcal{O}(b^{-3})$ for $b \gg 1$ and so only the first term must be regulated. This is done in a way that is analogous to \Eq{main_dimreg}, taking
\begin{equation}  \label{1reg_app}
\int_{1}^\infty \exd b  \to \int_{1}^\infty \exd b  \left( \frac{bk}{\mu} \right)^{\varepsilon} \simeq - 1 +\mathcal{O}(\varepsilon)  \ .
\end{equation}
Notice that the result is independent of the dimensional regulator $\varepsilon$ here, as is always the case when an integral is power-law divergent in a momentum UV cutoff. The rest of the terms converge giving
\begin{equation} 
\mathcal{C}_1 = \frac{16}{\cs^6 \Lambda_1^2} \int_0^1 \exd a \; \bigg[ - 3 + \frac{2}{a^2}+\frac{2 }{a^3} \left(a^4-1\right) \tanh ^{-1}(a) \bigg] = - \frac{48}{\cs^6 \Lambda_1^2}
\end{equation}
as quoted in \Eq{C1_text} of the main text. The computations for \Eqs{C2_text} and~\eqref{C3_text} are more cumbersome but follow through similarly, using the regularizations
\begin{eqnarray}  
\int_{1}^\infty \exd b \; b^{n} & \to & \int_{1}^\infty \exd b\; b^{n} \left( \frac{bk}{\mu} \right)^{\varepsilon} \simeq - \frac{1}{n+1} +\mathcal{O}(\varepsilon)  \qquad \mathrm{for\ } n=1,2 \\
\mathrm{and} \quad \int_{1}^\infty  \frac{\exd b}{b \pm a } & \to & \int_{1}^\infty \frac{\exd b}{b \pm a } \left( \frac{bk}{\mu} \right)^{\varepsilon} \simeq - \frac{1}{\varepsilon} + \log\left[ \frac{\mu}{(1\pm a)k} \right] +\mathcal{O}(\varepsilon) 
\end{eqnarray}
in addition to \Eqs{main_dimreg} and (\ref{1reg_app}). 

\bibliographystyle{JHEP}
\bibliography{biblio}

\providecommand{\href}[2]{#2}\begingroup\raggedright\begin{thebibliography}{100}

\bibitem{Koks:1996ga}
D.~Koks, A.~Matacz and B.L.~Hu, \emph{{Entropy and uncertainty of squeezed
  quantum open systems}},
  \href{https://doi.org/10.1103/PhysRevD.55.5917}{\emph{Phys. Rev. D}
  {\bfseries 55} (1997) 5917}
  [\href{https://arxiv.org/abs/quant-ph/9612016}{{\ttfamily
  quant-ph/9612016}}].

\bibitem{Rosenhaus:2014woa}
V.~Rosenhaus and M.~Smolkin, \emph{{Entanglement Entropy: A Perturbative
  Calculation}}, \href{https://doi.org/10.1007/JHEP12(2014)179}{\emph{JHEP}
  {\bfseries 12} (2014) 179} [\href{https://arxiv.org/abs/1403.3733}{{\ttfamily
  1403.3733}}].

\bibitem{Boyanovsky:2018fxl}
D.~Boyanovsky, \emph{{Information loss in effective field theory: entanglement
  and thermal entropies}},
  \href{https://doi.org/10.1103/PhysRevD.97.065008}{\emph{Phys. Rev. D}
  {\bfseries 97} (2018) 065008}
  [\href{https://arxiv.org/abs/1801.06840}{{\ttfamily 1801.06840}}].

\bibitem{Cheung:2023hkq}
C.~Cheung, T.~He and A.~Sivaramakrishnan, \emph{{Entropy growth in perturbative
  scattering}}, \href{https://doi.org/10.1103/PhysRevD.108.045013}{\emph{Phys.
  Rev. D} {\bfseries 108} (2023) 045013}
  [\href{https://arxiv.org/abs/2304.13052}{{\ttfamily 2304.13052}}].

\bibitem{Subba:2024mnl}
A.~Subba and R.~Rahaman, \emph{{On bipartite and tripartite entanglement at
  present and future particle colliders}},
  \href{https://arxiv.org/abs/2404.03292}{{\ttfamily 2404.03292}}.

\bibitem{Polarski:1995jg}
D.~Polarski and A.A.~Starobinsky, \emph{{Semiclassicality and decoherence of
  cosmological perturbations}},
  \href{https://doi.org/10.1088/0264-9381/13/3/006}{\emph{Class. Quant. Grav.}
  {\bfseries 13} (1996) 377}
  [\href{https://arxiv.org/abs/gr-qc/9504030}{{\ttfamily gr-qc/9504030}}].

\bibitem{Kiefer:1998qe}
C.~Kiefer, D.~Polarski and A.A.~Starobinsky, \emph{{Quantum to classical
  transition for fluctuations in the early universe}},
  \href{https://doi.org/10.1142/S0218271898000292}{\emph{Int. J. Mod. Phys. D}
  {\bfseries 7} (1998) 455}
  [\href{https://arxiv.org/abs/gr-qc/9802003}{{\ttfamily gr-qc/9802003}}].

\bibitem{Kiefer:2006je}
C.~Kiefer, I.~Lohmar, D.~Polarski and A.A.~Starobinsky, \emph{{Pointer states
  for primordial fluctuations in inflationary cosmology}},
  \href{https://doi.org/10.1088/0264-9381/24/7/002}{\emph{Class. Quant. Grav.}
  {\bfseries 24} (2007) 1699}
  [\href{https://arxiv.org/abs/astro-ph/0610700}{{\ttfamily
  astro-ph/0610700}}].

\bibitem{Kiefer:2008ku}
C.~Kiefer and D.~Polarski, \emph{{Why do cosmological perturbations look
  classical to us?}}, \href{https://doi.org/10.1166/asl.2009.1023}{\emph{Adv.
  Sci. Lett.} {\bfseries 2} (2009) 164}
  [\href{https://arxiv.org/abs/0810.0087}{{\ttfamily 0810.0087}}].

\bibitem{Sudarsky:2009za}
D.~Sudarsky, \emph{{Shortcomings in the Understanding of Why Cosmological
  Perturbations Look Classical}},
  \href{https://doi.org/10.1142/S0218271811018937}{\emph{Int. J. Mod. Phys. D}
  {\bfseries 20} (2011) 509} [\href{https://arxiv.org/abs/0906.0315}{{\ttfamily
  0906.0315}}].

\bibitem{Burgess:2014eoa}
C.P.~Burgess, R.~Holman, G.~Tasinato and M.~Williams, \emph{{EFT Beyond the
  Horizon: Stochastic Inflation and How Primordial Quantum Fluctuations Go
  Classical}}, \href{https://doi.org/10.1007/JHEP03(2015)090}{\emph{JHEP}
  {\bfseries 03} (2015) 090} [\href{https://arxiv.org/abs/1408.5002}{{\ttfamily
  1408.5002}}].

\bibitem{Martin:2015qta}
J.~Martin and V.~Vennin, \emph{{Quantum Discord of Cosmic Inflation: Can we
  Show that CMB Anisotropies are of Quantum-Mechanical Origin?}},
  \href{https://doi.org/10.1103/PhysRevD.93.023505}{\emph{Phys. Rev. D}
  {\bfseries 93} (2016) 023505}
  [\href{https://arxiv.org/abs/1510.04038}{{\ttfamily 1510.04038}}].

\bibitem{Martin:2021znx}
J.~Martin, A.~Micheli and V.~Vennin, \emph{{Discord and decoherence}},
  \href{https://doi.org/10.1088/1475-7516/2022/04/051}{\emph{JCAP} {\bfseries
  04} (2022) 051} [\href{https://arxiv.org/abs/2112.05037}{{\ttfamily
  2112.05037}}].

\bibitem{Chandran:2023ogt}
S.M.~Chandran, K.~Rajeev and S.~Shankaranarayanan, \emph{{Real-space
  quantum-to-classical transition of time dependent background fluctuations}},
  \href{https://doi.org/10.1103/PhysRevD.109.023503}{\emph{Phys. Rev. D}
  {\bfseries 109} (2024) 023503}
  [\href{https://arxiv.org/abs/2307.13611}{{\ttfamily 2307.13611}}].

\bibitem{Campo:2005sv}
D.~Campo and R.~Parentani, \emph{{Inflationary spectra and violations of Bell
  inequalities}}, \href{https://doi.org/10.1103/PhysRevD.74.025001}{\emph{Phys.
  Rev. D} {\bfseries 74} (2006) 025001}
  [\href{https://arxiv.org/abs/astro-ph/0505376}{{\ttfamily
  astro-ph/0505376}}].

\bibitem{Maldacena:2015bha}
J.~Maldacena, \emph{{A model with cosmological Bell inequalities}},
  \href{https://doi.org/10.1002/prop.201500097}{\emph{Fortsch. Phys.}
  {\bfseries 64} (2016) 10} [\href{https://arxiv.org/abs/1508.01082}{{\ttfamily
  1508.01082}}].

\bibitem{Martin:2016tbd}
J.~Martin and V.~Vennin, \emph{{Bell inequalities for continuous-variable
  systems in generic squeezed states}},
  \href{https://doi.org/10.1103/PhysRevA.93.062117}{\emph{Phys. Rev. A}
  {\bfseries 93} (2016) 062117}
  [\href{https://arxiv.org/abs/1605.02944}{{\ttfamily 1605.02944}}].

\bibitem{Choudhury:2016cso}
S.~Choudhury, S.~Panda and R.~Singh, \emph{{Bell violation in the Sky}},
  \href{https://doi.org/10.1140/epjc/s10052-016-4553-3}{\emph{Eur. Phys. J. C}
  {\bfseries 77} (2017) 60} [\href{https://arxiv.org/abs/1607.00237}{{\ttfamily
  1607.00237}}].

\bibitem{Choudhury:2016pfr}
S.~Choudhury, S.~Panda and R.~Singh, \emph{{Bell violation in primordial
  cosmology}}, \href{https://doi.org/10.3390/universe3010013}{\emph{Universe}
  {\bfseries 3} (2017) 13} [\href{https://arxiv.org/abs/1612.09445}{{\ttfamily
  1612.09445}}].

\bibitem{Martin:2017zxs}
J.~Martin and V.~Vennin, \emph{{Obstructions to Bell CMB Experiments}},
  \href{https://doi.org/10.1103/PhysRevD.96.063501}{\emph{Phys. Rev. D}
  {\bfseries 96} (2017) 063501}
  [\href{https://arxiv.org/abs/1706.05001}{{\ttfamily 1706.05001}}].

\bibitem{Ando:2020kdz}
K.~Ando and V.~Vennin, \emph{{Bipartite temporal Bell inequalities for two-mode
  squeezed states}},
  \href{https://doi.org/10.1103/PhysRevA.102.052213}{\emph{Phys. Rev. A}
  {\bfseries 102} (2020) 052213}
  [\href{https://arxiv.org/abs/2007.00458}{{\ttfamily 2007.00458}}].

\bibitem{Espinosa-Portales:2022yok}
L.~Espinosa-Portal\'es and V.~Vennin, \emph{{Real-space Bell inequalities in
  de~Sitter}}, \href{https://doi.org/10.1088/1475-7516/2022/07/037}{\emph{JCAP}
  {\bfseries 07} (2022) 037}
  [\href{https://arxiv.org/abs/2203.03505}{{\ttfamily 2203.03505}}].

\bibitem{Tejerina-Perez:2024opu}
P.~Tejerina-P\'erez, D.~Bertacca and R.~Jimenez, \emph{{An Entangled
  Universe}},  \href{https://arxiv.org/abs/2403.15742}{{\ttfamily 2403.15742}}.

\bibitem{Sou:2024tjv}
C.M.~Sou, J.~Wang and Y.~Wang, \emph{{Cosmological Bell Tests with Decoherence
  Effects}},  \href{https://arxiv.org/abs/2405.07141}{{\ttfamily 2405.07141}}.

\bibitem{Fukuma:2013uxa}
M.~Fukuma, Y.~Sakatani and S.~Sugishita, \emph{{Master equation for the
  Unruh-DeWitt detector and the universal relaxation time in de Sitter space}},
  \href{https://doi.org/10.1103/PhysRevD.89.064024}{\emph{Phys. Rev. D}
  {\bfseries 89} (2014) 064024}
  [\href{https://arxiv.org/abs/1305.0256}{{\ttfamily 1305.0256}}].

\bibitem{Choudhury:2017qyl}
S.~Choudhury and S.~Panda, \emph{{Quantum entanglement in de Sitter space from
  stringy axion: An analysis using $\alpha$ vacua}},
  \href{https://doi.org/10.1016/j.nuclphysb.2019.03.018}{\emph{Nucl. Phys. B}
  {\bfseries 943} (2019) 114606}
  [\href{https://arxiv.org/abs/1712.08299}{{\ttfamily 1712.08299}}].

\bibitem{Boyanovsky:2018soy}
D.~Boyanovsky, \emph{{Imprint of entanglement entropy in the power spectrum of
  inflationary fluctuations}},
  \href{https://doi.org/10.1103/PhysRevD.98.023515}{\emph{Phys. Rev. D}
  {\bfseries 98} (2018) 023515}
  [\href{https://arxiv.org/abs/1804.07967}{{\ttfamily 1804.07967}}].

\bibitem{Akhtar:2019qdn}
S.~Akhtar, S.~Choudhury, S.~Chowdhury, D.~Goswami, S.~Panda and A.~Swain,
  \emph{{Open Quantum Entanglement: A study of two atomic system in static
  patch of de Sitter space}},
  \href{https://doi.org/10.1140/epjc/s10052-020-8302-2}{\emph{Eur. Phys. J. C}
  {\bfseries 80} (2020) 748}
  [\href{https://arxiv.org/abs/1908.09929}{{\ttfamily 1908.09929}}].

\bibitem{Kaplanek:2020iay}
G.~Kaplanek and C.P.~Burgess, \emph{{Qubits on the Horizon: Decoherence and
  Thermalization near Black Holes}},
  \href{https://doi.org/10.1007/JHEP01(2021)098}{\emph{JHEP} {\bfseries 01}
  (2021) 098} [\href{https://arxiv.org/abs/2007.05984}{{\ttfamily
  2007.05984}}].

\bibitem{Martin:2021xml}
J.~Martin and V.~Vennin, \emph{{Real-space entanglement of quantum fields}},
  \href{https://doi.org/10.1103/PhysRevD.104.085012}{\emph{Phys. Rev. D}
  {\bfseries 104} (2021) 085012}
  [\href{https://arxiv.org/abs/2106.14575}{{\ttfamily 2106.14575}}].

\bibitem{Martin:2021qkg}
J.~Martin and V.~Vennin, \emph{{Real-space entanglement in the Cosmic Microwave
  Background}},
  \href{https://doi.org/10.1088/1475-7516/2021/10/036}{\emph{JCAP} {\bfseries
  10} (2021) 036} [\href{https://arxiv.org/abs/2106.15100}{{\ttfamily
  2106.15100}}].

\bibitem{Kaplanek:2021fnl}
G.~Kaplanek, C.P.~Burgess and R.~Holman, \emph{{Qubit heating near a hotspot}},
  \href{https://doi.org/10.1007/JHEP08(2021)132}{\emph{JHEP} {\bfseries 08}
  (2021) 132} [\href{https://arxiv.org/abs/2106.10803}{{\ttfamily
  2106.10803}}].

\bibitem{Brahma:2021mng}
S.~Brahma, A.~Berera and J.~Calder\'on-Figueroa, \emph{{Universal signature of
  quantum entanglement across cosmological distances}},
  \href{https://doi.org/10.1088/1361-6382/aca066}{\emph{Class. Quant. Grav.}
  {\bfseries 39} (2022) 245002}
  [\href{https://arxiv.org/abs/2107.06910}{{\ttfamily 2107.06910}}].

\bibitem{Brahma:2023hki}
S.~Brahma, J.~Calder\'on-Figueroa, M.~Hassan and X.~Mi, \emph{{Momentum-space
  entanglement entropy in de Sitter spacetime}},
  \href{https://doi.org/10.1103/PhysRevD.108.043522}{\emph{Phys. Rev. D}
  {\bfseries 108} (2023) 043522}
  [\href{https://arxiv.org/abs/2302.13894}{{\ttfamily 2302.13894}}].

\bibitem{Belfiglio:2023moe}
A.~Belfiglio, O.~Luongo and S.~Mancini, \emph{{Superhorizon entanglement from
  inflationary particle production}},
  \href{https://arxiv.org/abs/2312.11419}{{\ttfamily 2312.11419}}.

\bibitem{Brandenberger:1990bx}
R.H.~Brandenberger, R.~Laflamme and M.~Mijic, \emph{{Classical Perturbations
  From Decoherence of Quantum Fluctuations in the Inflationary Universe}},
  \href{https://doi.org/10.1142/S0217732390002651}{\emph{Mod. Phys. Lett. A}
  {\bfseries 5} (1990) 2311}.

\bibitem{Barvinsky:1998cq}
A.O.~Barvinsky, A.Y.~Kamenshchik, C.~Kiefer and I.V.~Mishakov,
  \emph{Decoherence in quantum cosmology at the onset of inflation},
  \href{https://doi.org/10.1016/S0550-3213(99)00208-4}{\emph{Nucl. Phys. B}
  {\bfseries 551} (1999) 374}.

\bibitem{Lombardo:2004fr}
F.C.~Lombardo, \emph{{Influence functional approach to decoherence during
  inflation}},
  \href{https://doi.org/10.1590/S0103-97332005000300005}{\emph{Braz. J. Phys.}
  {\bfseries 35} (2005) 391}
  [\href{https://arxiv.org/abs/gr-qc/0412069}{{\ttfamily gr-qc/0412069}}].

\bibitem{Lombardo:2005iz}
F.C.~Lombardo and D.~Lopez~Nacir, \emph{{Decoherence during inflation: The
  Generation of classical inhomogeneities}},
  \href{https://doi.org/10.1103/PhysRevD.72.063506}{\emph{Phys. Rev. D}
  {\bfseries 72} (2005) 063506}
  [\href{https://arxiv.org/abs/gr-qc/0506051}{{\ttfamily gr-qc/0506051}}].

\bibitem{Martineau:2006ki}
P.~Martineau, \emph{{On the decoherence of primordial fluctuations during
  inflation}}, \href{https://doi.org/10.1088/0264-9381/24/23/006}{\emph{Class.
  Quant. Grav.} {\bfseries 24} (2007) 5817}
  [\href{https://arxiv.org/abs/astro-ph/0601134}{{\ttfamily
  astro-ph/0601134}}].

\bibitem{Prokopec:2006fc}
T.~Prokopec and G.I.~Rigopoulos, \emph{{Decoherence from Isocurvature
  perturbations in Inflation}},
  \href{https://doi.org/10.1088/1475-7516/2007/11/029}{\emph{JCAP} {\bfseries
  11} (2007) 029} [\href{https://arxiv.org/abs/astro-ph/0612067}{{\ttfamily
  astro-ph/0612067}}].

\bibitem{Burgess:2006jn}
C.P.~Burgess, R.~Holman and D.~Hoover, \emph{{Decoherence of inflationary
  primordial fluctuations}},
  \href{https://doi.org/10.1103/PhysRevD.77.063534}{\emph{Phys. Rev. D}
  {\bfseries 77} (2008) 063534}
  [\href{https://arxiv.org/abs/astro-ph/0601646}{{\ttfamily
  astro-ph/0601646}}].

\bibitem{Sharman:2007gi}
J.W.~Sharman and G.D.~Moore, \emph{{Decoherence due to the Horizon after
  Inflation}}, \href{https://doi.org/10.1088/1475-7516/2007/11/020}{\emph{JCAP}
  {\bfseries 11} (2007) 020} [\href{https://arxiv.org/abs/0708.3353}{{\ttfamily
  0708.3353}}].

\bibitem{Campo:2008ju}
D.~Campo and R.~Parentani, \emph{{Decoherence and entropy of primordial
  fluctuations. I: Formalism and interpretation}},
  \href{https://doi.org/10.1103/PhysRevD.78.065044}{\emph{Phys. Rev. D}
  {\bfseries 78} (2008) 065044}
  [\href{https://arxiv.org/abs/0805.0548}{{\ttfamily 0805.0548}}].

\bibitem{Anastopoulos:2013zya}
C.~Anastopoulos and B.L.~Hu, \emph{{A Master Equation for Gravitational
  Decoherence: Probing the Textures of Spacetime}},
  \href{https://doi.org/10.1088/0264-9381/30/16/165007}{\emph{Class. Quant.
  Grav.} {\bfseries 30} (2013) 165007}
  [\href{https://arxiv.org/abs/1305.5231}{{\ttfamily 1305.5231}}].

\bibitem{Nelson:2016kjm}
E.~Nelson, \emph{{Quantum Decoherence During Inflation from Gravitational
  Nonlinearities}},
  \href{https://doi.org/10.1088/1475-7516/2016/03/022}{\emph{JCAP} {\bfseries
  03} (2016) 022} [\href{https://arxiv.org/abs/1601.03734}{{\ttfamily
  1601.03734}}].

\bibitem{Martin:2018zbe}
J.~Martin and V.~Vennin, \emph{{Observational constraints on quantum
  decoherence during inflation}},
  \href{https://doi.org/10.1088/1475-7516/2018/05/063}{\emph{JCAP} {\bfseries
  05} (2018) 063} [\href{https://arxiv.org/abs/1801.09949}{{\ttfamily
  1801.09949}}].

\bibitem{Martin:2018lin}
J.~Martin and V.~Vennin, \emph{{Non Gaussianities from Quantum Decoherence
  during Inflation}},
  \href{https://doi.org/10.1088/1475-7516/2018/06/037}{\emph{JCAP} {\bfseries
  06} (2018) 037} [\href{https://arxiv.org/abs/1805.05609}{{\ttfamily
  1805.05609}}].

\bibitem{Danielson:2022tdw}
D.L.~Danielson, G.~Satishchandran and R.M.~Wald, \emph{{Black holes decohere
  quantum superpositions}},
  \href{https://doi.org/10.1142/S0218271822410036}{\emph{Int. J. Mod. Phys. D}
  {\bfseries 31} (2022) 2241003}
  [\href{https://arxiv.org/abs/2205.06279}{{\ttfamily 2205.06279}}].

\bibitem{Danielson:2022sga}
D.L.~Danielson, G.~Satishchandran and R.M.~Wald, \emph{{Killing horizons
  decohere quantum superpositions}},
  \href{https://doi.org/10.1103/PhysRevD.108.025007}{\emph{Phys. Rev. D}
  {\bfseries 108} (2023) 025007}
  [\href{https://arxiv.org/abs/2301.00026}{{\ttfamily 2301.00026}}].

\bibitem{Oppenheim:2022xjr}
J.~Oppenheim, C.~Sparaciari, B.~\v{S}oda and Z.~Weller-Davies,
  \emph{{Gravitationally induced decoherence vs space-time diffusion: testing
  the quantum nature of gravity}},
  \href{https://doi.org/10.1038/s41467-023-43348-2}{\emph{Nature Commun.}
  {\bfseries 14} (2023) 7910}
  [\href{https://arxiv.org/abs/2203.01982}{{\ttfamily 2203.01982}}].

\bibitem{Colas:2022kfu}
T.~Colas, J.~Grain and V.~Vennin, \emph{{Quantum recoherence in the early
  universe}}, \href{https://doi.org/10.1209/0295-5075/acdd94}{\emph{EPL}
  {\bfseries 142} (2023) 69002}
  [\href{https://arxiv.org/abs/2212.09486}{{\ttfamily 2212.09486}}].

\bibitem{DaddiHammou:2022itk}
A.~Daddi~Hammou and N.~Bartolo, \emph{{Cosmic decoherence: primordial power
  spectra and non-Gaussianities}},
  \href{https://doi.org/10.1088/1475-7516/2023/04/055}{\emph{JCAP} {\bfseries
  04} (2023) 055} [\href{https://arxiv.org/abs/2211.07598}{{\ttfamily
  2211.07598}}].

\bibitem{Burgess:2022nwu}
C.P.~Burgess, R.~Holman, G.~Kaplanek, J.~Martin and V.~Vennin, \emph{{Minimal
  decoherence from inflation}},
  \href{https://doi.org/10.1088/1475-7516/2023/07/022}{\emph{JCAP} {\bfseries
  07} (2023) 022} [\href{https://arxiv.org/abs/2211.11046}{{\ttfamily
  2211.11046}}].

\bibitem{Sharifian:2023jem}
M.~Sharifian, M.~Zarei, M.~Abdi, N.~Bartolo and S.~Matarrese, \emph{{Open
  quantum system approach to the gravitational decoherence of spin-1/2
  particles}},  \href{https://arxiv.org/abs/2309.07236}{{\ttfamily
  2309.07236}}.

\bibitem{Ning:2023ybc}
S.~Ning, C.M.~Sou and Y.~Wang, \emph{{On the decoherence of primordial
  gravitons}}, \href{https://doi.org/10.1007/JHEP06(2023)101}{\emph{JHEP}
  {\bfseries 06} (2023) 101}
  [\href{https://arxiv.org/abs/2305.08071}{{\ttfamily 2305.08071}}].

\bibitem{Biggs:2024dgp}
A.~Biggs and J.~Maldacena, \emph{{Comparing the decoherence effects due to
  black holes versus ordinary matter}},
  \href{https://arxiv.org/abs/2405.02227}{{\ttfamily 2405.02227}}.

\bibitem{Danielson:2024yru}
D.L.~Danielson, G.~Satishchandran and R.M.~Wald, \emph{{Local Description of
  Decoherence of Quantum Superpositions by Black Holes and Other Bodies}},
  \href{https://arxiv.org/abs/2407.02567}{{\ttfamily 2407.02567}}.

\bibitem{Zurek:1981xq}
W.H.~Zurek, \emph{Pointer {Basis} of {Quantum} {Apparatus}: {Into} {What}
  {Mixture} {Does} the {Wave} {Packet} {Collapse}?},
  \href{https://doi.org/10.1103/PhysRevD.24.1516}{\emph{Phys. Rev. D}
  {\bfseries 24} (1981) 1516}.

\bibitem{Zurek:1982ii}
W.H.~Zurek, \emph{Environment induced superselection rules},
  \href{https://doi.org/10.1103/PhysRevD.26.1862}{\emph{Phys. Rev. D}
  {\bfseries 26} (1982) 1862}.

\bibitem{Joos:1984uk}
E.~Joos and H.~Zeh, \emph{The {Emergence} of classical properties through
  interaction with the environment},
  \href{https://doi.org/10.1007/BF01725541}{\emph{Z. Phys. B} {\bfseries 59}
  (1985) 223}.

\bibitem{breuerTheoryOpenQuantum2002}
H.P.~Breuer and F.~Petruccione, \emph{The theory of Open Quantum Systems},
  Oxford University Press (2002),
  \href{https://doi.org/10.1093/acprof:oso/9780199213900.001.0001}{10.1093/acprof:oso/9780199213900.001.0001}.

\bibitem{Burgess:2007pt}
C.P.~Burgess, \emph{Introduction to {Effective} {Field} {Theory}},
  \href{https://doi.org/10.1146/annurev.nucl.56.080805.140508}{\emph{Ann. Rev.
  Nucl. Part. Sci.} {\bfseries 57} (2007) 329}.

\bibitem{Burgess:2022rdo}
C.P.~Burgess and G.~Kaplanek, \emph{{Gravity, Horizons and Open EFTs}},
  \href{https://arxiv.org/abs/2212.09157}{{\ttfamily 2212.09157}}.

\bibitem{Colas:2023wxa}
T.~Colas, \emph{{Open Effective Field Theories for primordial cosmology :
  dissipation, decoherence and late-time resummation of cosmological
  inhomogeneities}}, Ph.D. thesis, Institut d'astrophysique spatiale, France,
  AstroParticule et Cosmologie, France, APC, Paris, 2023.

\bibitem{Colas:2024lse}
T.~Colas, \emph{{Open Effective Field Theories for cosmology}},  in \emph{{58th
  Rencontres de Moriond on Cosmology}}, 5, 2024
  [\href{https://arxiv.org/abs/2405.09639}{{\ttfamily 2405.09639}}].

\bibitem{Boyanovsky:2015xoa}
D.~Boyanovsky, \emph{{Effective Field Theory out of Equilibrium: Brownian
  quantum fields}},
  \href{https://doi.org/10.1088/1367-2630/17/6/063017}{\emph{New J. Phys.}
  {\bfseries 17} (2015) 063017}
  [\href{https://arxiv.org/abs/1503.00156}{{\ttfamily 1503.00156}}].

\bibitem{Boyanovsky:2015jen}
D.~Boyanovsky, \emph{{Effective field theory during inflation. II. Stochastic
  dynamics and power spectrum suppression}},
  \href{https://doi.org/10.1103/PhysRevD.93.043501}{\emph{Phys. Rev. D}
  {\bfseries 93} (2016) 043501}
  [\href{https://arxiv.org/abs/1511.06649}{{\ttfamily 1511.06649}}].

\bibitem{Boyanovsky:2015tba}
D.~Boyanovsky, \emph{{Effective field theory during inflation: Reduced density
  matrix and its quantum master equation}},
  \href{https://doi.org/10.1103/PhysRevD.92.023527}{\emph{Phys. Rev. D}
  {\bfseries 92} (2015) 023527}
  [\href{https://arxiv.org/abs/1506.07395}{{\ttfamily 1506.07395}}].

\bibitem{Hollowood:2017bil}
T.J.~Hollowood and J.I.~McDonald, \emph{{Decoherence, discord and the quantum
  master equation for cosmological perturbations}},
  \href{https://doi.org/10.1103/PhysRevD.95.103521}{\emph{Phys. Rev. D}
  {\bfseries 95} (2017) 103521}
  [\href{https://arxiv.org/abs/1701.02235}{{\ttfamily 1701.02235}}].

\bibitem{Shandera:2017qkg}
S.~Shandera, N.~Agarwal and A.~Kamal, \emph{{Open quantum cosmological
  system}}, \href{https://doi.org/10.1103/PhysRevD.98.083535}{\emph{Phys. Rev.
  D} {\bfseries 98} (2018) 083535}
  [\href{https://arxiv.org/abs/1708.00493}{{\ttfamily 1708.00493}}].

\bibitem{Choudhury:2018ppd}
S.~Choudhury and S.~Panda, \emph{{Cosmological Spectrum of Two-Point
  Correlation Function from Vacuum Fluctuation of Stringy Axion Field in De
  Sitter Space: A Study of the Role of Quantum Entanglement}},
  \href{https://doi.org/10.3390/universe6060079}{\emph{Universe} {\bfseries 6}
  (2020) 79} [\href{https://arxiv.org/abs/1809.02905}{{\ttfamily 1809.02905}}].

\bibitem{Burrage:2018pyg}
C.~Burrage, C.~K\"ading, P.~Millington and J.~Min\'a\v{r}, \emph{{Open quantum
  dynamics induced by light scalar fields}},
  \href{https://doi.org/10.1103/PhysRevD.100.076003}{\emph{Phys. Rev. D}
  {\bfseries 100} (2019) 076003}
  [\href{https://arxiv.org/abs/1812.08760}{{\ttfamily 1812.08760}}].

\bibitem{Cheung:2018cwt}
C.~Cheung, J.~Liu and G.N.~Remmen, \emph{{Proof of the Weak Gravity Conjecture
  from Black Hole Entropy}},
  \href{https://doi.org/10.1007/JHEP10(2018)004}{\emph{JHEP} {\bfseries 10}
  (2018) 004} [\href{https://arxiv.org/abs/1801.08546}{{\ttfamily
  1801.08546}}].

\bibitem{Brahma:2020zpk}
S.~Brahma, O.~Alaryani and R.~Brandenberger, \emph{{Entanglement entropy of
  cosmological perturbations}},
  \href{https://doi.org/10.1103/PhysRevD.102.043529}{\emph{Phys. Rev. D}
  {\bfseries 102} (2020) 043529}
  [\href{https://arxiv.org/abs/2005.09688}{{\ttfamily 2005.09688}}].

\bibitem{Rai:2020edx}
M.~Rai and D.~Boyanovsky, \emph{{Origin of entropy of gravitationally produced
  dark matter: The entanglement entropy}},
  \href{https://doi.org/10.1103/PhysRevD.102.063532}{\emph{Phys. Rev. D}
  {\bfseries 102} (2020) 063532}
  [\href{https://arxiv.org/abs/2007.09196}{{\ttfamily 2007.09196}}].

\bibitem{Burgess:2021luo}
C.P.~Burgess, R.~Holman and G.~Kaplanek, \emph{{Quantum Hotspots: Mean Fields,
  Open EFTs, Nonlocality and Decoherence Near Black Holes}},
  \href{https://doi.org/10.1002/prop.202200019}{\emph{Fortsch. Phys.}
  {\bfseries 70} (2022) 2200019}
  [\href{https://arxiv.org/abs/2106.10804}{{\ttfamily 2106.10804}}].

\bibitem{Banerjee:2021lqu}
S.~Banerjee, S.~Choudhury, S.~Chowdhury, J.~Knaute, S.~Panda and K.~Shirish,
  \emph{{Thermalization in quenched open quantum cosmology}},
  \href{https://doi.org/10.1016/j.nuclphysb.2023.116368}{\emph{Nucl. Phys. B}
  {\bfseries 996} (2023) 116368}
  [\href{https://arxiv.org/abs/2104.10692}{{\ttfamily 2104.10692}}].

\bibitem{Brahma:2022yxu}
S.~Brahma, A.~Berera and J.~Calder\'on-Figueroa, \emph{{Quantum corrections to
  the primordial tensor spectrum: open EFTs \& Markovian decoupling of UV
  modes}}, \href{https://doi.org/10.1007/JHEP08(2022)225}{\emph{JHEP}
  {\bfseries 08} (2022) 225}
  [\href{https://arxiv.org/abs/2206.05797}{{\ttfamily 2206.05797}}].

\bibitem{Kaplanek:2022xrr}
G.~Kaplanek and E.~Tjoa, \emph{{Effective master equations~for two accelerated
  qubits}}, \href{https://doi.org/10.1103/PhysRevA.107.012208}{\emph{Phys. Rev.
  A} {\bfseries 107} (2023) 012208}
  [\href{https://arxiv.org/abs/2207.13750}{{\ttfamily 2207.13750}}].

\bibitem{Kaplanek:2022opa}
G.~Kaplanek, \emph{{Some Applications of Open Effective Field Theories to
  Gravitating Quantum Systems}}, Ph.D. thesis, McMaster U., 2022.

\bibitem{Cao:2022kjn}
S.~Cao and D.~Boyanovsky, \emph{{Nonequilibrium dynamics of axionlike
  particles: The quantum master equation}},
  \href{https://doi.org/10.1103/PhysRevD.107.063518}{\emph{Phys. Rev. D}
  {\bfseries 107} (2023) 063518}
  [\href{https://arxiv.org/abs/2212.05161}{{\ttfamily 2212.05161}}].

\bibitem{Prudhoe:2022pte}
S.~Prudhoe and S.~Shandera, \emph{{Classifying the non-time-local and
  entangling dynamics of an open qubit system}},
  \href{https://doi.org/10.1007/JHEP02(2023)007}{\emph{JHEP} {\bfseries 02}
  (2023) 007} [\href{https://arxiv.org/abs/2201.07080}{{\ttfamily
  2201.07080}}].

\bibitem{Kading:2022jjl}
C.~K\"ading and M.~Pitschmann, \emph{{New method for directly computing reduced
  density matrices}},
  \href{https://doi.org/10.1103/PhysRevD.107.016005}{\emph{Phys. Rev. D}
  {\bfseries 107} (2023) 016005}
  [\href{https://arxiv.org/abs/2204.08829}{{\ttfamily 2204.08829}}].

\bibitem{Kading:2022hhc}
C.~K\"ading and M.~Pitschmann, \emph{{Density Matrix Formalism for Interacting
  Quantum Fields}},
  \href{https://doi.org/10.3390/universe8110601}{\emph{Universe} {\bfseries 8}
  (2022) 601} [\href{https://arxiv.org/abs/2210.06991}{{\ttfamily
  2210.06991}}].

\bibitem{Alicki:2023tfz}
R.~Alicki, G.~Barenboim and A.~Jenkins, \emph{{The irreversible relaxation of
  inflation}},  \href{https://arxiv.org/abs/2307.04803}{{\ttfamily
  2307.04803}}.

\bibitem{Alicki:2023rfv}
R.~Alicki, G.~Barenboim and A.~Jenkins, \emph{{Quantum thermodynamics of de
  Sitter space}},  \href{https://arxiv.org/abs/2307.04800}{{\ttfamily
  2307.04800}}.

\bibitem{Kading:2023mdk}
C.~K\"ading, M.~Pitschmann and C.~Voith, \emph{{Dilaton-induced open quantum
  dynamics}}, \href{https://doi.org/10.1140/epjc/s10052-023-11939-4}{\emph{Eur.
  Phys. J. C} {\bfseries 83} (2023) 767}
  [\href{https://arxiv.org/abs/2306.10896}{{\ttfamily 2306.10896}}].

\bibitem{Creminelli:2023aly}
P.~Creminelli, S.~Kumar, B.~Salehian and L.~Santoni, \emph{{Dissipative
  inflation via scalar production}},
  \href{https://doi.org/10.1088/1475-7516/2023/08/076}{\emph{JCAP} {\bfseries
  08} (2023) 076} [\href{https://arxiv.org/abs/2305.07695}{{\ttfamily
  2305.07695}}].

\bibitem{Pelliconi:2023ojb}
P.~Pelliconi and J.~Sonner, \emph{{The Influence Functional in open holography:
  entanglement and R\'enyi entropies}},
  \href{https://arxiv.org/abs/2310.13047}{{\ttfamily 2310.13047}}.

\bibitem{Colas:2024xjy}
T.~Colas, C.~de~Rham and G.~Kaplanek, \emph{{Decoherence out of fire: purity
  loss in expanding and contracting universes}},
  \href{https://doi.org/10.1088/1475-7516/2024/05/025}{\emph{JCAP} {\bfseries
  05} (2024) 025} [\href{https://arxiv.org/abs/2401.02832}{{\ttfamily
  2401.02832}}].

\bibitem{Keefe:2024cia}
A.~Keefe, N.~Agarwal and A.~Kamal, \emph{{Quantifying spectral signatures of
  non-Markovianity beyond Born-Redfield master equation}},
  \href{https://arxiv.org/abs/2405.01722}{{\ttfamily 2405.01722}}.

\bibitem{Bowen:2024emo}
B.~Bowen, N.~Agarwal and A.~Kamal, \emph{{Open system dynamics in interacting
  quantum field theories}},  \href{https://arxiv.org/abs/2403.18907}{{\ttfamily
  2403.18907}}.

\bibitem{Bhattacharyya:2024duw}
A.~Bhattacharyya, S.~Brahma, S.S.~Haque, J.S.~Lund and A.~Paul, \emph{{The
  Early Universe as an Open Quantum System: Complexity and Decoherence}},
  \href{https://arxiv.org/abs/2401.12134}{{\ttfamily 2401.12134}}.

\bibitem{Salcedo:2024smn}
S.A.~Salcedo, T.~Colas and E.~Pajer, \emph{{The Open Effective Field Theory of
  Inflation}},  \href{https://arxiv.org/abs/2404.15416}{{\ttfamily
  2404.15416}}.

\bibitem{Belfiglio:2024qsa}
A.~Belfiglio, O.~Luongo, S.~Mancini and S.~Tomasi, \emph{{Entanglement entropy
  in quantum black holes}},  \href{https://arxiv.org/abs/2404.00715}{{\ttfamily
  2404.00715}}.

\bibitem{Burgess:2015ajz}
C.P.~Burgess, R.~Holman and G.~Tasinato, \emph{{Open EFTs, IR effects \&
  late-time resummations: systematic corrections in stochastic inflation}},
  \href{https://doi.org/10.1007/JHEP01(2016)153}{\emph{JHEP} {\bfseries 01}
  (2016) 153} [\href{https://arxiv.org/abs/1512.00169}{{\ttfamily
  1512.00169}}].

\bibitem{Kaplanek:2019vzj}
G.~Kaplanek and C.P.~Burgess, \emph{{Hot Cosmic Qubits: Late-Time de Sitter
  Evolution and Critical Slowing Down}},
  \href{https://doi.org/10.1007/JHEP02(2020)053}{\emph{JHEP} {\bfseries 02}
  (2020) 053} [\href{https://arxiv.org/abs/1912.12955}{{\ttfamily
  1912.12955}}].

\bibitem{Kaplanek:2019dqu}
G.~Kaplanek and C.P.~Burgess, \emph{{Hot Accelerated Qubits: Decoherence,
  Thermalization, Secular Growth and Reliable Late-time Predictions}},
  \href{https://doi.org/10.1007/JHEP03(2020)008}{\emph{JHEP} {\bfseries 03}
  (2020) 008} [\href{https://arxiv.org/abs/1912.12951}{{\ttfamily
  1912.12951}}].

\bibitem{Colas:2022hlq}
T.~Colas, J.~Grain and V.~Vennin, \emph{{Benchmarking the cosmological master
  equations}},
  \href{https://doi.org/10.1140/epjc/s10052-022-11047-9}{\emph{Eur. Phys. J. C}
  {\bfseries 82} (2022) 1085}
  [\href{https://arxiv.org/abs/2209.01929}{{\ttfamily 2209.01929}}].

\bibitem{Chaykov:2022zro}
S.~Chaykov, N.~Agarwal, S.~Bahrami and R.~Holman, \emph{{Loop corrections in
  Minkowski spacetime away from equilibrium. Part I. Late-time resummations}},
  \href{https://doi.org/10.1007/JHEP02(2023)093}{\emph{JHEP} {\bfseries 02}
  (2023) 093} [\href{https://arxiv.org/abs/2206.11288}{{\ttfamily
  2206.11288}}].

\bibitem{Burgess:2024eng}
C.P.~Burgess, T.~Colas, R.~Holman, G.~Kaplanek and V.~Vennin, \emph{{Cosmic
  Purity Lost: Perturbative and Resummed Late-Time Inflationary Decoherence}},
  \href{https://arxiv.org/abs/2403.12240}{{\ttfamily 2403.12240}}.

\bibitem{Donath:2024utn}
Y.~Donath and E.~Pajer, \emph{{The In-Out Formalism for In-In Correlators}},
  \href{https://arxiv.org/abs/2402.05999}{{\ttfamily 2402.05999}}.

\bibitem{Serafini:2003ke}
A.~Serafini, F.~Illuminati and S.~De~Siena, \emph{{Von Neumann entropy, mutual
  information and total correlations of Gaussian states}},
  \href{https://doi.org/10.1088/0953-4075/37/2/L02}{\emph{J. Phys. B}
  {\bfseries 37} (2004) L21}
  [\href{https://arxiv.org/abs/quant-ph/0307073}{{\ttfamily
  quant-ph/0307073}}].

\bibitem{Cheung:2007st}
C.~Cheung, P.~Creminelli, A.L.~Fitzpatrick, J.~Kaplan and L.~Senatore,
  \emph{{The Effective Field Theory of Inflation}},
  \href{https://doi.org/10.1088/1126-6708/2008/03/014}{\emph{JHEP} {\bfseries
  03} (2008) 014} [\href{https://arxiv.org/abs/0709.0293}{{\ttfamily
  0709.0293}}].

\bibitem{Tolley:2009fg}
A.J.~Tolley and M.~Wyman, \emph{{The Gelaton Scenario: Equilateral
  non-Gaussianity from multi-field dynamics}},
  \href{https://doi.org/10.1103/PhysRevD.81.043502}{\emph{Phys. Rev. D}
  {\bfseries 81} (2010) 043502}
  [\href{https://arxiv.org/abs/0910.1853}{{\ttfamily 0910.1853}}].

\bibitem{Chen:2009zp}
X.~Chen and Y.~Wang, \emph{{Quasi-Single Field Inflation and
  Non-Gaussianities}},
  \href{https://doi.org/10.1088/1475-7516/2010/04/027}{\emph{JCAP} {\bfseries
  04} (2010) 027} [\href{https://arxiv.org/abs/0911.3380}{{\ttfamily
  0911.3380}}].

\bibitem{Assassi:2013gxa}
V.~Assassi, D.~Baumann, D.~Green and L.~McAllister, \emph{{Planck-Suppressed
  Operators}}, \href{https://doi.org/10.1088/1475-7516/2014/01/033}{\emph{JCAP}
  {\bfseries 01} (2014) 033} [\href{https://arxiv.org/abs/1304.5226}{{\ttfamily
  1304.5226}}].

\bibitem{Jazayeri:2022kjy}
S.~Jazayeri and S.~Renaux-Petel, \emph{{Cosmological bootstrap in slow
  motion}}, \href{https://doi.org/10.1007/JHEP12(2022)137}{\emph{JHEP}
  {\bfseries 12} (2022) 137}
  [\href{https://arxiv.org/abs/2205.10340}{{\ttfamily 2205.10340}}].

\bibitem{Jazayeri:2023xcj}
S.~Jazayeri, S.~Renaux-Petel and D.~Werth, \emph{{Shapes of the Cosmological
  Low-Speed Collider}},  \href{https://arxiv.org/abs/2307.01751}{{\ttfamily
  2307.01751}}.

\bibitem{Weinberg:2005vy}
S.~Weinberg, \emph{{Quantum contributions to cosmological correlations}},
  \href{https://doi.org/10.1103/PhysRevD.72.043514}{\emph{Phys. Rev. D}
  {\bfseries 72} (2005) 043514}
  [\href{https://arxiv.org/abs/hep-th/0506236}{{\ttfamily hep-th/0506236}}].

\bibitem{Weinberg:2006ac}
S.~Weinberg, \emph{{Quantum contributions to cosmological correlations. II. Can
  these corrections become large?}},
  \href{https://doi.org/10.1103/PhysRevD.74.023508}{\emph{Phys. Rev. D}
  {\bfseries 74} (2006) 023508}
  [\href{https://arxiv.org/abs/hep-th/0605244}{{\ttfamily hep-th/0605244}}].

\bibitem{Chen:2017ryl}
X.~Chen, Y.~Wang and Z.-Z.~Xianyu, \emph{{Schwinger-Keldysh Diagrammatics for
  Primordial Perturbations}},
  \href{https://doi.org/10.1088/1475-7516/2017/12/006}{\emph{JCAP} {\bfseries
  12} (2017) 006} [\href{https://arxiv.org/abs/1703.10166}{{\ttfamily
  1703.10166}}].

\bibitem{Adshead:2009cb}
P.~Adshead, R.~Easther and E.A.~Lim, \emph{The 'in-in' {Formalism} and
  {Cosmological} {Perturbations}},
  \href{https://doi.org/10.1103/PhysRevD.80.083521}{\emph{Phys. Rev. D}
  {\bfseries 80} (2009) 083521}.

\bibitem{Kaya:2018jdo}
A.~Kaya, \emph{{On $i\epsilon$ Prescription in Cosmology}},
  \href{https://doi.org/10.1088/1475-7516/2019/04/002}{\emph{JCAP} {\bfseries
  04} (2019) 002} [\href{https://arxiv.org/abs/1810.12324}{{\ttfamily
  1810.12324}}].

\bibitem{Albayrak:2023hie}
S.~Albayrak, P.~Benincasa and C.D.~Pueyo, \emph{{Perturbative Unitarity and the
  Wavefunction of the Universe}},
  \href{https://arxiv.org/abs/2305.19686}{{\ttfamily 2305.19686}}.

\bibitem{renyi1961measures}
A.~R{\'e}nyi, \emph{On measures of entropy and information},  in
  \emph{Proceedings of the fourth Berkeley symposium on mathematical statistics
  and probability, volume 1: contributions to the theory of statistics},
  vol.~4, pp.~547--562, University of California Press, 1961.

\bibitem{Kudler-Flam:2022zgm}
J.~Kudler-Flam, \emph{{R\'enyi Mutual Information in Quantum Field Theory}},
  \href{https://doi.org/10.1103/PhysRevLett.130.021603}{\emph{Phys. Rev. Lett.}
  {\bfseries 130} (2023) 021603}
  [\href{https://arxiv.org/abs/2211.01392}{{\ttfamily 2211.01392}}].

\bibitem{Kudler-Flam:2023kph}
J.~Kudler-Flam, L.~Nie and A.~Vijay, \emph{{R\'enyi mutual information in
  quantum field theory, tensor networks, and gravity}},
  \href{https://arxiv.org/abs/2308.08600}{{\ttfamily 2308.08600}}.

\bibitem{Martin:2022kph}
J.~Martin, A.~Micheli and V.~Vennin, \emph{{Comparing quantumness criteria}},
  \href{https://doi.org/10.1209/0295-5075/acc3be}{\emph{EPL} {\bfseries 142}
  (2023) 18001} [\href{https://arxiv.org/abs/2211.10114}{{\ttfamily
  2211.10114}}].

\bibitem{Colas:2021llj}
T.~Colas, J.~Grain and V.~Vennin, \emph{{Four-mode squeezed states: two-field
  quantum systems and the symplectic group $\mathrm {Sp}(4,{\mathbb {R}})$}},
  \href{https://doi.org/10.1140/epjc/s10052-021-09922-y}{\emph{Eur. Phys. J. C}
  {\bfseries 82} (2022) 6} [\href{https://arxiv.org/abs/2104.14942}{{\ttfamily
  2104.14942}}].

\bibitem{Maldacena:2002vr}
J.M.~Maldacena, \emph{Non-{Gaussian} features of primordial fluctuations in
  single field inflationary models},
  \href{https://doi.org/10.1088/1126-6708/2003/05/013}{\emph{JHEP} {\bfseries
  05} (2003) 013}.

\bibitem{Burgess:2018sou}
C.P.~Burgess, J.~Hainge, G.~Kaplanek and M.~Rummel, \emph{{Failure of
  Perturbation Theory Near Horizons: the Rindler Example}},
  \href{https://doi.org/10.1007/JHEP10(2018)122}{\emph{JHEP} {\bfseries 10}
  (2018) 122} [\href{https://arxiv.org/abs/1806.11415}{{\ttfamily
  1806.11415}}].

\bibitem{Akhmedov:2020haq}
E.T.~Akhmedov and O.~Diatlyk, \emph{{Secularly growing loop corrections in
  scalar wave background}},
  \href{https://doi.org/10.1007/JHEP10(2020)027}{\emph{JHEP} {\bfseries 10}
  (2020) 027} [\href{https://arxiv.org/abs/2004.01544}{{\ttfamily
  2004.01544}}].

\bibitem{Tsamis:1993ub}
N.C.~Tsamis and R.P.~Woodard, \emph{{The Physical basis for infrared
  divergences in inflationary quantum gravity}},
  \href{https://doi.org/10.1088/0264-9381/11/12/012}{\emph{Class. Quant. Grav.}
  {\bfseries 11} (1994) 2969}.

\bibitem{Starobinsky:1994bd}
A.A.~Starobinsky and J.~Yokoyama, \emph{Equilibrium state of a selfinteracting
  scalar field in the {De} {Sitter} background},
  \href{https://doi.org/10.1103/PhysRevD.50.6357}{\emph{Phys. Rev.} {\bfseries
  D50} (1994) 6357}.

\bibitem{Onemli:2002hr}
V.K.~Onemli and R.P.~Woodard, \emph{{Superacceleration from massless, minimally
  coupled phi**4}},
  \href{https://doi.org/10.1088/0264-9381/19/17/311}{\emph{Class. Quant. Grav.}
  {\bfseries 19} (2002) 4607}
  [\href{https://arxiv.org/abs/gr-qc/0204065}{{\ttfamily gr-qc/0204065}}].

\bibitem{Tsamis:2005hd}
N.C.~Tsamis and R.P.~Woodard, \emph{{Stochastic quantum gravitational
  inflation}},
  \href{https://doi.org/10.1016/j.nuclphysb.2005.06.031}{\emph{Nucl. Phys. B}
  {\bfseries 724} (2005) 295}
  [\href{https://arxiv.org/abs/gr-qc/0505115}{{\ttfamily gr-qc/0505115}}].

\bibitem{Riotto:2008mv}
A.~Riotto and M.S.~Sloth, \emph{{On Resumming Inflationary Perturbations beyond
  One-loop}}, \href{https://doi.org/10.1088/1475-7516/2008/04/030}{\emph{JCAP}
  {\bfseries 04} (2008) 030} [\href{https://arxiv.org/abs/0801.1845}{{\ttfamily
  0801.1845}}].

\bibitem{Burgess:2009bs}
C.P.~Burgess, L.~Leblond, R.~Holman and S.~Shandera, \emph{{Super-Hubble de
  Sitter Fluctuations and the Dynamical RG}},
  \href{https://doi.org/10.1088/1475-7516/2010/03/033}{\emph{JCAP} {\bfseries
  03} (2010) 033} [\href{https://arxiv.org/abs/0912.1608}{{\ttfamily
  0912.1608}}].

\bibitem{Marolf:2010zp}
D.~Marolf and I.A.~Morrison, \emph{{The IR stability of de Sitter: Loop
  corrections to scalar propagators}},
  \href{https://doi.org/10.1103/PhysRevD.82.105032}{\emph{Phys. Rev. D}
  {\bfseries 82} (2010) 105032}
  [\href{https://arxiv.org/abs/1006.0035}{{\ttfamily 1006.0035}}].

\bibitem{Burgess:2010dd}
C.P.~Burgess, R.~Holman, L.~Leblond and S.~Shandera, \emph{{Breakdown of
  Semiclassical Methods in de Sitter Space}},
  \href{https://doi.org/10.1088/1475-7516/2010/10/017}{\emph{JCAP} {\bfseries
  10} (2010) 017} [\href{https://arxiv.org/abs/1005.3551}{{\ttfamily
  1005.3551}}].

\bibitem{Gorbenko:2019rza}
V.~Gorbenko and L.~Senatore, \emph{{$\lambda \phi^4$ in dS}},
  \href{https://arxiv.org/abs/1911.00022}{{\ttfamily 1911.00022}}.

\bibitem{Green:2020txs}
D.~Green and A.~Premkumar, \emph{{Dynamical RG and Critical Phenomena in de
  Sitter Space}}, \href{https://doi.org/10.1007/JHEP04(2020)064}{\emph{JHEP}
  {\bfseries 04} (2020) 064}
  [\href{https://arxiv.org/abs/2001.05974}{{\ttfamily 2001.05974}}].

\bibitem{Cespedes:2023aal}
S.~C\'espedes, A.-C.~Davis and D.-G.~Wang, \emph{{On the IR divergences in de
  Sitter space: loops, resummation and the semi-classical wavefunction}},
  \href{https://doi.org/10.1007/JHEP04(2024)004}{\emph{JHEP} {\bfseries 04}
  (2024) 004} [\href{https://arxiv.org/abs/2311.17990}{{\ttfamily
  2311.17990}}].

\bibitem{Akhmedov:2015xwa}
E.T.~Akhmedov, H.~Godazgar and F.K.~Popov, \emph{{Hawking radiation and
  secularly growing loop corrections}},
  \href{https://doi.org/10.1103/PhysRevD.93.024029}{\emph{Phys. Rev. D}
  {\bfseries 93} (2016) 024029}
  [\href{https://arxiv.org/abs/1508.07500}{{\ttfamily 1508.07500}}].

\bibitem{breuerTimelocalMasterEquations2002}
H.-P.~Breuer, A.~Ma and F.~Petruccione, \emph{Time-local master equations:
  influence functional and cumulant expansion},
  \href{https://arxiv.org/abs/quant-ph/0209153}{{\ttfamily quant-ph/0209153}}.

\bibitem{2010ChPhB..19d0308C}
X.-Y.~{Chen}, \emph{{Perturbation theory of von Neumann entropy}},
  \href{https://doi.org/10.1088/1674-1056/19/4/040308}{\emph{Chinese Physics B}
  {\bfseries 19} (2010) 040308}
  [\href{https://arxiv.org/abs/0902.4733}{{\ttfamily 0902.4733}}].

\bibitem{Balasubramanian:2011wt}
V.~Balasubramanian, M.B.~McDermott and M.~Van~Raamsdonk, \emph{{Momentum-space
  entanglement and renormalization in quantum field theory}},
  \href{https://doi.org/10.1103/PhysRevD.86.045014}{\emph{Phys. Rev. D}
  {\bfseries 86} (2012) 045014}
  [\href{https://arxiv.org/abs/1108.3568}{{\ttfamily 1108.3568}}].

\bibitem{Wong:2013gua}
G.~Wong, I.~Klich, L.A.~Pando~Zayas and D.~Vaman, \emph{{Entanglement
  Temperature and Entanglement Entropy of Excited States}},
  \href{https://doi.org/10.1007/JHEP12(2013)020}{\emph{JHEP} {\bfseries 12}
  (2013) 020} [\href{https://arxiv.org/abs/1305.3291}{{\ttfamily 1305.3291}}].

\bibitem{Rodrigues:2019cnb}
S.~Rodrigues, Franklin L.~\, G.~De~Chiara, M.~Paternostro and G.T.~Landi,
  \emph{{Thermodynamics of Weakly Coherent Collisional Models}},
  \href{https://doi.org/10.1103/PhysRevLett.123.140601}{\emph{Phys. Rev. Lett.}
  {\bfseries 123} (2019) 140601}.

\bibitem{Tomaras:2019sjq}
T.N.~Tomaras and N.~Toumbas, \emph{{IR dynamics and entanglement entropy}},
  \href{https://doi.org/10.1103/PhysRevD.101.065006}{\emph{Phys. Rev. D}
  {\bfseries 101} (2020) 065006}
  [\href{https://arxiv.org/abs/1910.07847}{{\ttfamily 1910.07847}}].

\bibitem{Chen:2020ild}
Y.~Chen, L.~Hackl, R.~Kunjwal, H.~Moradi, Y.K.~Yazdi and M.~Zilh\~ao,
  \emph{{Towards spacetime entanglement entropy for interacting theories}},
  \href{https://doi.org/10.1007/JHEP11(2020)114}{\emph{JHEP} {\bfseries 11}
  (2020) 114} [\href{https://arxiv.org/abs/2002.00966}{{\ttfamily
  2002.00966}}].

\bibitem{Dadras:2020xfl}
P.~Dadras and A.~Kitaev, \emph{{Perturbative calculations of entanglement
  entropy}}, \href{https://doi.org/10.1007/JHEP03(2021)198}{\emph{JHEP}
  {\bfseries 03} (2021) 198}
  [\href{https://arxiv.org/abs/2011.09622}{{\ttfamily 2011.09622}}].

\bibitem{Grace:2021jsf}
M.R.~Grace and S.~Guha, \emph{{Perturbation Theory for Quantum Information}},
  \href{https://arxiv.org/abs/2106.05533}{{\ttfamily 2106.05533}}.

\bibitem{Fedida:2024dwc}
S.~Fedida, A.~Mazumdar, S.~Bose and A.~Serafini, \emph{{Entanglement entropy in
  scalar quantum electrodynamics}},
  \href{https://doi.org/10.1103/PhysRevD.109.065028}{\emph{Phys. Rev. D}
  {\bfseries 109} (2024) 065028}
  [\href{https://arxiv.org/abs/2401.10332}{{\ttfamily 2401.10332}}].

\bibitem{Holzhey:1994we}
C.~Holzhey, F.~Larsen and F.~Wilczek, \emph{{Geometric and renormalized entropy
  in conformal field theory}},
  \href{https://doi.org/10.1016/0550-3213(94)90402-2}{\emph{Nucl. Phys. B}
  {\bfseries 424} (1994) 443}
  [\href{https://arxiv.org/abs/hep-th/9403108}{{\ttfamily hep-th/9403108}}].

\bibitem{Calabrese:2004eu}
P.~Calabrese and J.L.~Cardy, \emph{{Entanglement entropy and quantum field
  theory}}, \href{https://doi.org/10.1088/1742-5468/2004/06/P06002}{\emph{J.
  Stat. Mech.} {\bfseries 0406} (2004) P06002}
  [\href{https://arxiv.org/abs/hep-th/0405152}{{\ttfamily hep-th/0405152}}].

\bibitem{Creminelli:2006xe}
P.~Creminelli, M.A.~Luty, A.~Nicolis and L.~Senatore, \emph{{Starting the
  Universe: Stable Violation of the Null Energy Condition and Non-standard
  Cosmologies}},
  \href{https://doi.org/10.1088/1126-6708/2006/12/080}{\emph{JHEP} {\bfseries
  12} (2006) 080} [\href{https://arxiv.org/abs/hep-th/0606090}{{\ttfamily
  hep-th/0606090}}].

\bibitem{Senatore:2010wk}
L.~Senatore and M.~Zaldarriaga, \emph{{The Effective Field Theory of Multifield
  Inflation}}, \href{https://doi.org/10.1007/JHEP04(2012)024}{\emph{JHEP}
  {\bfseries 04} (2012) 024} [\href{https://arxiv.org/abs/1009.2093}{{\ttfamily
  1009.2093}}].

\bibitem{Piazza:2013coa}
F.~Piazza and F.~Vernizzi, \emph{{Effective Field Theory of Cosmological
  Perturbations}},
  \href{https://doi.org/10.1088/0264-9381/30/21/214007}{\emph{Class. Quant.
  Grav.} {\bfseries 30} (2013) 214007}
  [\href{https://arxiv.org/abs/1307.4350}{{\ttfamily 1307.4350}}].

\bibitem{Noumi:2012vr}
T.~Noumi, M.~Yamaguchi and D.~Yokoyama, \emph{{Effective field theory approach
  to quasi-single field inflation and effects of heavy fields}},
  \href{https://doi.org/10.1007/JHEP06(2013)051}{\emph{JHEP} {\bfseries 06}
  (2013) 051} [\href{https://arxiv.org/abs/1211.1624}{{\ttfamily 1211.1624}}].

\bibitem{Tong:2017iat}
X.~Tong, Y.~Wang and S.~Zhou, \emph{{On the Effective Field Theory for
  Quasi-Single Field Inflation}},
  \href{https://doi.org/10.1088/1475-7516/2017/11/045}{\emph{JCAP} {\bfseries
  11} (2017) 045} [\href{https://arxiv.org/abs/1708.01709}{{\ttfamily
  1708.01709}}].

\bibitem{An:2017hlx}
H.~An, M.~McAneny, A.K.~Ridgway and M.B.~Wise, \emph{{Quasi Single Field
  Inflation in the non-perturbative regime}},
  \href{https://doi.org/10.1007/JHEP06(2018)105}{\emph{JHEP} {\bfseries 06}
  (2018) 105} [\href{https://arxiv.org/abs/1706.09971}{{\ttfamily
  1706.09971}}].

\bibitem{Kim:2021pbr}
S.~Kim, T.~Noumi, K.~Takeuchi and S.~Zhou, \emph{{Perturbative unitarity in
  quasi-single field inflation}},
  \href{https://doi.org/10.1007/JHEP07(2021)018}{\emph{JHEP} {\bfseries 07}
  (2021) 018} [\href{https://arxiv.org/abs/2102.04101}{{\ttfamily
  2102.04101}}].

\bibitem{Pimentel:2022fsc}
G.L.~Pimentel and D.-G.~Wang, \emph{{Boostless cosmological collider
  bootstrap}}, \href{https://doi.org/10.1007/JHEP10(2022)177}{\emph{JHEP}
  {\bfseries 10} (2022) 177}
  [\href{https://arxiv.org/abs/2205.00013}{{\ttfamily 2205.00013}}].

\bibitem{Wang:2022eop}
D.-G.~Wang, G.L.~Pimentel and A.~Ach\'ucarro, \emph{{Bootstrapping multi-field
  inflation: non-Gaussianities from light scalars revisited}},
  \href{https://doi.org/10.1088/1475-7516/2023/05/043}{\emph{JCAP} {\bfseries
  05} (2023) 043} [\href{https://arxiv.org/abs/2212.14035}{{\ttfamily
  2212.14035}}].

\bibitem{Arkani-Hamed:2015bza}
N.~Arkani-Hamed and J.~Maldacena, \emph{{Cosmological Collider Physics}},
  \href{https://arxiv.org/abs/1503.08043}{{\ttfamily 1503.08043}}.

\bibitem{Mukhanov:1985rz}
V.F.~Mukhanov, \emph{{Gravitational Instability of the Universe Filled with a
  Scalar Field}}, {\emph{JETP Lett.} {\bfseries 41} (1985) 493}.

\bibitem{Mukhanov:1988jd}
V.F.~Mukhanov, \emph{{Quantum Theory of Gauge Invariant Cosmological
  Perturbations}}, {\emph{Sov. Phys. JETP} {\bfseries 67} (1988) 1297}.

\bibitem{Wei:2004xx}
H.~Wei, R.-G.~Cai and A.~Wang, \emph{{Second-order corrections to the power
  spectrum in the slow-roll expansion with a time-dependent sound speed}},
  \href{https://doi.org/10.1016/j.physletb.2004.10.034}{\emph{Phys. Lett. B}
  {\bfseries 603} (2004) 95}
  [\href{https://arxiv.org/abs/hep-th/0409130}{{\ttfamily hep-th/0409130}}].

\bibitem{Planck:2018jri}
{\scshape Planck} collaboration, \emph{{Planck 2018 results. X. Constraints on
  inflation}}, \href{https://doi.org/10.1051/0004-6361/201833887}{\emph{Astron.
  Astrophys.} {\bfseries 641} (2020) A10}
  [\href{https://arxiv.org/abs/1807.06211}{{\ttfamily 1807.06211}}].

\bibitem{Braglia:2024zsl}
M.~Braglia and L.~Pinol, \emph{{No time to derive: unraveling total time
  derivatives in in-in perturbation theory}},
  \href{https://arxiv.org/abs/2403.14558}{{\ttfamily 2403.14558}}.

\bibitem{Bunch:1978yq}
T.S.~Bunch and P.C.W.~Davies, \emph{Quantum {Field} {Theory} in de {Sitter}
  {Space}: {Renormalization} by {Point} {Splitting}},
  \href{https://doi.org/10.1098/rspa.1978.0060}{\emph{Proc. Roy. Soc. Lond.}
  {\bfseries A360} (1978) 117}.

\bibitem{Grain:2019vnq}
J.~Grain and V.~Vennin, \emph{{Canonical transformations and squeezing
  formalism in cosmology}},
  \href{https://doi.org/10.1088/1475-7516/2020/02/022}{\emph{JCAP} {\bfseries
  02} (2020) 022} [\href{https://arxiv.org/abs/1910.01916}{{\ttfamily
  1910.01916}}].

\bibitem{Mirbabayi:2015hva}
M.~Mirbabayi and M.~Simonovi\'c, \emph{{Effective Theory of Squeezed
  Correlation Functions}},
  \href{https://doi.org/10.1088/1475-7516/2016/03/056}{\emph{JCAP} {\bfseries
  03} (2016) 056} [\href{https://arxiv.org/abs/1507.04755}{{\ttfamily
  1507.04755}}].

\bibitem{Cohen:2020php}
T.~Cohen and D.~Green, \emph{{Soft de Sitter Effective Theory}},
  \href{https://doi.org/10.1007/JHEP12(2020)041}{\emph{JHEP} {\bfseries 12}
  (2020) 041} [\href{https://arxiv.org/abs/2007.03693}{{\ttfamily
  2007.03693}}].

\bibitem{Tong:2023krn}
X.~Tong, Y.~Wang, C.~Zhang and Y.~Zhu, \emph{{BCS in the Sky: Signatures of
  Inflationary Fermion Condensation}},
  \href{https://arxiv.org/abs/2304.09428}{{\ttfamily 2304.09428}}.

\bibitem{Anber:2009ua}
M.M.~Anber and L.~Sorbo, \emph{{Naturally inflating on steep potentials through
  electromagnetic dissipation}},
  \href{https://doi.org/10.1103/PhysRevD.81.043534}{\emph{Phys. Rev. D}
  {\bfseries 81} (2010) 043534}
  [\href{https://arxiv.org/abs/0908.4089}{{\ttfamily 0908.4089}}].

\bibitem{Bordin:2018pca}
L.~Bordin, P.~Creminelli, A.~Khmelnitsky and L.~Senatore, \emph{{Light
  Particles with Spin in Inflation}},
  \href{https://doi.org/10.1088/1475-7516/2018/10/013}{\emph{JCAP} {\bfseries
  10} (2018) 013} [\href{https://arxiv.org/abs/1806.10587}{{\ttfamily
  1806.10587}}].

\bibitem{Lee:2016vti}
H.~Lee, D.~Baumann and G.L.~Pimentel, \emph{{Non-Gaussianity as a Particle
  Detector}}, \href{https://doi.org/10.1007/JHEP12(2016)040}{\emph{JHEP}
  {\bfseries 12} (2016) 040}
  [\href{https://arxiv.org/abs/1607.03735}{{\ttfamily 1607.03735}}].

\bibitem{Peloso:2022ovc}
M.~Peloso and L.~Sorbo, \emph{{Instability in axion inflation with strong
  backreaction from gauge modes}},
  \href{https://doi.org/10.1088/1475-7516/2023/01/038}{\emph{JCAP} {\bfseries
  01} (2023) 038} [\href{https://arxiv.org/abs/2209.08131}{{\ttfamily
  2209.08131}}].

\bibitem{1983LNP...190.....K}
K.~{Kraus}, A.~{B{\"o}hm}, J.D.~{Dollard} and W.H.~{Wootters}, \emph{{States,
  Effects, and Operations Fundamental Notions of Quantum Theory}}, vol.~190,
  Springer (1983),
  \href{https://doi.org/10.1007/3-540-12732-1}{10.1007/3-540-12732-1}.

\bibitem{Green:2023ids}
D.~Green, Y.~Huang, C.-H.~Shen and D.~Baumann, \emph{{Positivity from
  Cosmological Correlators}},
  \href{https://doi.org/10.1007/JHEP04(2024)034}{\emph{JHEP} {\bfseries 04}
  (2024) 034} [\href{https://arxiv.org/abs/2310.02490}{{\ttfamily
  2310.02490}}].

\bibitem{Aoude:2024xpx}
R.~Aoude, G.~Elor, G.N.~Remmen and O.~Sumensari, \emph{{Positivity in
  Amplitudes from Quantum Entanglement}},
  \href{https://arxiv.org/abs/2402.16956}{{\ttfamily 2402.16956}}.

\bibitem{Baumann:2022jpr}
D.~Baumann, D.~Green, A.~Joyce, E.~Pajer, G.L.~Pimentel, C.~Sleight et~al.,
  \emph{{Snowmass White Paper: The Cosmological Bootstrap}},  in
  \emph{{Snowmass 2021}}, 3, 2022
  [\href{https://arxiv.org/abs/2203.08121}{{\ttfamily 2203.08121}}].

\bibitem{Goodhew:2020hob}
H.~Goodhew, S.~Jazayeri and E.~Pajer, \emph{{The Cosmological Optical
  Theorem}}, \href{https://doi.org/10.1088/1475-7516/2021/04/021}{\emph{JCAP}
  {\bfseries 04} (2021) 021}
  [\href{https://arxiv.org/abs/2009.02898}{{\ttfamily 2009.02898}}].

\bibitem{Goodhew:2021oqg}
H.~Goodhew, S.~Jazayeri, M.H.~Gordon~Lee and E.~Pajer, \emph{{Cutting
  cosmological correlators}},
  \href{https://doi.org/10.1088/1475-7516/2021/08/003}{\emph{JCAP} {\bfseries
  08} (2021) 003} [\href{https://arxiv.org/abs/2104.06587}{{\ttfamily
  2104.06587}}].

\bibitem{Cespedes:2020xqq}
S.~C\'espedes, A.-C.~Davis and S.~Melville, \emph{{On the time evolution of
  cosmological correlators}},
  \href{https://doi.org/10.1007/JHEP02(2021)012}{\emph{JHEP} {\bfseries 02}
  (2021) 012} [\href{https://arxiv.org/abs/2009.07874}{{\ttfamily
  2009.07874}}].

\bibitem{LopezNacir:2011kk}
D.~Lopez~Nacir, R.A.~Porto, L.~Senatore and M.~Zaldarriaga, \emph{{Dissipative
  effects in the Effective Field Theory of Inflation}},
  \href{https://doi.org/10.1007/JHEP01(2012)075}{\emph{JHEP} {\bfseries 01}
  (2012) 075} [\href{https://arxiv.org/abs/1109.4192}{{\ttfamily 1109.4192}}].

\bibitem{Baidya:2017eho}
A.~Baidya, C.~Jana, R.~Loganayagam and A.~Rudra, \emph{{Renormalization in open
  quantum field theory. Part I. Scalar field theory}},
  \href{https://doi.org/10.1007/JHEP11(2017)204}{\emph{JHEP} {\bfseries 11}
  (2017) 204} [\href{https://arxiv.org/abs/1704.08335}{{\ttfamily
  1704.08335}}].

\bibitem{Avinash:2019qga}
Avinash, C.~Jana and A.~Rudra, \emph{{Renormalisation in Open Quantum Field
  theory II: Yukawa theory and PV reduction}},
  \href{https://arxiv.org/abs/1906.10180}{{\ttfamily 1906.10180}}.

\bibitem{Agon:2014uxa}
C.~Agon, V.~Balasubramanian, S.~Kasko and A.~Lawrence, \emph{{Coarse Grained
  Quantum Dynamics}},
  \href{https://doi.org/10.1103/PhysRevD.98.025019}{\emph{Phys. Rev. D}
  {\bfseries 98} (2018) 025019}
  [\href{https://arxiv.org/abs/1412.3148}{{\ttfamily 1412.3148}}].

\bibitem{Agon:2017oia}
C.~Ag\'on and A.~Lawrence, \emph{{Divergences in open quantum systems}},
  \href{https://doi.org/10.1007/JHEP04(2018)008}{\emph{JHEP} {\bfseries 04}
  (2018) 008} [\href{https://arxiv.org/abs/1709.10095}{{\ttfamily
  1709.10095}}].

\bibitem{Winczewski:2021bpr}
M.~Winczewski and R.~Alicki, \emph{{Renormalization in the Theory of Open
  Quantum Systems via the Self-Consistency Condition}},
  \href{https://arxiv.org/abs/2112.11962}{{\ttfamily 2112.11962}}.

\bibitem{Correa:2023pwg}
L.A.~Correa and J.~Glatthard, \emph{{Potential renormalisation, Lamb shift and
  mean-force Gibbs state -- to shift or not to shift?}},
  \href{https://arxiv.org/abs/2305.08941}{{\ttfamily 2305.08941}}.

\bibitem{Crowder:2023zwq}
E.~Crowder, L.~Lampert, G.~Manchanda, B.~Shoffeitt, S.~Gadamsetty, Y.~Pei
  et~al., \emph{{Invalidation of the Bloch-Redfield equation~in the sub-Ohmic
  regime via a practical time-convolutionless fourth-order master equation}},
  \href{https://doi.org/10.1103/PhysRevA.109.052205}{\emph{Phys. Rev. A}
  {\bfseries 109} (2024) 052205}
  [\href{https://arxiv.org/abs/2310.15089}{{\ttfamily 2310.15089}}].

\bibitem{Melville:2021lst}
S.~Melville and E.~Pajer, \emph{{Cosmological Cutting Rules}},
  \href{https://doi.org/10.1007/JHEP05(2021)249}{\emph{JHEP} {\bfseries 05}
  (2021) 249} [\href{https://arxiv.org/abs/2103.09832}{{\ttfamily
  2103.09832}}].

\end{thebibliography}\endgroup

\end{document}